\documentclass[longauth]{aa}  

\usepackage{amsmath,amssymb}
\usepackage[nopatch]{microtype}
\usepackage{booktabs}
\usepackage{footmisc}

\bibpunct{(}{)}{;}{a}{}{,} 

\usepackage{booktabs} 
\usepackage{threeparttable,longtable}  
\usepackage{makecell}
\usepackage{cprotect}
\usepackage{longtable}
\usepackage{threeparttable}
\usepackage{multicol}
\usepackage{multirow}
\usepackage{gensymb}

\usepackage{upgreek}
\usepackage{orcidlink}

\usepackage{rotating}

\definecolor{dodgerblue}{RGB}{30, 144, 255}
\definecolor{cpink}{RGB}{243, 141, 252}
\newcommand{\XMM}[1]{\textit{XMM-Newton}}

\usepackage{xpatch}
\usepackage{hyperref}
\hypersetup{
	colorlinks=true,
	breaklinks=true,
	citecolor=blue,
	allcolors=blue,
	frenchlinks=true
}

\makeatletter
\xpatchcmd\NAT@citex
{%
	\@citea\NAT@hyper@{%
		\NAT@nmfmt{\NAT@nm}%
		\hyper@natlinkbreak{\NAT@aysep\NAT@spacechar}{\@citeb\@extra@b@citeb}%
		\NAT@date
	}%
}
{%
	\@citea
	\NAT@nmfmt{\NAT@nm}%
	\NAT@aysep\NAT@spacechar
	\NAT@hyper@{\NAT@date}%
}
{}{}
\xpatchcmd\NAT@citex
{%
	\@citea\NAT@hyper@{%
		\NAT@nmfmt{\NAT@nm}%
		\hyper@natlinkbreak{\NAT@spacechar\NAT@@open\if*#1*\else#1\NAT@spacechar\fi}%
		{\@citeb\@extra@b@citeb}%
		\NAT@date
	}%
}
{
	\@citea
	\NAT@nmfmt{\NAT@nm}%
	\NAT@spacechar\NAT@@open\if*#1*\else#1\NAT@spacechar\fi
	\NAT@hyper@{\NAT@date}%
}
{}{}
\makeatother

\makeatletter
\renewcommand*\aa@pageof{, page \thepage{} of \pageref*{LastPage}}
\makeatother

\begin{document}

\title{Relighting the fire in Hickson Compact Group (HCG) 15: magnetised fossil plasma revealed by the SKA Pathfinders \& Precursors}

   \author{C.~J.~Riseley\,\orcidlink{0000-0002-3369-1085}\,\inst{1,2,3},
        T.~Vernstrom\,\orcidlink{0000-0001-7093-3875}\,\inst{4,5},
        L.~Lovisari\,\orcidlink{0000-0002-3754-2415}\,\inst{6,7}
        E.~O'Sullivan\,\orcidlink{0000-0002-5671-6900}\,\inst{7},
        F.~Gastaldello\,\orcidlink{0000-0002-9112-0184}\,\inst{6},
        M.~Brienza\,\orcidlink{0000-0003-4120-9970}\,\inst{3},
        Prasanta~K.~Nayak\,\orcidlink{0000-0002-4638-1035}\,\inst{8},
        A.~Bonafede\,\orcidlink{0000-0002-5068-4581}\,\inst{2,3},
        E.~Carretti\,\orcidlink{0000-0002-3973-8403}\,\inst{3},
        S.~W.~Duchesne\,\orcidlink{0000-0002-3846-0315}\,\inst{4},
        S.~Giacintucci\,\orcidlink{0000-0002-1634-9886}\,\inst{9},
        A.~M.~Hopkins\,\orcidlink{0000-0002-6097-2747}\,\inst{10},
        B.~S.~Koribalski\,\orcidlink{0000-0003-4351-993X}\,\inst{11,12},
        F.~Loi\,\orcidlink{0000-0002-8627-6627}\,\inst{13},
        C.~Pfrommer\,\orcidlink{0000-0002-7275-3998}\,\inst{14},
        W.~Raja,\,\inst{11},
        K.~Ross\,\orcidlink{0000-0002-8666-6588}\,\inst{15},
        K.~Rubinur\,\orcidlink{0000-0001-5574-5104}\,\inst{16},
        M.~Ruszkowski\,\orcidlink{0009-0002-2669-9908}\,\inst{17},
        T.~W.~Shimwell\,\inst{18,19},
        M.~S.~de~Villiers\,\orcidlink{0000-0001-5628-0417}\,\inst{26},
        J.~West\,\orcidlink{0000-0001-7722-8458}\,\inst{20},
        H.~R.~M.~Zovaro\,\orcidlink{0000-0003-4334-9811}\,\inst{21,22},
        T.~Akahori\,\orcidlink{0000-0001-9399-5331}\,\inst{23},
        C.~S.~Anderson\,\orcidlink{0000-0002-6243-7879}\,\inst{22},
        D.~J.~Bomans\,\orcidlink{0000-0001-5126-5365}\,\inst{1},
        A.~Drabent\,\orcidlink{0000-0003-2792-1793}\,\inst{24},
        L.~Rudnick\,\orcidlink{0000-0001-5636-7213}\,\inst{25},
        R.~Santra\,\orcidlink{0009-0002-0373-570X}\,\inst{27}
    }
    \authorrunning{C. J. Riseley et al.}
    \titlerunning{Magnetised fossil plasma in HCG15}
    \institute{
        Astronomisches Institut der Ruhr-Universit\"{a}t Bochum (AIRUB), Universit\"{a}tsstra{\ss}e 150, 44801 Bochum, Germany\\
        \email{riseley@astro.ruhr-uni-bochum.de}
        \and 
        Dipartimento di Fisica e Astronomia, Universit\`a degli Studi di Bologna, via P. Gobetti 93/2, 40129 Bologna, Italy
        \and 
        INAF -- Istituto di Radioastronomia, via P. Gobetti 101, 40129 Bologna, Italy
        \and 
        CSIRO Space \& Astronomy, PO Box 1130, Bentley, WA 6102, Australia
        \and 
        ICRAR, The University of Western Australia, 35 Stirling Hw, 6009 Crawley, Australia\\
        \email{tessa.vernstrom@csiro.au}
        \and 
        INAF -- IASF Istituto di Astrofisica Spaziale e Fisica Cosmica di Milano, Via Alfonso Corti 12, 20133 Milano, Italy
        \and 
        Center for Astrophysics | Harvard \& Smithsonian, 60 Garden Street, Cambridge, MA 02138, USA
        \and 
        Instituto de Astrof\'{i}sica, Pontificia Universidad Cat\'{o}lica de Chile, Av. Vicu\~{n}a MacKenna 4860, 7820436, Santiago, Chile
        \and 
        Naval Research Laboratory, 4555 Overlook Avenue SW, Code 7213, Washington, DC 20375, USA
        \and 
        Australian Astronomical Optics, Macquarie University, 105 Delhi Rd, North Ryde, NSW 2113, Australia
        \and 
        Australia Telescope National Facility (ATNF), CSIRO, Space and Astronomy, PO Box 76, Epping, NSW 1710, Australia
        \and 
        School of Science, Western Sydney University, Locked Bag 1797, Penrith, NSW 2751, Australia
        \and 
        INAF -- Osservatorio Astronomico di Cagliari, Via della Scienza 5, I-09047 Selargius (CA), Italy
        \and 
        Leibniz Institute for Astrophysics Potsdam (AIP), An der Sternwarte 16, 14482 Potsdam, Germany
        \and 
        International Centre for Radio Astronomy Research (ICRAR), Curtin University, Bentley, WA 6102, Australia
        \and 
        Institute of Theoretical Astrophysics, University of Oslo, PO box 1029 Blindern, Oslo 0315, Norway
        \and 
        Department of Astronomy, University of Michigan, 1085 S. University Ave., 323 West Hall, Ann Arbor, MI 48109-1107, USA
        \and 
        ASTRON, The Netherlands Institute for Radio Astronomy, Postbus 2, 7990 AA Dwingeloo, The Netherlands
        \and 
        Leiden Observatory, Leiden University, PO Box 9513, 2300 RA Leiden, The Netherlands
        \and 
        National Research Council Canada, Herzberg Research Centre for Astronomy and Astrophysics, Dominion Radio Astrophysical Observatory, Penticton, Canada
        \and 
        Research School of Astronomy and Astrophysics, The Australian National University, Canberra, ACT 2611, Australia
        \and 
        ARC Centre of Excellence for All Sky Astrophysics in 3~Dimensions (ASTRO 3D), Canberra, ACT 2611, Australia
        \and 
        Mizusawa VLBI Observatory, National Astronomical Observatory of Japan, 2-21-1 Osawa, Mitaka, Tokyo 181-8588, Japan
        \and 
        Th\"{u}ringer Landessternwarte, Sternwarte 5, D-07778 Tautenburg
        \and 
        Minnesota Institute for Astrophysics, University of Minnesota, 116 Church St. SE, Minneapolis, MN 55455, USA
        \and 
        South African Radio Astronomy Observatory, 2 Fir Street, Black River Park, Observatory, 7925, RSA
        \and 
        National Centre for Radio Astrophysics, Tata Institute of Fundamental Research, Pune 411007, India
    }

   \date{Received: 4~Mar~2025; accepted: 10~Mar~2025; in original form: 24~Jan~2025}

\abstract
    {In the context of the life cycle and evolution of active galactic nuclei (AGN), the environment plays an important role. In particular, the over-dense environments of galaxy groups, where dynamical interactions and bulk motions have significant impact, offer an excellent but under-explored window into the life cycles of AGN and the processes that shape the evolution of relativistic plasma.
    Pilot Survey observations with the Australian Square Kilometre Array Pathfinder (ASKAP) Evolutionary Map of the Universe (EMU) survey recovered diffuse emission associated with the nearby ($z = 0.0228$) galaxy group HCG15, which was revealed to be strongly linearly polarised. We study the properties of this emission in unprecedented detail to settle open questions about its nature and its relation to the group-member galaxies.
    We perform a multi-frequency spectropolarimetric study of HCG15 incorporating our ASKAP EMU observations as well as new data from MeerKAT, the LOw-Frequency ARray (LOFAR), the Giant Metrewave Radio Telescope (GMRT), and the Karl G. Jansky Very Large Array (VLA), plus X-ray data from \textit{XMM-Newton} and optical spectra from the Himalayan Chandra Telescope (HCT).
    Our study confirms that the diffuse structure represents remnant emission from historic AGN activity, likely associated with HCG15-D, some $80 - 86$~Myr ago (based on ageing analysis). We detect significant highly linearly-polarised emission from a diffuse `ridge'-like structure with a highly ordered magnetic field. Our analysis suggests that this emission is generated by draping of magnetic field lines in the intra-group medium (IGrM), although further exploration with simulations would aid our understanding. We confirm that HCG15-C is a group-member galaxy. Finally, we report the detection of thermal emission associated with a background cluster at redshift $z \approx 0.87$ projected onto the IGrM of HCG15, which matches the position and redshift of the recent Sunyaev-Zel'dovich (SZ) detection of ACT-CL~J0207.8+0209.}


\maketitle

\section{Introduction}
Galaxies live in a variety of environments, from isolated solitude \citep[e.g.][]{Grogin2000_VoidGalaxies} to the over-dense and tempestuous environments of galaxy clusters \citep[see e.g.][for observational reviews]{vanWeeren2019_review,Paul2023_Clusters_review}. Between these extremes, we find the most common structures inhabited by galaxies: groups.

Galaxy groups represent an excellent opportunity to study a rich array of processes including galaxy interactions, the triggering and quenching of star formation, and the feeding and feedback of active galactic nuclei (AGN). In particular, compact galaxy groups---which contain dense configurations of a few galaxies that tend to be located in under-dense environments on larger scales \citep[e.g.][]{Hickson1982}---represent excellent targets for studying galaxy-galaxy interactions, as these over-dense environments make such events a more likely occurrence, visible through a variety of tracers \citep[see e.g.][and references therein]{Jones2023_HCG-HI}. Furthermore, the role of the AGN in the evolution of the intra-group medium (IGrM) is more prominent due to the shallower gravitational potential well; even relatively low-power epochs of AGN activity can input sufficient energy into the IGrM to drive material out of the group into the larger-scale ambient environment. See for example \cite{Eckert_2021_groups_review} or \cite{Oppenheimer2021_GalaxyGroups_Simulations} for reviews of AGN feedback in galaxy groups.

Additionally, such groups provide insight into the effect of the environment on the evolution of jets and lobes from the AGN that reside therein. These jets and lobes are a recurring phenomenon during the lifetime of a radio galaxy: in the traditional picture of radio galaxy life cycles, young compact AGN host small-scale jets that may eventually break out of the ambient medium of the host galaxy. These young AGN grow into larger-scale `adult' radio galaxies,  which may be active for $\sim10^7$ to $10^8$~yr, with jets pushing far beyond the host galaxy and expanding into lobes. After some period of activity, the fuel supply to the central engine ceases, causing the jets switch off, and the source becomes a `remnant' radio galaxy where the spectral and morphological evolution is dominated by energy losses and dynamics in the environment. After some period of time, the central AGN may receive a new fuel supply due to accretion of matter or galaxy-galaxy interactions, for example, in which case it may reactivate, becoming a `restarted' AGN. See for example \cite{Morganti2021_AGNLifeCycles} and \cite{Morganti2024_AGNLifeCycles} and references therein for a more detailed picture.

In the radio regime, the synchrotron spectrum is quantified by spectral index $\alpha$, where we take the convention that flux density $S$ is related to frequency $\nu$ as $S \propto \nu^{\alpha}$. Typical active galaxies have a radio spectral index around $\alpha \approx -0.8$ \citep[e.g.][]{Condon1998,deGasperin2018_TGSS-NVSS}, but as electron energy losses by synchrotron emission are proportional to the square of the electron energy, the cosmic ray electron population responsible for powering radio lobes preferentially loses energy at the highest energies, i.e. the highest frequencies \citep{Pacholczyk1970}. Thus, once a radio galaxy enters its remnant phase, the remnant lobes fade quickly at higher frequencies and the spectrum steepens above a break frequency which is proportional to the age of the source. 

It is during this remnant phase, in which the lobes are no longer fed by the central AGN and are cooling passively, that the environment plays an increasingly important role. Remnant radio galaxies outside overdense environments tend to age passively, with the lobes gradually relaxing and fading as they radiate away their energy and the spectrum steepens in accordance with ageing models \citep[e.g.][]{deGasperin2014_NGC5580,Brienza2016_Remnant,Shulevski2017_B0924+30,Duchesne2019_NGC1534,Brienza2020_3C388,Quici2021_Remnants}. However, those residing in dense, dynamic environments like groups and clusters are often perturbed. 

Dynamical interactions such as mergers, bulk motions in the IGrM, and accretion of material and/or galaxies can drive energy into the IGrM, re-distributing fossil plasma from remnant lobes onto larger scales, as well as depositing energy into the fossil plasma, re-accelerating the cosmic ray electrons. This energy deposition can be mediated by a variety of processes such as sloshing, compression, and/or shocks, but the end result is broadly similar: fossil plasma is redistributed and re-lit, becoming detectable once more at radio wavelengths. As such, many remnant sources hosted in overdense environments of groups and clusters demonstrate complicated morphologies and complex spectral features \citep[e.g.][]{Brienza2021_NEST200047,Brienza2022_NGC507,Riseley2022_Abell3266,Candini2023_NGC6086,Shulevski2024_A1318_remnant}. Hence, in order to complete the picture of the life cycles of radio galaxies, the importance of multi-frequency spectral coverage cannot be over-emphasised \citep[see in particular][for examples]{Jurlin2021_Remnants,Quici2021_Remnants}.

In this paper we present a detailed spectropolarimetric analysis of one particular galaxy group: Hickson Compact Group (HCG) 15. This group was visually identified in data from the Australian Square Kilometre Array Pathfinder \citep[ASKAP;][]{Johnston2007,DeBoer2009,Hotan2021}, which began survey science operations in late 2022 following several years of early science and Pilot Surveys. 

\begin{figure*}
\centering
\includegraphics[width=0.9\textwidth]{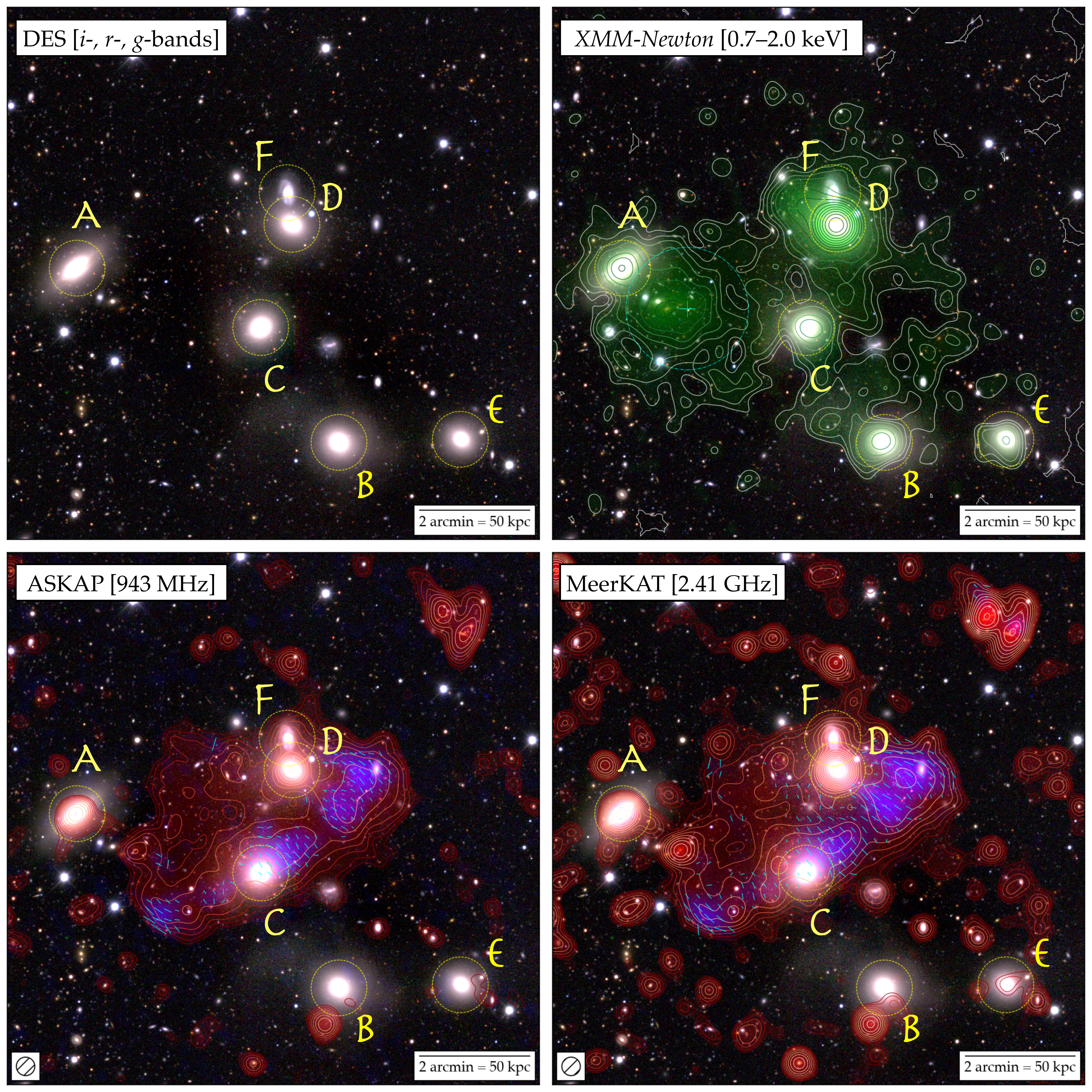}
\caption{Colour-composite images of HCG15. The optical RGB image is constructed using $i$-, $r$- and $g$-bands from Data Release 2 of the Dark Energy Survey (DES). Green colour and contours denote our \textit{XMM-Newton} X-ray data in the $0.7-2.0$~keV range, smoothed with a Gaussian of 7~pixels (14~arcsec) FWHM. Radio data from ASKAP at 943\,MHz (\textit{lower-left}) and MeerKAT at 2.4\,GHz (\textit{lower-right}) are shown, with red colour and contours denoting continuum emission and blue colour denoting linearly-polarised emission. All radio data is shown at 20~arcsec resolution, as indicated by the hatched circle in the lower-left corner. Contours start at $3\sigma$ and scale by a factor of $\sqrt{2}$. Cyan vectors denote the orientation of the magnetic field, with vector size proportional to the fractional polarisation. Yellow circles and labels identify the six group-member galaxies; cyan `+' sign in the top-right panel denotes the SZ position of a background cluster at $z \simeq 0.88$ reported by \protect\cite{Klein2024_ACT-DR5_MCMF}, with the circle tracing a 500~kpc radius at that redshift.}
\label{fig:HCG15_Composite}
\end{figure*}

The deep hemispheric continuum survey with ASKAP is the Evolutionary Map of the Universe \citep[EMU;][]{Norris2011,Norris2021_EMUPilot} Survey. During inspection of the large-area mosaic taken as part of the EMU Pilot II Survey, unusual diffuse emission associated with HCG15 was visually identified. Subsequent investigation revealed this emission to be strongly polarised, motivating the detailed analysis presented in this paper. We present a colour-composite image of HCG15 in Fig.~\ref{fig:HCG15_Composite} using data from the Dark Energy Survey (DES, \textit{top left panel}), overlaid with X-ray data from \textit{XMM-Newton} (green colour and contours, \textit{top right panel}), radio continuum data (in red colour and contours) and polarisation data (blue colour and vectors) from both ASKAP-EMU at 943~MHz (\textit{lower left panel}) and MeerKAT at 2.41~GHz (\textit{lower right panel}).

Alongside our ASKAP-EMU data, we employ new data from the LOw-Frequency ARray \citep[LOFAR;][]{vanHaarlem2013}, MeerKAT, the Karl G. Jansky Very Large Array \citep[VLA;][]{Perley2011_EVLA}, and the Giant Metrewave Radio Telescope \citep[GMRT;][]{Swarup1990_GMRT}. 

This paper is structured as follows: we discuss the observations and data reduction in \S\ref{sec:observations}, we present our results on the group-member galaxies in \S\ref{sec:results:galaxies} and on the diffuse emission in \S\ref{sec:results:diffuse}. Our analysis and interpretation is presented in \S\ref{sec:analysis}, and we draw our conclusions in \S\ref{sec:conclusions}. Throughout, we assume a $\Lambda$CDM cosmology of H$_0 = 73 ~ \rm{km} ~ \rm{s}^{-1} ~ \rm{Mpc}^{-1}$, $\Omega_{\rm{m}} = 0.27$, $\Omega_{\Lambda} = 0.73$. At the representative redshift of HCG15 \citep[$z = 0.0228$;][]{Hickson1992} the angular scale to linear size conversion is 1~arcsec to 440~pc, with our cosmology. We quote all uncertainties at the $1\upsigma$ level.\\

\noindent\textbf{HCG15} is a somewhat atypical compact galaxy group. In particular, unlike the majority of X-ray luminous groups, as seen in Figure~\ref{fig:HCG15_Composite} it is not dominated by a single massive elliptical galaxy but comprises six known group members of similar brightness \citep[e.g.][]{Hickson1982,Hickson1989,Hickson1992}. These six galaxies show a mixed elliptical and spiral population, with a broad range of stellar masses from $M_{*} = 2.1\times10^{9} \, {\rm M}_{\odot}$ to $1.1 \times 10^{11} \, {\rm M}_{\odot}$ \citep{Bitsakis2011}. We summarise some of the observational properties of these galaxies in Table~\ref{tab:hcg15_galaxies}. Most galaxies show negligible star formation, although the spiral galaxy HCG15-F shows a high specific star formation rate relative to the other group members. 

The group has a total mass of $M_{200} \sim 2 \times 10^{13}$\,M$_{\odot}$, derived from NFW \citep{Navarro1997_NFW} fits to the derived group mass profile \citep[see][]{Rasmussen2008_HCGs}. For comparison, similar halo profile fits to galaxy groups detected in the Galaxy and Mass Assembly \citep[GAMA;][]{Driver2009_GAMA,Driver2011_GAMA,Liske2015_GAMA} survey yield a mean mass of $M_{200} \sim 2.57 \times 10^{13}$\,M$_{\odot}$ \citep{Riggs2021_GAMA_Groups}, suggesting that HCG15 is not a particularly low-mass group.

While HCG15 shows a strong deficiency in neutral gas, with a depletion factor around $76\%$ \citep{VerdesMontenegro2001,Jones2023_HCG-HI} and large amounts of diffuse intra-group light 
(IGL) with the IGL representing some $\sim20\%$ of the total light from the group \citep{DaRocha2008_IGL}, these are relatively typical characteristics of compact groups. HCG15 does host a somewhat irregular extended X-ray halo \citep{Mulchaey2003_XrayGroups}, and a diffuse radio `ridge' \citep[][]{Giacintucci2011_GalaxyGroups}, which are more unusual features in such systems \citep[see also e.g.][]{Coziol2004_CompactGroups}. All evidence indicates that it is a dynamically-disturbed system.

\begin{table*}
\small
\centering
\caption{Observational properties of the six galaxies in HCG15. SFR and sSFR respectively stand for the star formation rate and specific star formation rate, derived by \cite{Bitsakis2011} from UV-to-IR data. Here we quote the original radial velocity measurements from \cite{Hickson1992}, although we derive new radial velocities for HCG15-C and HCG15-F based on our optical spectral later in this paper. These new velocities are shown here in brackets. \label{tab:hcg15_galaxies}}
\renewcommand{\arraystretch}{1.2}
\begin{tabular}{lcccccccc}
\hline
Galaxy   &  Right Ascension  &   Declination     &   Type$^1$  &   B$^1$    &   Stellar mass$^2$     &   SFR$^2$     &   sSFR$^2$    &   Radial velocity$^3$ \\
         &   (J2000)         &   (J2000)         &          &     $[\rm mags]$      & $[\times10^{10} \, \rm M_{\odot}]$  & $[\rm M_{\odot} / yr]$  &  $[\rm \times10^{-2} / Gyr]$  &   $[\rm km \, s^{-1}]$    \\
\hline\hline
HCG15-A  &  02:07:53.03      &  +02:10:03.3      &   Sa     &   14.87   &   8.13  &  0.10 &   0.11  &   $6967 \pm 30$   \\
HCG15-B  &  02:07:34.12      &  +02:06:54.7      &   E0     &   15.31   &   4.07  &  0.05 &   0.11  &   $7117 \pm 36$   \\
\multirow{2}{*}{HCG15-C}  &  \multirow{2}{*}{02:07:39.77}      &  \multirow{2}{*}{+02:08:59.0}      &   \multirow{2}{*}{E0}     &   \multirow{2}{*}{14.91}   &   \multirow{2}{*}{10.72} &  \multirow{2}{*}{0.02} &   \multirow{2}{*}{0.02}  &   $7222 \pm 30$  \\
   &         &         &         &    &     &    &      &   $(6803 \pm 12)$   \\
HCG15-D  &  02:07:37.52      &  +02:10:50.8      &   E0     &   15.60   &   2.46  &  0.01 &   0.32  &   $6244 \pm 36$   \\
HCG15-E  &  02:07:25.36      &  +02:06:58.1      &   Sa     &   15.94   &   2.57  &  0.06 &   0.22  &   $7197 \pm 32$   \\
\multirow{2}{*}{HCG15-F}  &  \multirow{2}{*}{02:07:37.87}      &  \multirow{2}{*}{+02:11:25.0}      &   \multirow{2}{*}{Sbc}    &   \multirow{2}{*}{16.73}   &   \multirow{2}{*}{0.21}  &  \multirow{2}{*}{0.28} &   \multirow{2}{*}{14.03} &   $6242 \pm 103$  \\
 &         &         &        &       &      &    &    &   $(6133 \pm 98)$  \\
\hline
\multicolumn{9}{l}{References: $^1$ \cite{Hickson1989}, $^2$ \cite{Bitsakis2011}, $^3$ \cite{Hickson1992}.}\\
\end{tabular}
\end{table*}

The nature of the radio ridge has been the subject of much debate, with various scenarios proposed from turbulence-driven (mini-)halo-type emission, fossil AGN emission, and/or shock-driven relic-type emission \citep[e.g.][]{Giacintucci2011_GalaxyGroups,NikielWroczynski2017_GalaxyGroups,NikielWroczynski2021_GalaxyGroups}. However, these studies have proven largely inconclusive due to the paucity of historical data, and the nature of the diffuse radio ridge remains uncertain.

\section{Observations and Data Reduction}\label{sec:observations}
In this section we detail the new and archival data from the Australian SKA Pathfinder (\S\ref{sec:askap}), MeerKAT (\S\ref{sec:meerkat}), LOFAR (\S\ref{sec:lofar}), the VLA (\S\ref{sec:vla}), and the ATCA (\S\ref{sec:atca}) at radio wavelengths, \textit{XMM-Newton} (\S\ref{sec:xmm}) at X-ray wavelengths, and the HCT (\S\ref{sec:hct}) at optical wavelengths. We summarise the observations in Table~\ref{tab:observations}, but details are provided in each section.

\begin{table*}

\centering
\caption{Summary of key details for new observations of HCG15 presented in this paper. Note that the GMRT observations on 2023~Nov.~04 and 06 also took simultaneous data with the wide-band feeds, covering $250 - 500$~MHz and $550 - 900$~MHz; however, these data are not used in this paper. \label{tab:observations}}
\renewcommand{\arraystretch}{1.2}
\small
\begin{tabular}{lcccc}
\hline
Telescope   &   Observation Date    &   Configuration/Mode   &   Spectral Coverage   &   On-source time \\
\hline
\multirow{11}{*}{LOFAR}       & 2018~Jul.~12     &  HBA\_DUAL\_INNER     &  $120 - 168$~MHz     & 3~hours     \\
            & 2018~Jul.~13     &  HBA\_DUAL\_INNER     &  $120 - 168$~MHz     & 4~hours     \\
            & 2018~Jul.~19     &  HBA\_DUAL\_INNER     &  $120 - 168$~MHz     & 3~hours     \\
            & 2018~Sept.~01    &  HBA\_DUAL\_INNER     &  $120 - 168$~MHz     & 3~hours     \\
            & 2018~Sept.~11    &  HBA\_DUAL\_INNER     &  $120 - 168$~MHz     & 4~hours     \\
            & 2018~Oct.~28     &  HBA\_DUAL\_INNER     &  $120 - 168$~MHz     & 4~hours     \\
            & 2019~Jul.~15     &  HBA\_DUAL\_INNER     &  $120 - 168$~MHz     & 3~hours     \\
            & 2020~Jul.~02     &  HBA\_DUAL\_INNER     &  $120 - 168$~MHz     & 3~hours     \\
            & 2020~Jul.~04     &  HBA\_DUAL\_INNER     &  $120 - 168$~MHz     & 3~hours     \\        
            & 2020~Jul.~31     &  HBA\_DUAL\_INNER     &  $120 - 168$~MHz     & 3~hours     \\
            & 2020~Aug.~01     &  HBA\_DUAL\_INNER     &  $120 - 168$~MHz     & 3~hours     \\
\hline
\multirow{5}{*}{GMRT}   & 2010~Apr.~25     &  $-$     &  $592 - 624$~MHz     & 9.3~hours   \\
            & 2010~May~23      &  $-$     &  $592 - 624$~MHz     & 10.6~hours  \\
            & 2010~Jul.~23     &  $-$     &  $307 - 340$~MHz     & 10.1~hours  \\ 
            & 2023~Nov.~04     &  $-$     &  $307 - 340$~MHz     & 9.3~hours  \\
            & 2023~Nov.~06     &  $-$     &  $592 - 624$~MHz     & 7~hours  \\
            
\hline
\multirow{2}{*}{ASKAP}         & 2021~Nov.~05     &  $-$     &  $800 - 1088$~MHz    & 5~hours     \\
            & 2021~Nov.~10     &  $-$     &  $800 - 1088$~MHz    & 5~hours     \\
\hline
\multirow{2}{*}{MeerKAT}       & 2023~Jul.~09     &  $-$     &  $1.968 - 2.843$~GHz & 4~hours     \\
            & 2023~Jul.~10     &  $-$     &  $1.968 - 2.843$~GHz     & 4~hours     \\
\hline
ATCA        & 2024~May~31      &  H168    &  $4.5 - 6.5$~GHz \& $8 - 10$~GHz   & 2.2~hours   \\
\hline
\multirow{5}{*}{VLA}           & 2013~Feb.~23     &  D       &  $4 - 6$~GHz \& $6 - 8$~GHz        & 49~minutes   \\
            & 2013~Mar.~09     &  D       &  $4 - 6$~GHz \& $6 - 8$~GHz        & 49~minutes   \\
            & 2013~Mar.~10     &  D       &  $4 - 6$~GHz \& $6 - 8$~GHz        & 49~minutes   \\
            & 2024~Feb.~03     &  C       &  $4 - 6$~GHz \& $6 - 8$~GHz        & 1.9~hours   \\
            & 2024~Mar.~03     &  C       &  $4 - 6$~GHz \& $6 - 8$~GHz        & 6.9~hours   \\
\hline
HCT         & 2024~Jan.~02     &   HFOSC  & $3800 - 6840$~\AA{} \& $5800 - 8350$~\AA{}  &  4~hours \\
\hline
\textit{XMM-Newton} &   2014~Aug.~06    &   MOS \& pn & $0.15 - 12$~keV  &    140~ks \\
\hline
\end{tabular}
\end{table*}

\subsection{Australian SKA Pathfinder}\label{sec:askap}
HCG15 was observed with ASKAP during the EMU Pilot II Survey, performed in 2021. ASKAP is comprised of 36 twelve-metre dishes located in the Murchison Radioastronomy Observatory (MRO) in Western Australia. The telescope is capable of observing between 700\,MHz and 1.8\,GHz, with an instantaneous bandwidth of up to 288\,MHz. As a result of the unique phased-array feeds \citep[PAF;][]{Hotan2014,McConnell2016}, ASKAP is capable of simultaneously forming up to 36 independent beams, covering $\sim30$\,deg$^2$, rendering ASKAP a powerful survey telescope. With this large field of view (FOV), operating at low frequencies, and with baselines between 22\,m and 6\,km, ASKAP has excellent sensitivity to extended faint emission, such as the features associated with galaxy groups and clusters.

Observations were performed using ASKAP's Band~1 receivers with a central frequency of 943\,MHz and using the full instantaneous bandwidth of 288\,MHz. The beams were arranged in the standard `closepack-36' beam configuration. Due to the equatorial nature of the EMU tile observed (tile EMU$\_$0208$+$00), observations were performed in two five-hour sessions; the scheduling blocks are SB33275 and SB33442, carried out on 2021~Nov.~05 and 2021~Nov.~10 respectively.

\subsubsection{Initial processing}
ASKAP uses the standard calibrator PKS~B1934$-$638 for bandpass and absolute flux density scale calibration, observed daily on a per-beam basis. Following standard practice the data are processed using the \textsc{ASKAPsoft} pipeline \citep{Guzman2019}, incorporating RFI flagging, bandpass calibration, averaging and multiple rounds of direction-independent (DI) amplitude and phase self-calibration\footnote{\url{https://www.atnf.csiro.au/computing/software/askapsoft/sdp/docs/current/pipelines/introduction.html}}. Following self-calibration, the data are averaged to 1\,MHz spectral resolution. During this process, each beam is processed independently before being co-added in the image plane.

While in general the equatorial nature of the field means that artefacts from bright sources appear more pronounced than in the Pilot I Survey \citep{Norris2021_EMUPilot}, the region around HCG15 is largely free from contamination. Consequently we did not need to employ the direction-dependent (DD) calibration routine demonstrated by \cite{Riseley2022_Abell3266}, although some post-processing was required to improve the image fidelity and reduce contamination from artefacts. 

HCG15 is close to the phase centre of a single beam (beam~32) from the EMU tile, and we retrieved the visibilities for this beam from the CSIRO\footnote{Commonwealth Scientific \& Industrial Research Organisation} ASKAP Science Data Archive \citep[CASDA\footnote{\url{https://research.csiro.au/casda/}}; e.g.][]{Chapman2017,Huynh2020} in order to perform further processing.

\subsubsection{Post-processing}
Initially, we used the \textsc{FixMS} \texttt{v0.2.3} library\footnote{\textsc{FixMS} can be accessed at \url{https://fixms.readthedocs.io/en/latest/}} to apply three corrections. First, due to ASKAP's roll axis, the visibilities must be rotated from the telescope frame to the sky frame. Second, we updated the header in the visibilities to point to the beam phase centre rather than the default phase centre of the ASKAP tile footprint. Third, before imaging outside of \textsc{ASKAPSoft} a correction factor is required to convert the Stokes definitions from those used by \textsc{ASKAPSoft} (e.g. Stokes $I = (XX + YY)$) to those following the IAU convention as needed by \textsc{WSclean} (e.g. Stokes $I = \frac{1}{2}(XX + YY)$), and similarly for other Stokes parameters.

Subsequently, we performed two rounds of imaging and self-calibration using \textsc{WSclean} \texttt{v3.3} \citep{Offringa2014,Offringa2017}\footnote{WSclean is available at \url{https://gitlab.com/aroffringa/wsclean}} and \textsc{killMS} \citep{Tasse2014,Smirnov2015} respectively. During these imaging rounds, images were made out to a radius of around 3.5~degrees from the phase centre, encompassing the first sidelobe of the ASKAP primary beam, in order to ensure particularly bright sources outside the region of interest were included in our model. After two rounds of self-calibration processing for our wide-field image had converged, and so we could proceed to extract a smaller dataset.

We then subtracted all sources outside a radius of 0.61~degree from the beam phase centre (a radius sufficiently large to include HCG15) and performed further self-calibration, initially in phase-only mode and subsequently in amplitude-and-phase mode, on this extracted dataset. After four rounds of self-calibration, our calibration had converged and we proceeded to generate our final images.

\subsubsection{Continuum imaging}
Our final radio continuum images of HCG15 were generated using \textsc{WSclean}. We used \texttt{-multiscale} clean to more accurately model the extended emission from our target, and used Briggs weighting \citep{Briggs1995} with a \texttt{robust} of $-0.5$ to balance sensitivity and resolution. We also employed a combination of automated and manual masking during the deconvolution process.

Additionally, to isolate the diffuse emission from the group, we re-imaged our data using an inner \textit{uv}-cut of 5k$\lambda$ (tuned to filter out the diffuse emission present) in order to generate a model of the embedded and background radio sources in the vicinity of HCG15. We used these high-resolution images to measure the integrated flux densities of all radio counterparts to the group-member galaxies. These sources were then subtracted from our data, which were re-imaged with a \textit{uv}-taper and convolved in the image plane to achieve our target resolution of 20 arcsec.

\subsubsection{Polarisation imaging}
We also imaged our data in linear polarisation using \textsc{WSclean}. We employed the \texttt{-join-channels} option to leverage the sensitivity of the full bandwidth while cleaning at high spectral resolution, and the \texttt{-join-polarizations} option to search for peaks in linear polarisation (i.e. $\sqrt{Q^2 + U^2}$) rather than in each Stokes parameter separately. We generated images at the full spectral resolution of our calibrated dataset, 288 channels of 1~MHz width covering a frequency range from 800~MHz to 1088~MHz.

We then convolved the resulting image cube to a common resolution of 20~arcsec. Five channels were unable to reach this target resolution due to severe flagging, and so were discarded; a further six channel images were excised due to having an off-source noise larger than $5\sigma_{QU}$. In order to correct for instrumental leakage and the primary beam response, we used the frequency-dependent ASKAP beam model provided with each observation in conjunction with the \texttt{linmos} task within \textsc{Yandasoft}\footnote{\url{https://github.com/ATNF/yandasoft}} \texttt{v1.11.0}; we note that the primary beam correction was also applied to our radio continuum images using \texttt{linmos}.

Finally we ran Rotation Measure (RM) synthesis\footnote{See e.g. \citet{Burn1966} or \citet{Brentjens2005} for a detailed description of RM synthesis, or \citet{Heald2009_RMclean} for an overview of the key concepts} and RM-clean on our polarisation cube using the \textsc{RM-tools} software suite tasks \texttt{rmsynth3d} and \texttt{rmclean3d}. The three key Faraday-space parameters of interest when performing RM-synthesis and RM-clean are the resolution in Faraday space $\Delta({\rm RM})$, the maximum recoverable RM $|{\rm RM_{max}}|$ and the maximum recoverable scale in Faraday space ${\rm max.-scale}$, which are related to the frequency coverage of the input Stokes $Q,U$ cubes as follows:

\begin{subequations}
    \begin{gather}
    \label{eq:rm_drm}
        \Delta({\rm RM}) = 2 \sqrt{3} / \Delta( \lambda^2) \\
    \label{eq:rm_maxrm}
        |{\rm RM_{max}}| = \sqrt{3} / \delta{\lambda^2} \\
    \label{eq:rm_maxscale}
        {\rm max.-scale} = \pi / \lambda _{\rm min}^2
    \end{gather}
\end{subequations}
where $\Delta( \lambda^2)$ is the wavelength-squared difference across the observing bandwidth, $\delta{\lambda^2}$ is the wavelength-squared difference across channels in the input cube, and $\lambda _{\rm min}$ is the shortest wavelength in the input cube \citep[see e.g.][]{Brentjens2005}. For our ASKAP polarisation cubes, Equation~\eqref{eq:rm_drm} implies we have a Faraday-space resolution of $\Delta({\rm RM}) \simeq 61$~rad~m$^{-2}$ and Equation~\eqref{eq:rm_maxscale} indicates that the largest recoverable scale in Faraday space is ${\rm max.-scale} \simeq 41$~rad~m$^{-2}$. When viewed in concert, this means that we retain sensitivity to Faraday-thin structures only. Equation~\eqref{eq:rm_maxrm} implies we retain sensitivity to RMs up to $|{\rm RM_{max}}| \simeq 8130$~rad~m$^{-2}$.

When performing RM-synthesis, we initially opted to explore the majority of this range, generating RM cubes that cover $|{\rm RM_{max}}| \leq 3750$~rad~m$^{-2}$ at a resolution of $10$~rad~m$^{-2}$, as we do not anticipate extreme RMs in the environment of HCG15. After inspection of these RM-cubes, we found no significant components with $|{\rm RM}| \gtrsim 50$~rad~m$^{-2}$, and so we narrowed the focus of our RM-synthesis to cover $|{\rm RM_{max}}| \leq 500$~rad~m$^{-2}$ at a resolution of $2.5$~rad~m$^{-2}$ in order to better model the RM variation across our target. We then carried out RM-clean, searching for components down to a threshold of $4\sigma_{QU}$ following standard practice \citep[e.g.][]{Heald2009_RMclean,Heald2009_WSRT-SINGS} where $\sigma_{QU} = 24.7 \, \upmu$Jy~beam$^{-1}$ is the measured rms noise in our channel-wise Stokes Q and U cubes scaled by a factor $1 / \sqrt{n_{\rm chan}}$ and $n_{\rm chan} = 277$ is the number of channels in our polarisation cubes. After this process, our cleaned Faraday dispersion function (FDF) cubes, which trace the polarised flux density in Stokes Q \& U as well as linear polarisation as a function of RM, were ready for analysis.

We also sourced data from the Rapid ASKAP Continuum Survey \citep[RACS;][]{McConnell2020_RACS} for our analysis. While RACS possesses insufficient sensitivity to recover the diffuse emission associated with HCG15, the compact component associated with the brightest embedded radio galaxy HCG15-D is detected, and we extract flux density measurements from RACS-Low \citep[887\,MHz;][]{Hale2021_RACS-Low}, RACS-Mid \citep[1.367\,GHz;][]{Duchesne2024_RACS-Mid} and RACS-High \citep[1.655\,GHz;][]{Duchesne2025_RACS-High}.

\subsection{MeerKAT}\label{sec:meerkat}
We secured new MeerKAT observations (\citealt{Jonas2016}, though see also \citealt{Camilo2018} and \citealt{Mauch2020}) under Director's Discretionary Time (DDT) project code DDT-20230705-CR-01. Due to the equatorial location of HCG15, the field was observed for two five-hour observing runs under capture block IDs (CBIDs) 1688875155 and 1688961378, performed on 2023~Jul.~09 and 2023~Jul.~10 respectively. At the time of observation, the S-band receivers installation was ongoing, meaning that a total of 56 antennas were available. The total time on-source was of the order of eight hours.

Our observations were performed with the new S-band receiver system, using the S1 sub-band selection covering the frequency range 1968.75~MHz to 2843.75~MHz. On each day the primary calibrators PKS~B1934$-$638 and PKS~B0407-658 were used to derive delays and bandpass calibration solutions, with PKS~B1934$-$638 also being used to set the flux scale according to the \cite{Reynolds1994} scale. At the time of observation, the list of gain calibrators suitable for use with MeerKAT in S-band was still being verified; we selected the secondary gain calibrator J0217$+$017 (Right Ascension, Declination 02:17:48.95, $+$01:44:49.7) from the Australia Telescope Compact Array \citep[ATCA;][]{FraterBrooks1992} Calibrator Database\footnote{\url{https://www.narrabri.atnf.csiro.au/calibrators/calibrator_database.html}} as a potential calibrator, and this source was approved by the SARAO Operations team. This secondary calibrator was observed for two minutes at a half-hour cadence. In each observing run we also included two scans on the known polarisation calibrator 3C\,138, separated by a large parallactic angle range, to set the absolute polarisation position angle.

\subsubsection{First-generation calibration}
First-generation calibration was performed using the Common Astronomy Software Applications \citep[\textsc{CASA};][]{McMullin2007} \texttt{v5.5.0-149.el7}. After removing edge channels, we performed automated sum-threshold flagging on our calibrator visibilities using the \texttt{tfcrop} and \texttt{rflag} algorithms before proceeding to calibration. First we used the \texttt{setjy} task to set the models for PKS~B1934$-$638 (using the `Stevens-Reynolds 2016' scale) and PKS~B0407$-$658, which does not have a standard model in \textsc{CASA}. As such we followed the process described online in the MeerKAT External Service Desk\footnote{Available at \url{https://skaafrica.atlassian.net/wiki/spaces/ESDKB/pages/1481408634/Flux+and+bandpass+calibration}.} to set the model for PKS~B0407$-$658.

We performed continuum calibration according to standard practice, deriving the delay and bandpass corrections and time-dependent parallel-hand gain solutions. For polarisation calibration we followed techniques developed for the Karoo Array Telescope \citep[KAT-7;][]{Foley2016_KAT7} employed by \cite{Riseley2015} and discussed in detail by \cite{RiseleyThesis}. To summarise the process: we used the \texttt{qufromgain} task to estimate the polarisation properties of all sources present in our parallel-hand gain table. We then derived cross-hand delays and an initial estimate of the XY-phase and source polarisation (jointly) using the \texttt{gaincal} task with \texttt{gaintype = `XYf+QU'}, each using 3C\,138 and assuming an arbitrary non-zero polarisation model. This estimate of the cross-hand phase was then corrected for the $n\pi$-ambiguity using the observed polarisation model derived by \texttt{qufromgain} using the task \texttt{xyamb}. Finally, we re-derived our time-dependent gain table (applying parallactic angle correction on the fly) using the updated polarisation model for 3C\,138 (assuming zero polarisation for other sources) and then derived corrections for on-axis instrumental leakage using the \texttt{polcal} task with \texttt{poltype = `Dflls'}. Finally we derived flux scale corrections using PKS~B1934$-$638, before applying all calibration tables --- parallel-hand and cross-hand delays, bandpass solutions, XY-phase solution, instrumental leakage, and flux scale corrected time-dependent gains --- to our target field. We note that this calibration methodology is also present in the Containerized Automated Radio Astronomy Calibration (\texttt{CARACal}) pipeline\footnote{\url{https://github.com/caracal-pipeline/caracal}} \citep{Jozsa2020,Jozsa2021}, and as documented by \cite{Loi2025_MeerKAT_FornaxSurvey} achieves successful polarisation calibration at L-band, but S-band calibration capabilities were not available within the \texttt{CARACal} framework at the time of data processing.

As these observations represent the first non-commissioning MeerKAT S-band polarisation study, we feel it is prudent to report briefly on the calibration results. We discuss the results of our polarisation calibration in Appendix~\ref{appendix_meerkat_pol}, but to summarise briefly, the measured S-band polarisation properties for 3C\,138 are broadly consistent with what we would expect based on extrapolation from L-band measurements \citep[e.g.][]{Taylor2024_MeerKAT_polcals}.

We then employed \textsc{AOflagger} \citep{Offringa2010_AOflagger-ASCL,Offringa2012_AOflagger} \texttt{v3.0} to perform RFI excision on our data using a custom strategy. The overall flagged fraction of our data is low, around 15 per cent, and dominated by the excision of edge channels where the bandpass roll-off reduces the instrumental sensitivity. Finally we averaged by a factor 8 in frequency, yielding 454 output channels of 1.7\,MHz width, and proceeded to self-calibration.

\subsubsection{Second-generation calibration}
For our direction-independent (or ``second-generation'') self-calibration we used \textsc{WSclean} \texttt{v3.3} for imaging and \textsc{killMS} for self-calibration. We imaged a FOV around 0.83~degree radius around our target to capture all significant sources in the wider field and mitigate the impact of any artefacts on the diffuse emission from HCG15. We employed multi-scale multi-frequency-synthesis cleaning using 16 output channels and a third-order polynomial to describe the frequency behaviour of the clean components, as well as the in-built \texttt{wgridder} \citep{Arras2021_wgridder,Ye2021_wgridder} to better correct for wide-field effects. Finally, in every \textsc{WSclean} instance we employed a combination of auto-masking and manual masking to ensure recovery of the fainter sources in the field.

Self-calibration was performed in phase-only mode at first, with a decreasing solution interval in every round to better correct for time-dependent errors in our gains. We performed five rounds of phase-only self-calibration and subsequently three further rounds solving for gains in both amplitude and phase. Once the direction-independent self-calibration had converged, we extracted a small region of 0.4~degree radius around our target, by subtracting our best model of the wider-field sources, to enable efficient postprocessing. We attempted further self-calibration on this extracted dataset, although no further improvement was achieved. We corrected for the primary beam response using the holography measurements from \cite{deVilliers2023_MeerKAT_holography}.

\subsubsection{Polarisation imaging}
To generate our linear polarisation maps for use with RM Synthesis, we again employed \textsc{WSclean} in the same manner as for our ASKAP observations. We generated images at the full spectral resolution, i.e. 454 channels of 1.7~MHz width covering a frequency range from 2.012~GHz to 2.786~GHz. The image cube was then convolved to a common resolution of 20~arcsec (to match our ASKAP polarisation cube) before excising 46 channel images where the off-source noise was beyond $5\sigma_{QU}$. Finally we ran RM-synthesis and RM-clean on our polarisation cube using the \textsc{RM-tools} software suite tasks \texttt{rmsynth3d} and \texttt{rmclean3d}.

While MeerKAT's bandwidth is large, the relatively high frequency of S-band means that our Faraday-space FWHM is also fairly broad. From earlier, Equation~\ref{eq:rm_drm} yields $\Delta({\rm RM}) = 2 \sqrt{3} / \Delta( \lambda^2) \simeq 360$~rad~m$^{-2}$. As with our ASKAP observations, we are sensitive only to Faraday-thin structures as Equation~\ref{eq:rm_maxscale} gives ${\rm max.-scale} \simeq 271$~rad~m$^{-2}$. However, the high spectral resolution means that we retain sensitivity to large values of the RM, as Equation~\ref{eq:rm_maxrm} gives $|{\rm RM_{max}}| = \sqrt{3} / \delta{\lambda^2} \simeq 8 \times 10^4$~rad~m$^{-2}$. We covered a parameter space to $|{\rm RM_{max}}| \leq 5000$~rad~m$^{-2}$ at a resolution of $2.5$~rad~m$^{-2}$. When performing RM-clean, we cleaned down to a threshold of $4\sigma_{QU}$, where $\sigma_{QU} = 2.97 \, \upmu$Jy~beam$^{-1}$ is the measured rms noise in our channel-wise Stokes Q and U cubes scaled by a factor $1 / \sqrt{n_{\rm chan}}$ and $n_{\rm chan} = 408$ is the number of channels in our polarisation cubes.

\subsection{Low-Frequency Array}\label{sec:lofar}
HCG15 was observed with LOFAR as part of the ongoing LOFAR Two-metre Sky Survey \citep[LoTSS;][]{Shimwell2017_LoTSS_PaperI,Shimwell2019_LOTSS_PaperII,Shimwell2022_LOTSS_PaperIII}. Following standard practice for LoTSS, observations were performed using the full International LOFAR Telescope \citep[ILT;][]{vanHaarlem2013} in \texttt{HBA\_DUAL\_INNER} mode, covering the frequency range 120$-$168~MHz. In this work we make use of the data from the Dutch LOFAR array (Core and Remote stations, encompassing baselines out to $\sim80$~km).

A total of three LoTSS fields cover the region around HCG15: P031+01, P031+04, and P033+01. Due to the equatorial nature of these fields, each pointing was observed on either three or four occasions to achieve the target of 12\,hours on-source. P031+01 was observed on 2018~Jul.~19, 2019~Jul.~15, 2020~Jul.~02 and 2020~Jul.~04; P031+04 was observed on 2018 Jul.~13, 2018~Sept.~11, and 2018~Oct.~28; P033+01 was observed on 2018~Jul.~12, 2018~Sept.~01, 2020~Jul.~31, and 2020~Aug.~01. Hence, the total on-source time is 36\,hours, although HCG15 is relatively far from the phase centre of each pointing at distances of $1.39\degree$, $1.64\degree$ and $1.99\degree$ respectively for P031+01, P031+04 and P033+01.

The Dutch LOFAR array data were processed using the standard LoTSS pipeline\footnote{\url{github.com/mhardcastle/ddf-pipeline/}}, described in detail by \cite{Shimwell2019_LOTSS_PaperII,Shimwell2022_LOTSS_PaperIII} and \cite{Tasse2021}. To briefly summarise, the pipeline performs flagging, initial calibration, and both DI and DD self-calibration using \textsc{killMS} and \textsc{DDfacet}. An extracted dataset covering the region around HCG15 was then created using the process described by \cite{vanWeeren2021} to allow for efficient re-imaging with different weighting schemes and \textit{uv}-selection ranges. This process performs several rounds of DI self-cal on the extracted dataset to improve the quality of the data. Our final science images were generated using \textsc{WSclean} \texttt{v3.3}.

\subsubsection{LOFAR flux scaling}
In verifying the flux scale of our LOFAR data, we used the established procedure discussed by \cite{Hardcastle2016} and \cite{Shimwell2019_LOTSS_PaperII}. This process bootstraps the flux scale of an extracted dataset to that of LoTSS. We generated an image at 6\,arcsec resolution using \texttt{WSclean}, and then extracted a source catalogue using the Python Blob Detection and Source Finder \citep[\texttt{PyBDSF};][]{MohanRafferty2015} software. This extracted catalogue was then compared with the catalogues from the full-field LoTSS images of the three pointings that cover HCG15, with a filter applied to all catalogues to select only compact sources with a signal-to-noise ratio of at least seven. Finally, the routine performs a linear regression in the flux/flux plane to determine the best-fit flux density ratio between catalogues. To derive the final scaling factor, we took the mean of the flux density ratios derived for each pointing, yielding a value of 1.68. While this seems a significant correction, it is typical of LoTSS observations, particularly at low elevation (priv. comm. LOFAR Surveys Collaboration). Following \cite{Shimwell2022_LOTSS_PaperIII}, we adopt a representative 10 per cent uncertainty for the flux scale of our LOFAR images, and we fold in an additional 14 per cent uncertainty to account for the standard deviation of the scaling factors derived for each pointing.

\subsection{Karl G. Jansky Very Large Array}\label{sec:vla}
We sourced further higher-frequency data from the VLA to expand our frequency coverage. HCG15 has been observed with the VLA using the C-band receiver system covering the 4$-$8\,GHz range in D-configuration (project code 13A-278; P.I. Nikiel-Wroczy\'{n}ski) and C-configuration (project code 24A-063; P.I. Riseley). For the D-configuration observations, four scheduling blocks were executed, of which one was unusable due to severe flagging. The useable observations were performed on 2013~Feb.~23, 2013~Mar.~09 and 2013~Mar.~10, for a total on-source time of around 2.5~hours. For the C-configuration data, observations began on 2024~Feb.~03 but were aborted after around 2.5~hours due to high winds; HCG15 was re-observed on 2024~Mar.~03. The total on-source time in C-configuration was around 9~hours.

Initial calibration was performed using the VLA pipeline with \textsc{CASA} \texttt{v6.5.4-9}. This pipeline follows standard recipes to derive primary and secondary calibration solutions for calibration of the radio continuum data. We downloaded these pipeline-calibrated measurement sets from the NRAO VLA Data Archive and proceeded to perform self-calibration. Self-calibration was performed independently on the observations from each configuration, using standard techniques. 

Due to the faint steep-spectrum nature of the diffuse emission associated with HCG15, for the study of the diffuse emission associated with the IGrM we only make use of the lower band, which covers the 4$-$6\,GHz frequency range. However, for studying the spectral properties of the compact radio sources associated with the group, we use both the lower and upper bands, covering the frequency ranges 4$-$6\,GHz and 6$-$8\,GHz. We generated our final science images by jointly deconvolving the C- and D-configuration observations using \textsc{WSclean} \texttt{v3.3}, although we jointly deconvolved the datasets after excision of compact sources due to intrinsic compact source variability.

In addition to the C-band data, we also sourced archival VLA survey data to provide flux density measurements for embedded compact sources. We used 1.4\,GHz L-band data from the VLA's Faint Images of the Radio Sky at Twenty centimetres \citep[FIRST;][]{Becker1994_FIRST} survey as well as 3\,GHz data from all three epochs of the VLA Sky Survey \citep[VLASS;][]{Lacy2020}, both sourced via the Canadian Initiative for Radio Astronomy Data Analysis (CIRADA) image cutout server\footnote{\url{http://cutouts.cirada.ca/}}. Further, we sourced archival VLA observations from the NRAO VLA Archive Survey (NVAS\footnote{The NVAS can currently be browsed through \url{http://www.vla.nrao.edu/astro/nvas/}}). Three epochs were found: one A-configuration observation at 4.89\,GHz from 1987~Oct.~03 and two observations at 1.58\,GHz in A-configuration (1984~Dec.~11) and B-configuration (1984~Jan.~26)

\subsection{Australia Telescope Compact Array}\label{sec:atca}
New Green Time observations were sourced using the Australia Telescope Compact Array \citep[ATCA;][]{FraterBrooks1992} under project code CX573 to follow up on variability identified during analysis of the VLA data. HCG15 was observed using a hybrid array configuration (H168) where five of the six antennas are distributed over distances of up to 168\,metres and the sixth is fixed at a distance of around 6\,km. We used the 4\,cm receivers with the standard frequency selection of 5.5 and 9\,GHz, each with 2\,GHz of bandwidth.

HCG15 was observed for 134 minutes of on-source time on 2024~May~31. Primary calibration was provided by the standard calibrator PKS~B1934$-$638. The data were processed using standard techniques with the \textsc{Miriad} software package \citep{Sault1995_Miriad}. While antenna 6 was included for calibration purposes, it was excluded from imaging as per standard practice for hybrid-array data processing. Due to the compact nature of the H168 configuration and the short integration time, only the compact radio source associated with HCG15-D was visible in the data; as such, imaging was performed with \textsc{Miriad}. Self-calibration yielded no visible improvement in the data quality.

\subsection{Giant Metrewave Radio Telescope}\label{sec:gmrt}
We also sourced unpublished archival data from the GMRT to complement our broad-band data, from Cycle 18 (project code $18\_007$). During Cycle 18, the GMRT Software Backend \citep[GSB;][]{Roy2010_GSB} correlator was offered in parallel with the older GMRT Hardware Backend (GHB) correlator; in this work we make use of the GSB data only.

These observations were performed on 2010~Jul.~23 using the 325\,MHz receiver (on-source time around 10\,hours) as well as 2010~Apr.~25 and 2010~May~23 with the 610\,MHz receiver (total on-source time around 19\,hours). The bandwidth at each frequency was 32\,MHz.

Further, HCG15 was observed with the GMRT in 2023 (project code 45\_047; P.I. Riseley). Data were taken using the newer GMRT Wideband Backend \citep[GWB;][]{Reddy2017_GWB} in Band~3 and Band~4, covering 250 -- 500\,MHz and 550 -- 900\,MHz respectively, with simultaneous data captured using the legacy GSB. These observations were performed on 2023~Nov.~04 and~06, for around 9.3~hours and 7~hours on-source, respectively.

All GMRT data were processed using the Source Peeling and Atmospheric Modelling \citep[\textsc{SPAM};][]{Intema2009,Intema2017} pipeline, which performs both primary calibration and self-calibration (in both the DI and DD regimes) as well as iterative RFI flagging.

Due to systematic observing issues and severe ionospheric activity during Cycle~45, we found that the quality of the GSB data from this cycle was markedly worse than that of Cycle~18, and so we use the newer data for compact source flux density measurements only. Processing of the GWB data through SPAM did not yield science-quality image products; advanced techniques may mitigate some of these issues, although would require significant additional work, and so we leave the GWB data for future analysis, and in this study make use of GSB data only.

After the \textsc{SPAM} pipeline processing, the calibrated visibilities were exported for postprocessing using \textsc{WSclean}, to improve recovery of diffuse emission associated with HCG15. In general however, the quality of the GSB maps is far lower than our broad-band data from ASKAP, MeerKAT, and LOFAR. This is due to a combination of factors, including the observing conditions and the narrow fractional bandwidth which results in a far lower \textit{uv}-plane filling factor. We only show images generated from the Cycle~18 data, and while we use \textit{integrated} flux density measurements at both frequencies for exploration of the integrated spectrum, we do not use any GMRT data in the \textit{resolved} spectral study we perform later in this paper.

\subsection{XMM-Newton}\label{sec:xmm}
HCG15 was observed twice with {\it XMM-Newton}. 
For this work, we retrieved the longest observations, Obs. ID 0741440101 (140\,ks on-source time) from the {\it XMM-Newton} data archive.
This observation was performed in full frame mode and using the thin filters. 
The data files were processed with the {\it XMM-Newton} Science Analysis System (SAS) v19.1.0. 
The tasks \texttt{emchain} and \texttt{epchain} were used to generate calibrated event files from raw data. 
We only considered event patterns <13 for MOS and <5 for pn, and also performed bright pixels and hot columns removal (i.e., \texttt{FLAG==0}) and accounted for the contamination by pn Out-of-Time events. 
Additionally, we also excluded all the CCDs in the so-called anomalous state \citep[for more details, see][]{Kuntz_Snowden_2008}. 
The data were cleaned for periods of high background induced by solar flares using the tasks \texttt{mos-filter} and \texttt{pn-filter}. 
The remaining exposure times after cleaning are 106.5, 107.9, 37.8 ks for MOS1, MOS2, and pn, respectively. 
Point-like sources were detected using the task \texttt{edetect\_chain} and excluded from the event files.
All the background event files were cleaned by applying the same PATTERN selection, flare rejection criteria, and point-source removal as used for the observation events. 

The background-subtracted and vignetting-corrected X-ray image in the 0.7-2.0\,keV band presented in Figure~\ref{fig:HCG15_Xray} was obtained using a binning of 40 physical pixels (corresponding to a resolution of 2 arcsec). For visualisation purposes we refilled the point-source regions using the \textsc{CIAO} task {\it dmfilth}. 

The spectral analysis was performed with the \textsc{XSPEC} \citep{1996ASPC..101...17A} package \texttt{v12.12.0}, in the [0.5-12] keV and [0.5-14] keV energy range for MOS and pn, respectively. The X-ray emission was modelled with an APEC thermal plasma model \citep{2001ApJ...556L..91S} with metallicities from \cite{2009ARA&A..47..481A}, and using C-statistics. 
The absorption was fixed at the total (neutral and molecular; see \citealt{2013MNRAS.431..394W}) $N_H\rm=2.93\times10^{20} cm^{-2}$. 
The MOS and pn spectra were fitted simultaneously linking temperature and metallicity but allowing all normalizations to vary freely. 
The modelling of the background is quite complex and we defer the interested reader to \cite{2019MNRAS.483..540L} for a detailed description. The thermodynamic maps presented in Fig. \ref{fig:HCG15_Xray} were obtained following the method presented in \cite{2024A&A...682A..45L} with the regions obtained using the Weighted Voronoi Tessellation (WVT) binning algorithm by \cite{2006MNRAS.368..497D}, which is a generalization of the \cite{2003MNRAS.342..345C} Voronoi binning algorithm, by requiring a signal-to-noise S/N$\sim$20.

\subsection{Himalayan Chandra Telescope}\label{sec:hct}
HCG15 was observed on 2024~Jan.~02 with the 2.01-metre Himalayan Chandra Telescope \citep[HCT;][]{Prabhu_Anupama_2010_IAO} at Hanle Observatory, India\footnote{\url{https://www.iiap.res.in/centres_iao.htm}}, under proposal HCT-2024-C1-P38 (P.I. Riseley). Observations were performed using the Himalaya Faint Object Spectrograph and Camera (HFOSC) instrument, collecting spectroscopic data only for HCG15. We used the 167-micron slit, which has a width of 1.92~arcsec and a length of 11~arcmin. Two Grisms were used to obtain optical spectra in two different wavelength ranges: Grism-7, which covers the wavelength range 3800$-$6840\,\AA{} at a spectral resolution $R(\lambda / \delta\lambda) = 1330$, and Grism-8 which covers the wavelength range 5800$-$8350\,\AA{} at a resolution of 2190. Thus, with each Grism, our spectral resolution was of the order of 1.2$-$2.5\,\AA{}.

Due to scheduling constraints, it was only possible to obtain observations of two members of the HCG15 group with both grisms; we selected galaxies HCG15-C and HCG15-F. The observing conditions were good, with clear skies and a seeing of 2.3$-$2.5\,arcsec. Each galaxy was observed with each Grism for around one hour, divided into three exposures. 

Wavelength calibration was performed using the standard iron-argon (FeAr) lamp for Grism-7 and iron-neon (FeNe) lamp for Grism-8. The standard star Feige~34 was used for absolute flux calibration. Standard bias spectra and flat-field halogen spectra were taken periodically during the observing run.

Data processing was performed using the Image Reduction and Analysis Facility (\texttt{IRAF}) software through the Gemini Virtual Machine (GVM\footnote{For information on the GVM, see the documentation at \url{https://gemini-iraf-vm-tutorial.readthedocs.io/en/stable/index.html}}). We used an in-house data processing pipeline, which carries out the following steps. Firstly, all snapshots of each source (target, bias fields, flat fields, and lamps) are combined into a single file using the \texttt{zerocombine} task, before applying bias corrections and flat-field corrections using the \texttt{imarith} task. Secondly, the one-dimensional spectrum was extracted using the \texttt{apall} task, during which we defined the aperture, extracted the spectrum, and traced the spectrum along the dispersion axis. We then performed wavelength calibration by identifying features in the extracted lamp spectrum and cross-referenced them with reference spectra to fit for the channel-to-wavelength conversion, using the \texttt{identify} and \texttt{dispcor} functions. Finally, we then performed the absolute flux calibration on the wavelength-calibrated spectrum using the observations of Feige~34, and tasks \texttt {standard} and \texttt{calibrate}. This process was carried out on each Grism independently.

\section{Results \& Discussion: Group-Member Galaxies}\label{sec:results:galaxies}
\begin{figure*}
\begin{center}
\includegraphics[width=0.98\textwidth]{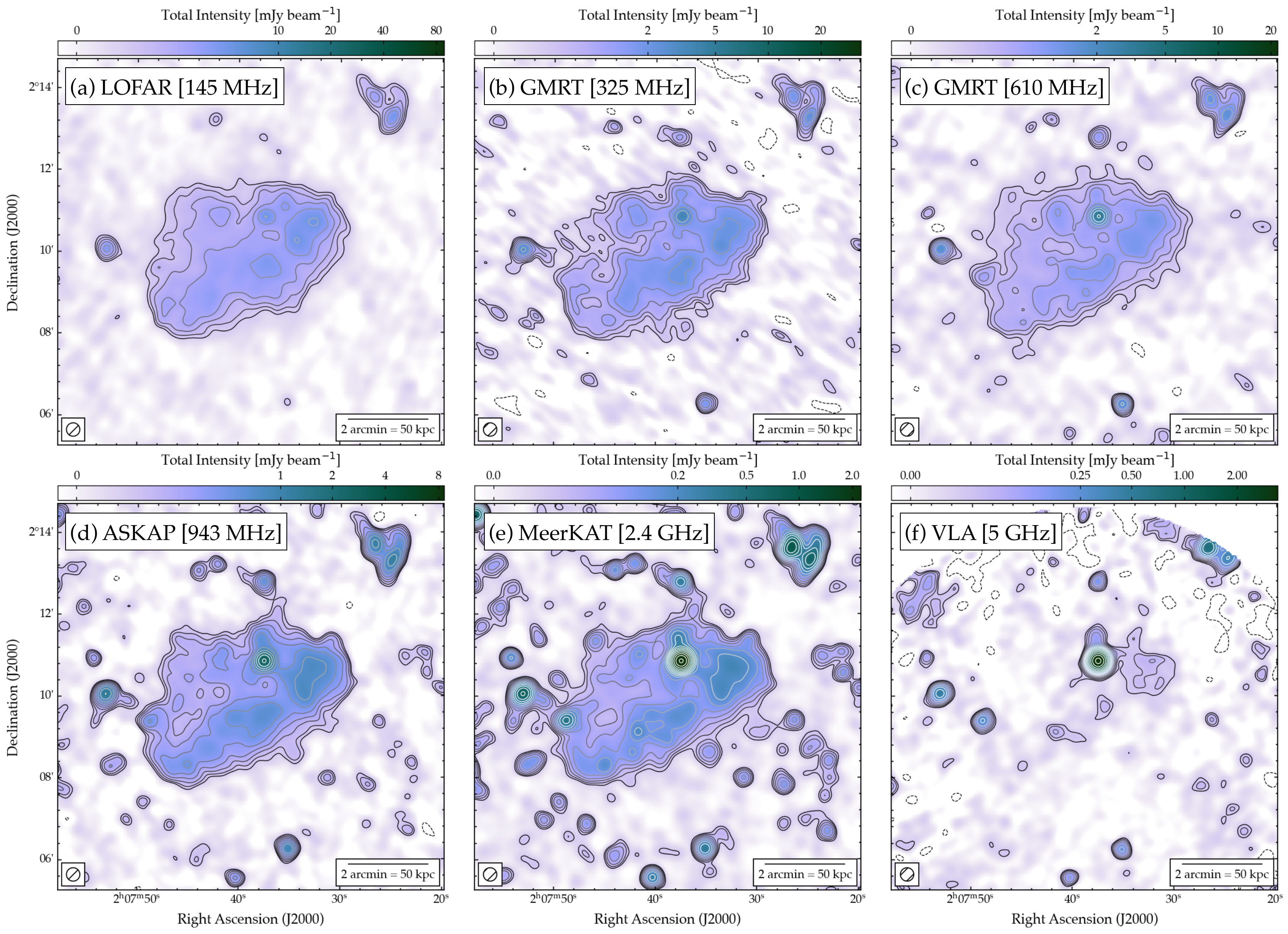}
\caption{Radio continuum maps of HCG15 from 145\,MHz to 5\,GHz, shown before subtraction of emission from discrete radio sources. From top-left, panels show (a) LOFAR at 145\,MHz, (b) GMRT at 325\,MHz, (c) GMRT at 610\,MHz, (d) ASKAP at 943\,MHz, (e) MeerKAT at 2.4\,GHz, and (f) VLA at 5\,GHz. All maps are shown at a resolution of 20\,arcsec, indicated by the hatched circle in the lower-left corner. Note that panel (f) shows the image produced from VLA-D configuration data only due to the variability of HCG15-D (see Section~\ref{sec:hcg15-d}). Contours start at $3\sigma$ and scale by a factor of $\sqrt{2}$, where $\sigma$ values are reported in Tab.\,\ref{tab:img_summary}. Colourmaps range from $-1\sigma$ to $300\sigma$ on an arcsinh stretch to emphasise faint emission.}
\label{fig:HCG15_Continuum}
\vspace{-5mm}
\end{center}
\end{figure*}

\begin{figure*}
\begin{center}
\includegraphics[width=0.98\textwidth]{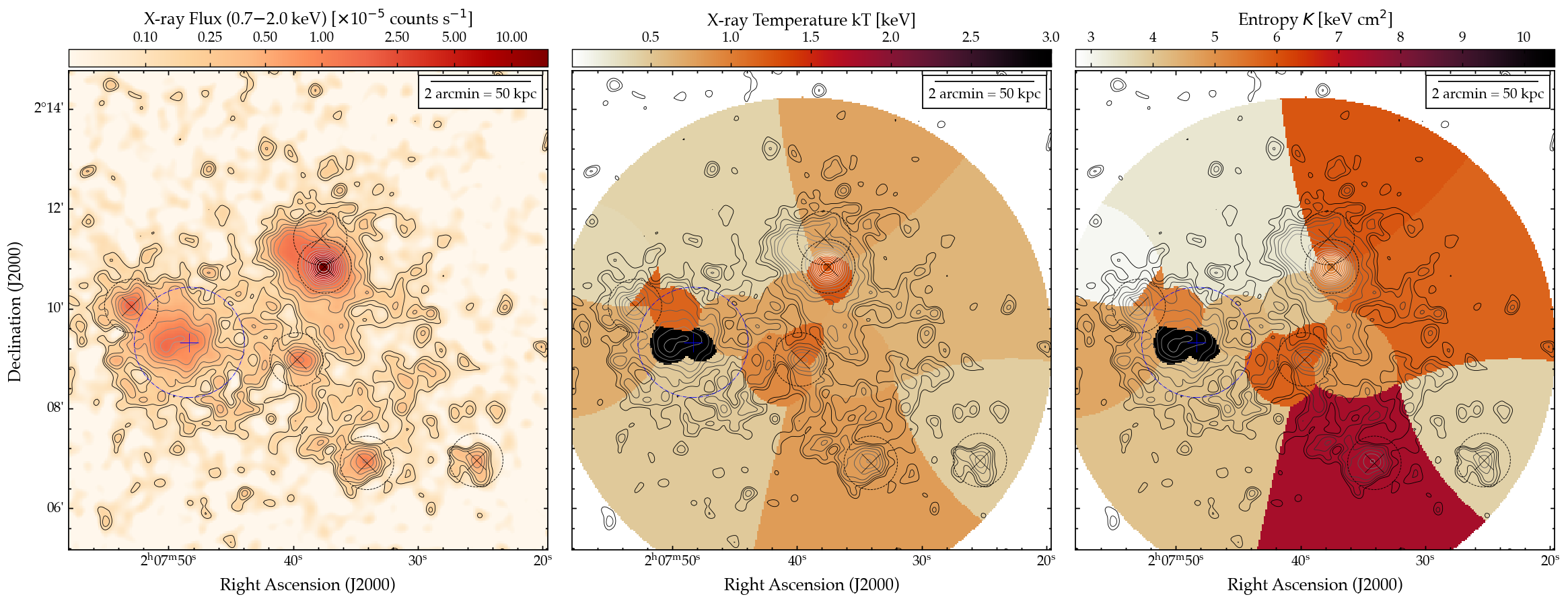}
\caption{Thermodynamic maps of HCG15 produced from \textit{XMM-Newton}. From left to right, panels show the X-ray surface brightness, X-ray temperature, and pseudo-entropy. Contours show the X-ray surface brightness as per the left panel, starting at a level of $1.2\times10^{-6}$ counts s$^{-1}$ ($4\sigma$) and scaling by a factor of $\sqrt{2}$. The `X' markers in circles denote the positions of the six group-member galaxies. The `+' indicates the SZ position of the background cluster at $z \simeq 0.88$ reported by \protect\cite{Klein2024_ACT-DR5_MCMF}, with the circle denoting a 500~kpc radius at redshift 0.88.}
\label{fig:HCG15_Xray}
\vspace{-5mm}
\end{center}
\end{figure*}

Figure~\ref{fig:HCG15_Continuum} shows our radio continuum maps of HCG15 between 145\,MHz and 5\,GHz at a common resolution of 20\,arcsec. Properties of these images are reported in Table\,\ref{tab:img_summary}. Figure~\ref{fig:HCG15_Xray} shows the X-ray surface brightness map derived from our \textit{XMM-Newton} data, smoothed with a Gaussian kernel equal to the on-axis PSF of \textit{XMM-Newton}. For the remainder of this section, we discuss our results on the group member galaxies; we discuss the diffuse radio emission separately in Section~\ref{sec:results:diffuse}.

\begin{table}
\centering
\caption{Summary of radio image properties for images of HCG15 at 20~arcsec resolution. RMS noise values reported here are the average of several off-source measurements in the vicinity of the phase centre. \label{tab:img_summary}}
\renewcommand{\arraystretch}{1.2}
\begin{tabular}{lccc}
\hline
Telescope   & Freq. & \texttt{Robust} & RMS noise     \\
  	    & $[$GHz$]$ & &  $[\upmu$Jy beam$^{-1}]$  \\
\hline\hline
\multirow{1}{*}{LOFAR}   & \multirow{1}{*}{0.145} & $-0.5$   & 294   \\
\hline
                         & \multirow{1}{*}{0.325} & $-0.5$   & 112   \\
\multirow{-2}{*}{GMRT}   & \multirow{1}{*}{0.610} & $-0.5$  &  72    \\
\hline
\multirow{1}{*}{ASKAP}   & \multirow{1}{*}{0.943} & $-0.5$   & 28.6  \\
\hline
\multirow{1}{*}{MeerKAT} & \multirow{1}{*}{2.412} & $-0.5$   & 7.4   \\
\hline
                         & \multirow{1}{*}{5.000} & $+1.0$  & 11.3  \\
\multirow{-2}{*}{VLA-D}  & \multirow{1}{*}{7.000} & $+1.0$  & 14.4  \\
\hline
                          & \multirow{1}{*}{5.000} & $0.0$  & 4.9   \\
\multirow{-2}{*}{VLA-C+D} & \multirow{1}{*}{7.000} & $0.0$  & 7.2   \\
\hline
\end{tabular}
\end{table}


\begin{figure*}
\centering
\includegraphics[width=0.8\textwidth]{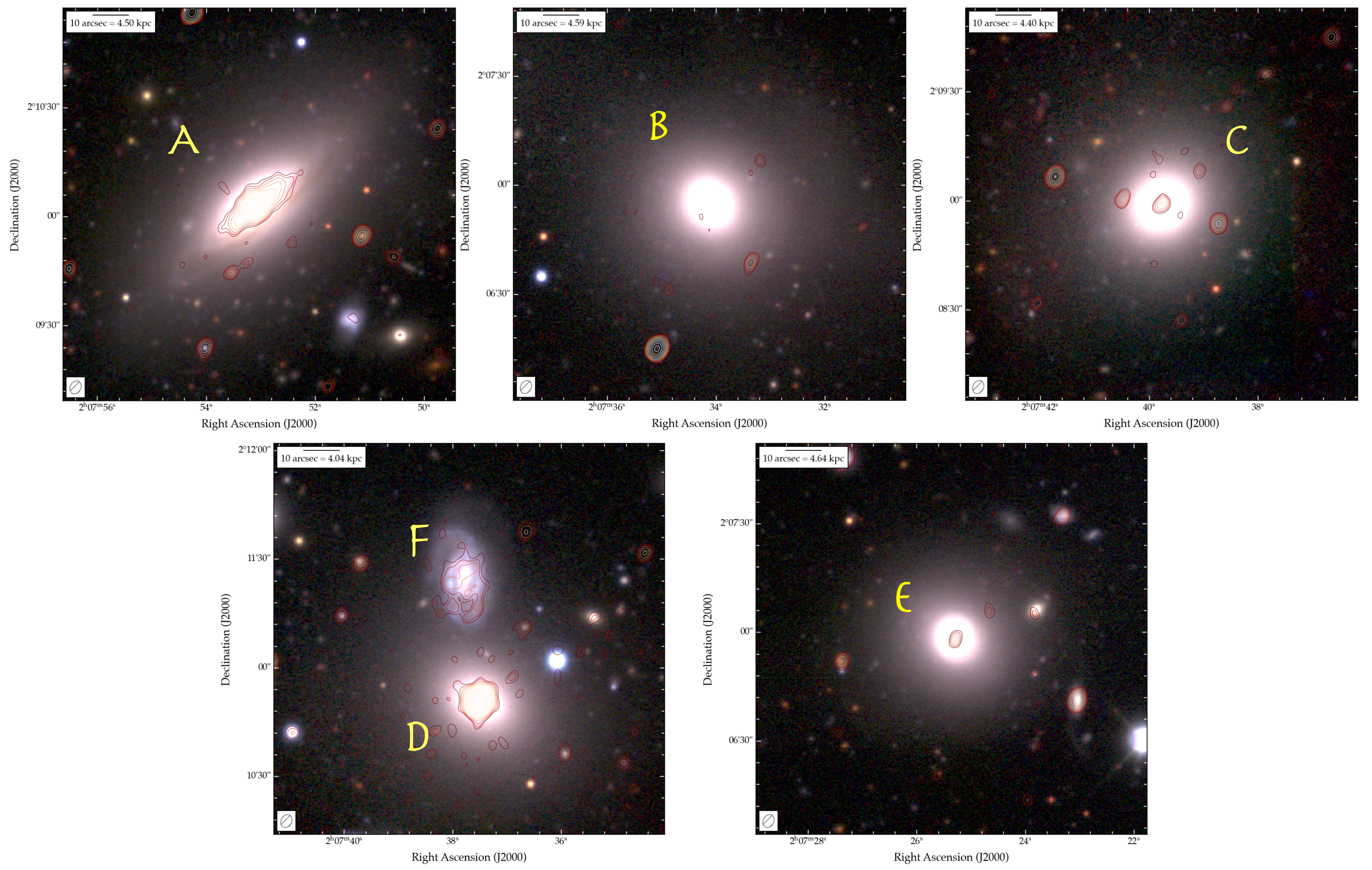}
\caption{Colour-composite images of the six group members in HCG15, overlaying radio contours on optical RGB. The optical RGB image is the same as Figure~\ref{fig:HCG15_Composite} (\textit{top left}); radio data shown is our high-resolution MeerKAT image at 2.4\,GHz ($3.73 \times 2.93$~arcsec$^2$). Contours start at $3\sigma$ and scale by a factor $\sqrt{2}$, where $\sigma = 3~\upmu$Jy~beam$^{-1}$. No radio emission was detected in association with HCG15-B by any of the telescopes involved in this work.}
\label{fig:HCG15_Composite_HighRes}
\end{figure*}

\begin{figure}
    \centering
    \includegraphics[width=0.95\linewidth]{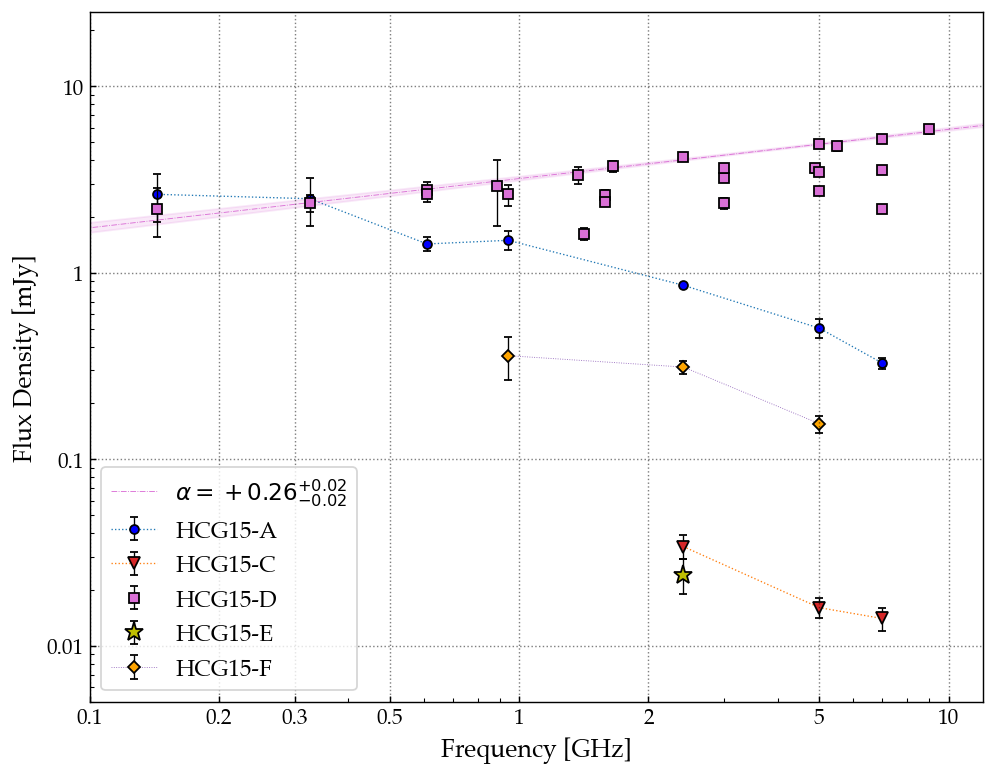}
\caption{Integrated radio SED for the galaxies HCG15-A, -C, -D, -E and -F. Dot-dashed pink line denotes power-law fit to datapoints for the spectrum of HCG15-D, which has a spectral index $\alpha = +0.26^{+0.02}_{-0.02}$ for the main trend trend. The shaded region marks the $1\sigma$ uncertainty. Datapoints for HCG15-A, -C and -F are connected by simple lines to guide the reader.}
\label{fig:hcg15_sed_galaxies}
\end{figure}

HCG15 contains six group member galaxies, HCG15-A through -F. All group members aside from HCG15-B are detected in our radio observations in at least one frequency. Figure~\ref{fig:HCG15_Composite_HighRes} presents colour-composite images of the group-member galaxies showing our high-resolution MeerKAT data at 2.4\,GHz overlaid on the DES optical RGB. Radio continuum flux density measurements for the five members detected at radio wavelengths are listed in Table~\ref{tab:galaxy_fluxes} and presented in Figure~\ref{fig:hcg15_sed_galaxies}.

\subsection{Galaxy HCG15-A}
HCG15-A is a Sa-type galaxy with B = 14.87\,mags \citep{Hickson1989} that lies to the east of the group. The radial velocity is $cz = 6967 \pm 30\,$km\,s$^{-1}$ \citep[$z = 0.02324$;][]{Hickson1992}. As is visible in Figure~\ref{fig:HCG15_Composite_HighRes}, HCG15-A exhibits a prominent dust lane, highlighting the morphology and inclination with respect to the observer. This galaxy also shows a wide H\textsc{i} detection with Arecibo from the ALFALFA survey \citep{Haynes2018_ALFALFA}. This H\textsc{i} emission has a systemic velocity of $v_{\rm H\textsc{i}} = 7010$~km~s$^{-1}$ and a flux density of $S_{\rm H\textsc{i} } = 1.59 \pm 0.14$~Jy~km~s$^{-1}$, corresponding to an H\textsc{i} mass of $M_{\rm H\textsc{i} } = (3.54 \pm 0.02) \times 10^9 ~ {\rm M}_{\odot}$ \citep[see][]{Haynes2018_ALFALFA}.

At radio wavelengths, the emission is dominated by a compact component potentially associated with an AGN in the centre of the galaxy, although at some frequencies we recover a symmetric extended component that largely follows the optical disk, as seen in Figure~\ref{fig:HCG15_Composite_HighRes} and noted in Table~\ref{tab:galaxy_fluxes} where appropriate. The suggestion that HCG15-A hosts a central AGN is supported by our X-ray data, as \textit{XMM-Newton} detects a compact component\footnote{The on-axis PSF of \textit{XMM-Newton} is 6~arcsec FWHM.} with an X-ray flux of $F_{\rm (X, ~ [0.7-2.0\,keV])} = (3.42 \pm 0.34) \times 10^{-15}$~ergs~cm$^{-2}$~s$^{-1}$, equivalent to an X-ray luminosity of $L_{\rm X} = (3.87 \pm 0.38) \times 10^{39}$~ergs~s$^{-1}$.

The $k$-corrected radio power $P_{\nu}$ at frequency $\nu$ is expressed as:
\begin{equation}\label{eq:radio_lum}
    P_{\nu} = 4 \pi \,  D_{\rm L}^2 \, S_{\nu} \, (1 + z)^{-(1 + \alpha)} \, ,
\end{equation}
where $D_{\rm L}$ is the luminosity distance to the object and $S_{\nu}$ is the flux density at frequency $\nu$. Given our assumed cosmology, the luminosity distance of HCG15-A is 97.2\,Mpc. Thus, interpolating our flux density measurements from ASKAP and MeerKAT, we derive a 1.4\,GHz radio luminosity of $P_{\rm 1.4 \, GHz} = (1.45 \pm 0.13) \times 10^{21}$\,W\,Hz$^{-1}$ from Equation~\ref{eq:radio_lum}.

As we see both a central compact component as well as an extended disk in the radio, an important consideration is the balance between AGN activity and star formation. We can use known scaling relations between radio luminosity and star formation rates to predict the radio luminosity due to star formation; we use the updated scaling relations derived by \cite{Gurkan2018_LOFAR-HATLAS} from the LOFAR/\textit{Herschel}-ATLAS field. Inserting the stellar masses and star formation rates in our Table~\ref{tab:hcg15_galaxies}, and the coefficients in their table 5 into their equation 3, we predict a star-formation-driven radio luminosity of $P_{\rm 1.4 \, GHz ,\, SF} \simeq 1.71 \times 10^{21}$\,W\,Hz$^{-1}$ for HCG15-A, consistent with our observed radio luminosity.

We can also compare our observed X-ray luminosity with known scaling relations as a further test of whether an AGN may be present in HCG15-A. Using the completeness-corrected scaling relations from the eROSITA Final Equatorial Depth Survey \citep[eFEDS;][]{Brunner2022_eFEDS} derived by \cite{Riccio2023_eFEDS_SFR} (their table 4), we predict a star-formation driven X-ray luminosity of $L_{\rm X, \, SF} \simeq 3.39 \times 10^{39}$~ergs~s$^{-1}$, consistent with our measured X-ray luminosity. As such, both the observed radio and X-ray luminosities are consistent with those predicted from star formation, negating the requirement for an AGN to be present in HCG15-A.

\begin{table}
\footnotesize
\centering
\caption{Radio flux density measurements for the galaxies in HCG15. Note that no radio emission associated with HCG15-B is detected by any of the telescopes involved in this work, and as such it is omitted from this table. For HCG15-A, we note which components are detected: a point source associated with the central AGN, plus in some cases extended emission associated with a likely star-forming disk. \label{tab:galaxy_fluxes}}
\renewcommand{\arraystretch}{1.2}
\begin{tabular}{lccl}
\hline
Galaxy                      & Frequency & Flux density        & Notes \& Comments    \\
  	                    & $[$GHz$]$ & $[$mJy$]$           &          \\
\hline\hline
                            & 0.145     & $2.632 \pm 0.764$   & AGN       \\
                            & 0.325     & $2.493 \pm 0.709$   & AGN       \\
                            & 0.610     & $1.426 \pm 0.125$   & AGN \& partial disk \\
                            & 0.943     & $1.492 \pm 0.174$   & AGN \& disk  \\
                            & 2.412     & $0.856 \pm 0.008$   & AGN \& disk  \\
                            & 5.000     & $0.504 \pm 0.059$   & AGN \& disk  \\
\multirow{-7}{*}{A}         & 7.000     & $0.328 \pm 0.023$   & AGN, near cutoff       \\
\hline
                            & 2.412     & $0.034 \pm 0.005$   &          \\
                            & 5.000     & $0.016 \pm 0.002$   &          \\
\multirow{-3}{*}{C}         & 7.000     & $0.014 \pm 0.002$   &          \\
\hline
                            & 0.145     & $2.194 \pm 0.649$   &          \\
                            & 0.325     & $2.361 \pm 0.247$   & 18\_007 (2010/06)    \\
                            & 0.610     & $2.790 \pm 0.281$   & 18\_007 (2010/04)    \\
                            & 0.610     & $2.657 \pm 0.268$   & 45\_047 (2023/11)         \\
                            & 0.888     & $2.907 \pm 1.135$   & RACS-Low \\
                            & 0.943     & $2.630 \pm 0.340$   & EMU Pilot II         \\
                            & 1.368     & $3.340 \pm 0.335$   & RACS-Mid \\
                            & 1.412     & $1.620 \pm 0.114$   & VLA FIRST          \\
                            & 1.580     & $2.398 \pm 0.131$   & VLA-B (1984/01)         \\
                            & 1.580     & $2.615 \pm 0.144$   & VLA-A (1984/12)         \\
                            & 1.655     & $3.729 \pm 0.254$   & RACS-High                  \\
                            & 2.412     & $4.179 \pm 0.085$   &          \\
                            & 3.000$^{\dag}$     & $3.226 \pm 0.169$   & VLASS 1 (2017/11)  \\
                            & 3.000$^{\dag}$     & $2.357 \pm 0.161$   & VLASS 2 (2020/07)  \\
                            & 3.000$^{\dag}$     & $3.655 \pm 0.171$   & VLASS 3 (2023/07)  \\
                            & 4.890$^{\dag}$     & $3.630 \pm 0.210$   & VLA-A (1987/10)  \\
                            & 5.000$^{\dag}$     & $4.874 \pm 0.100$   & VLA-D (2013)  \\
                            & 5.000$^{\dag}$     & $2.743 \pm 0.139$   & VLA-C (2024/03)  \\
                            & 5.000$^{\dag}$     & $3.480 \pm 0.174$   & VLA-C (2024/04)  \\
                            & 5.500              & $4.760 \pm 0.250$   & ATCA \\
                            & 7.000$^{\dag}$     & $5.235 \pm 0.111$   & VLA-D (2013)  \\
                            & 7.000$^{\dag}$     & $2.187 \pm 0.109$   & VLA-C (2024/03)  \\
                            & 7.000$^{\dag}$     & $3.541 \pm 0.177$   & VLA-C (2024/03)  \\
\multirow{-23}{*}{D}        & 9.000              & $5.860 \pm 0.300$   & ATCA \\
\hline
E                           & 2.412     & $0.024 \pm 0.005$   &          \\
\hline
                            & 0.943     & $0.359 \pm 0.092$   &          \\
                            & 2.412     & $0.312 \pm 0.026$   &          \\
\multirow{-3}{*}{F}         & 5.000     & $0.155 \pm 0.016$   & (Partial recovery)   \\
\hline
\multicolumn{4}{p{8cm}}{$^{\dag}$ The radio counterpart to HCG15-D was found to exhibit variability at high frequencies. Flux density measurements from all observational epochs are reported.}
\end{tabular}
\end{table}

\subsection{Galaxy HCG15-B}
HCG15-B is a ${\rm B} = 15.31\, {\rm mags}$ elliptical galaxy with a radial velocity of $cz = 7117 \pm 36$\,km\,s$^{-1}$ \citep[$z = 0.02373$;][]{Hickson1989,Hickson1992}. \cite{Mendes_de_Oliveira_Hickson_1994} comment that HCG15-B exhibits a strong central dust lane, although no such lane is visible in the DES imaging shown in Figure~\ref{fig:HCG15_Composite_HighRes}. No radio emission is detected from this group member in any of the images produced by the instruments involved in this study, although \textit{XMM-Newton} detects a compact component with an X-ray flux of $F_{\rm (X, ~ [0.7-2.0\,keV])} = (2.16 \pm 0.26) \times 10^{-15}$~ergs~cm$^{-2}$~s$^{-1}$,$(L_{\rm X} = (2.55 \pm 0.31) \times 10^{39}$~ergs~s$^{-1})$. While this might suggest the presence of an AGN, the scaling relations derived from eFEDS predict a star-formation-driven X-ray luminosity of $L_{\rm X, \, SF} \simeq 2.36 \times 10^{39}$~ergs~s$^{-1}$, consistent with our measured X-ray luminosity. Similarly to HCG15-A, this argues against the presence of an AGN in HCG15-B.

\subsection{Galaxy HCG15-C}
HCG15-C is an elliptical galaxy with ${\rm B} = 14.91 \, {\rm mags}$ \citep{Hickson1989} located roughly at the centroid of the group. HCG15-C hosts a faint compact radio source associated with the galaxy centre, which we identify as a likely AGN. This suggestion is supported by the detection of compact emission by \textit{XMM-Newton}, which measures an X-ray flux of $F_{\rm (X, ~ [0.7-2.0\,keV])} = (5.51 \pm 0.27) \times 10^{-15}$~ergs~cm$^{-2}$~s$^{-1}$ $(L_{\rm X} = (5.93 \pm 0.29) \times 10^{39}$~ergs~s$^{-1})$. The recession velocity of this galaxy (and thus its membership of the group) has been the subject of discussion in the literature, which we address using our new optical spectroscopy in Section~\ref{sec:spectra}.

The compact radio AGN in HCG15-C is detected only by MeerKAT at 2.41\,GHz and the VLA at both 5 and 7\,GHz. Our flux density measurements suggest a spectrum that flattens toward higher frequencies, as we find two-point spectral indices of $\alpha_{\rm 2.4\,GHz}^{\rm 5\,GHz} = -1.03 \pm 0.26$ and $\alpha_{\rm 5\,GHz}^{\rm 7\,GHz} = -0.39 \pm 0.56$. From Equation~\ref{eq:radio_lum} we derive a 2.41\,GHz radio power of $P_{\rm 2.41 \, GHz} = (3.66 \pm 0.54) \times 10^{19}$\,W\,Hz$^{-1}$ or $P_{\rm 1.4 \, GHz} = (6.30 \pm 0.92) \times 10^{19}$\,W\,Hz$^{-1}$ extrapolating with the observed $\alpha = -1.03$.

As before, using the scaling relations from \cite{Gurkan2018_LOFAR-HATLAS} we estimate a predicted star-formation-driven radio luminosity of $P_{\rm 1.4 \, GHz,\, SF} = 8.4 \times 10^{20}$\,W\,Hz$^{-1}$ for HCG15-C. This is comparable to the total observed radio luminosity, and as such star formation could explain all the observed radio emission. In this case, the flattening spectrum between 5 and 7 GHz may represent the emerging free-free emission component due to star formation \citep[e.g.][]{Condon1992_Review,Galvin2016_StarburstSEDs}. However, given the compact radio morphology, such star formation would have to be taking place in the nuclear region of HCG15-C, and our new optical spectrum does not show evidence of ongoing star formation (see \S\ref{sec:spectra}).

Similarly, using the scaling relation from eFEDS, the star formation rate of HCG15-C would suggest star-formation-driven X-ray luminosity of $L_{\rm X, \, SF} \simeq 1.47 \times 10^{39}$~ergs~s$^{-1}$. Our measured X-ray luminosity is around a factor four higher, which would suggest that some AGN contribution is necessary to explain our measured X-ray luminosity. Full spectral decomposition to separate the contributions from AGN and star formation would be insightful, but is beyond the scope of this work.

\subsection{Galaxy HCG15-D}\label{sec:hcg15-d}
HCG15-D does not stand out particularly compared to the other group members in the optical, as it is an elliptical galaxy with ${\rm B = 15.60~mags}$ \citep{Hickson1989}. The radial velocity is $cz = 6244 \pm 36$\,km\,s$^{-1}$ \citep[dimensionless redshift $z = 0.0208$;][]{Hickson1992}. It is also detected as a compact source by \textit{XMM-Newton}, which measures an X-ray flux of $F_{\rm (X, ~ [0.7-2.0\,keV])} = (5.51 \pm 0.12) \times 10^{-14}$~ergs~cm$^{-2}$~s$^{-1}$ $(L_{\rm X} = (4.97 \pm 1.08) \times 10^{40}$~ergs~s$^{-1})$.

However, the compact radio component associated with this galaxy is far more standout. As seen in Figure~\ref{fig:HCG15_Continuum}, it is increasingly dominant toward higher frequencies, suggesting an inverted spectrum, and it is compact at the resolution of all our instruments (see e.g. Figure~\ref{fig:HCG15_Composite_HighRes}) down to a limiting angular size of $\lesssim 2$~arcsec (linear size $\lesssim 0.9$~kpc).

Indeed, the flux density measurements listed in Table~\ref{tab:galaxy_fluxes} and presented in Figure~\ref{fig:hcg15_sed_galaxies} clearly demonstrate that the compact AGN in HCG15-D shows an inverted spectrum. Further, the broad-band SED shows significant \textit{temporal} variability at high frequencies, confirmed frequencies above 3\,GHz. Other compact sources detected in all VLA maps show flux density measurements consistent to within the measurement uncertainty between epochs.

Fitting a simple power-law relation to \textit{all} measurements, we find a spectral index of $\alpha = +0.21 \pm 0.02$; omitting those measurements that show significant departure from the overall trend, we find a somewhat flatter spectrum of $\alpha = +0.26 \pm 0.02$. It is this latter fit that is shown in Figure~\ref{fig:hcg15_sed_galaxies}.

For HCG15-D, the redshift of $z = 0.0208$ corresponds to a $D_{\rm L} = 86.8$~Mpc given our cosmology; thus, Equation~\ref{eq:radio_lum} suggests a radio power of $P_{\rm 150 \, MHz} = (1.58 \pm 0.09) \times 10^{21}$~W~Hz$^{-1}$ and $P_{\rm 1.4 \, GHz} = (2.55 \pm 0.05) \times 10^{21}$~W~Hz$^{-1}$. For completeness, we compared our observed radio and X-ray luminosities with those predicted from the observed star formation rates, respectively $P_{\rm 1.4 \, GHz,\, SF} = 3.5 \times 10^{20}$\,W\,Hz$^{-1}$ and $L_{\rm X, \, SF} \simeq 1.02 \times 10^{39}$~ergs~s$^{-1}$. Our observed luminosities are around a factor seven higher in the case of radio luminosity and a factor 40 higher in the case of X-ray luminosity than expected from star formation, providing further support for the presence of an AGN.

The variability we observe can be quantified using various methods, e.g. the debiased variability index \citep[DVI; e.g.][]{Barvainis2005_variability}, which quantifies the variability in excess of the measurement uncertainties and is defined as:
\begin{equation}
    {\rm DVI} = \sqrt{ \frac{ \sigma^2_{S, \nu} - \langle \delta^2_S \rangle }{ \langle S_{\nu} \rangle^2 } }
\end{equation}
where $\sigma^2_{S, \nu}$ is the variance of the source lightcurve at frequency $\nu$, $\langle \delta^2_S \rangle$ is the mean squared flux density error, and $\langle S_{\nu} \rangle$ is the mean flux density at frequency $\nu$. 

Given the measurements presented in Table~\ref{tab:galaxy_fluxes}, the value of DVI increases toward higher frequencies from 16\% at 3\,GHz to 42\% at 7\,GHz, implying increasing variability with frequency. The timescales of this variability --- from months to years --- are inconsistent with interstellar scintillation (ISS), which would predict variability on the order of hours \citep[e.g.][]{Walker1998}, and cannot explain the change in spectral shape in the VLA C-band observations, thus supporting the interpretation that the variability of HCG15-D is driven by astrophysical processes within the compact source (i.e. linked to the AGN) rather than propagation effects. Our data may suggest long-term variability around 1.4\,GHz as the measurements from archival VLA data from the NVAS and FIRST survey \citep{Becker1994} show strong discrepancy with more recent observations from RACS-Mid/RACS-High \citep{Duchesne2024_RACS-Mid,Duchesne2025_RACS-High}. However, there is no variability at lower frequencies, as GMRT data from 2010 and 2023, as well as our ASKAP-EMU and RACS-Low \citep{Hale2021_RACS-Low} measurements (from 2021 and 2019 respectively), are consistent.

Based on established taxonomy, the compact radio AGN associated with HCG15-D most fits the class of peaked-spectrum objects known as `high-frequency peaked' (HFP) sources \citep[e.g.][]{ODea1998_CSS_GPS,Dallacasa2000_HFP,Tinti2005_HFP}, which show a peak in the SED at high frequencies (typically $\gtrsim$5\,GHz). Such peaked-spectrum (PS) sources represent an opportunity to follow the evolution of AGN from compact PS sources to extended radio galaxies, as these pathways remain unclear. Competing theories suggest that they may be `young' (age $\lesssim 10^5$\,yr) and thus simply have not yet grown to their full size \citep[e.g.][]{ODea_Baum_1997} or may be `frustrated' and confined on small spatial scales by the dense ISM \citep[e.g.][]{ODea1991_GPS}. Further, some recent studies show peaked-spectrum cores embedded within remnant lobes, suggesting restarted AGN activity \citep[e.g.][]{HernandezGarcia2019}. Such a scenario appears consistent with our observations of HCG15.

The nature of the AGN in HCG15-D remains unanswered by our existing data, and the physical origins of the variability found at higher frequencies requires both further monitoring and high resolution observations. Recent studies have identified PS sources as a more variable population than typical AGN \citep[e.g.][]{Ross2021_PS_variability} yet distinguishing the physical mechanism that is causing the observed variability relies on spectral \textit{and} temporal coverage. Both characterisation of the spectral variability and high resolution imaging have been used to differentiate between the ``youth'' and ``frustration'' scenarios \citep[e.g.][]{Phillips1982,Tzioumis2010,Tingay2015,Ross2022_PS_variability}. For example, variability below the spectral turnover can often only be explained by variations in the optical depth due to an inhomogeneous surrounding medium, pointing towards the ``frustration'' scenario. However, in the case of HCG15-D, the lack of variability at even lower frequencies suggests this explanation is insufficient and a complex interaction or physical process may be involved. Likewise, the variability of HCG15-D may be explained by a young evolving ejecta, like a tidal disruption event (TDE), but any confidence in this conclusion would rely on monitoring of the changing spectra of the evolving ejecta or direct imaging from high resolution observations. HCG15-D may be revisited in future work, as our team has recently been awarded Very Long Baseline Interferometry (VLBI) observations with the enhanced Multi Element Remotely Linked Interferometer Network (eMERLIN) to probe this variability at subarcsecond resolution (project CY18006; PI: Riseley).

\subsection{Galaxy HCG15-E}
HCG15-E lies to the south-west of the group. Although initially reported as an Sa-type galaxy \citep{Hickson1989}, in the DES imaging (Figure~\ref{fig:HCG15_Composite_HighRes}) it more resembles a classical elliptical galaxy akin to HCG15-B, -C and -D. It is a faint galaxy with $\rm B = 15.94 \, mags$ \citep{Hickson1989} and a radial velocity $cz = 7197 \pm 32$\,km\,s$^{-1}$ \citep[dimensionless redshift $z = 0.02401$;][]{Hickson1992}. We detect a very faint compact radio source associated with this galaxy in our MeerKAT data only, $S_{\rm 2.41\,GHz} = 24.0 \pm 0.5 ~ \upmu$Jy; Assuming a typical synchrotron spectral index of $\alpha = -0.8$, we estimate a radio luminosity of $P_{\rm 1.4 \, GHz} = (4.46 \pm 0.09) \times 10^{19}$~W~Hz$^{-1}$. We also note the presence of an X-ray component detected by \textit{XMM-Newton}, with a luminosity of $F_{\rm (X, ~ [0.7-2.0\,keV])} = (1.39 \pm 0.22) \times 10^{-15}$~ergs~cm$^{-2}$~s$^{-1}$) $(L_{\rm X} = (1.68 \pm 0.27) \times 10^{39}$~ergs~s$^{-1})$.

From scaling relations, we predict star-formation-driven luminosities of $P_{\rm 1.4 \, GHz,\, SF} = 8.5 \times 10^{20}$\,W\,Hz$^{-1}$ and $L_{\rm X, \, SF} \simeq 2.60 \times 10^{39}$~ergs~s$^{-1}$, consistent with our observed luminosities. As such, the observed X-ray and radio emission could be powered by star formation activity, although given the location and morphology of the radio component detected, such star formation would have to be occurring in the nuclear region of HCG15-E.

\subsection{Galaxy HCG15-F}
HCG15-F is a Sbc galaxy of $\rm B = 16.73 \, mags$ \citep{Hickson1989} close to HCG15-D. The radial velocity measurement from \cite{Hickson1992} is $cz = 6242 \pm 102$\,km\,s$^{-1}$ has a large uncertainty, likely due to the faintness of this galaxy. The imaging study of \cite{Mendes_de_Oliveira_Hickson_1994}, reports a clearly disturbed morphology, also visible in the DES imaging (Figure~\ref{fig:HCG15_Composite_HighRes}).

In the radio, we detect no compact components but extended diffuse emission that largely traces the inner region of the spiral. This emission is recovered by ASKAP at 943\,MHz, MeerKAT at 2.41\,GHz and the VLA at 5\,GHz; at lower frequencies it is not detectable in high-resolution images, and at lower resolution it is not possible to separate this emission from the larger-scale diffuse emission associated with the IGrM, and so we report only flux density measurements from ASKAP, MeerKAT and the VLA in Table~\ref{tab:galaxy_fluxes}, although we note that this emission is only partially recovered in our VLA-C observations. The emission exhibits a flat spectrum, with a two-point spectral index of $\alpha_{\rm 943\,MHz}^{\rm 2.41\,GHz} = -0.15 \pm 0.29$, perhaps suggestive of a thermal origin.

While some X-ray emission is visible at the location of HCG15-F in Figure~\ref{fig:HCG15_Xray}, the morphology indicates that this traces extended emission from the IGrM, and so HCG15-F is the only galaxy in the group to not have a detectable compact X-ray counterpart. 

However, in H\textsc{i} imaging by \cite{Jones2023_HCG-HI}, it is the only group member to show a significant H\textsc{i} detection, although the resolution of their study was too poor analyse the morphology in detail or to perform component separation. Similarly, HCG15-F is also catalogued by \cite{Haynes2018_ALFALFA} as part of the ALFALFA survey, at a systemic velocity $v_{\rm H\textsc{i}} = 6257$~km~s$^{-1}$. The measured H\textsc{i} line flux density is $S_{\rm H\textsc{i} } = 0.78 \pm 0.07$~Jy~km~s$^{-1}$, corresponding to a mass of $M_{\rm H\textsc{i} } = (1.41 \pm 0.01) \times 10^9 ~ {\rm M}_{\odot}$.

Overall, given the shallow radio spectrum, lack of an X-ray counterpart, the high sSFR for this galaxy, the detection of significant H\textsc{i} and the presence of emission lines associated with H$\alpha$ and $[$N\textsc{ii}$]$ (see Section~\ref{sec:spectra}) it is likely that the observed radio emission traces ongoing star formation.

\subsection{Additional group members}
\begin{figure}
\centering
\includegraphics[width=0.675\linewidth]{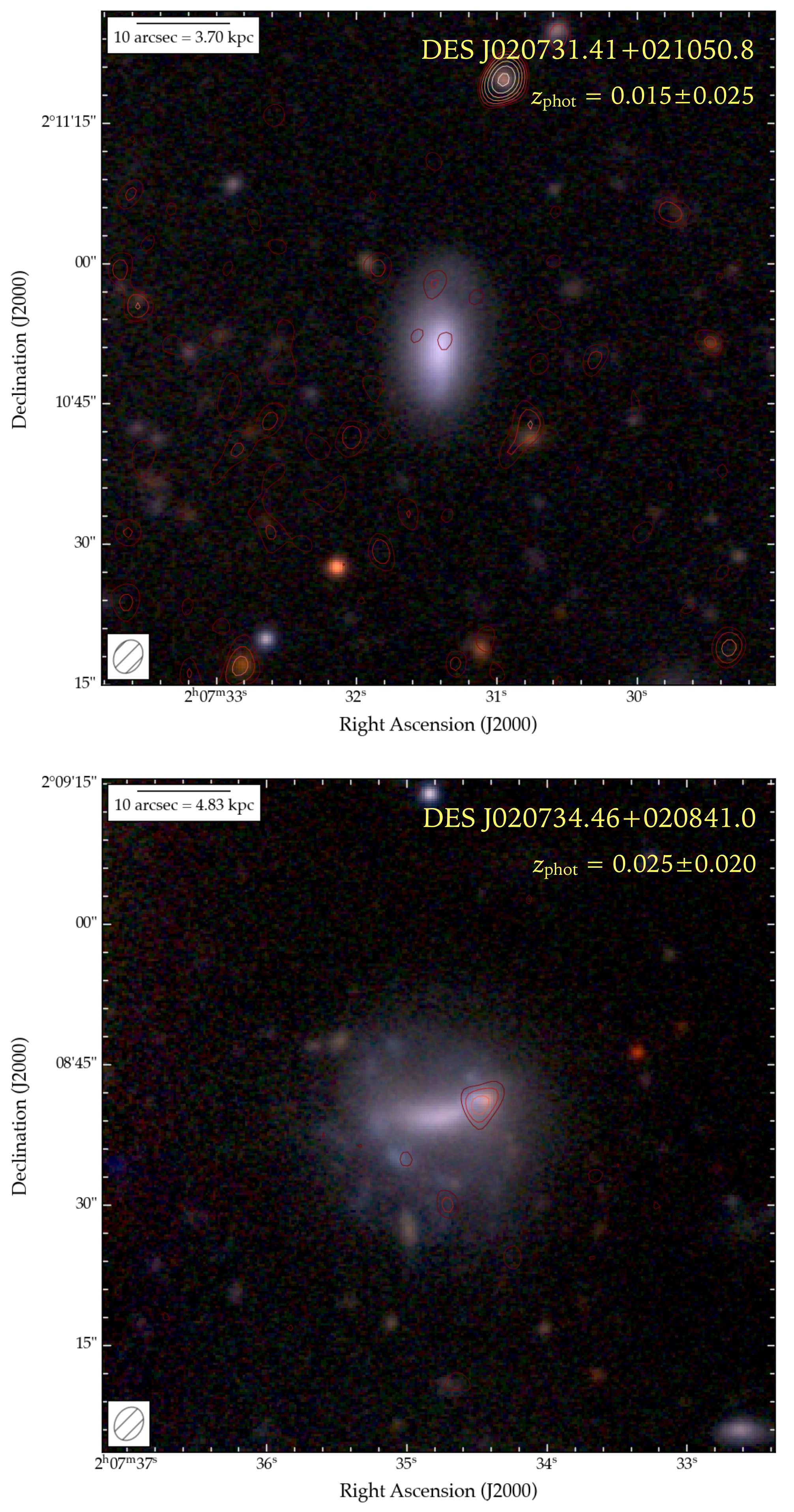}
\includegraphics[width=0.675\linewidth]{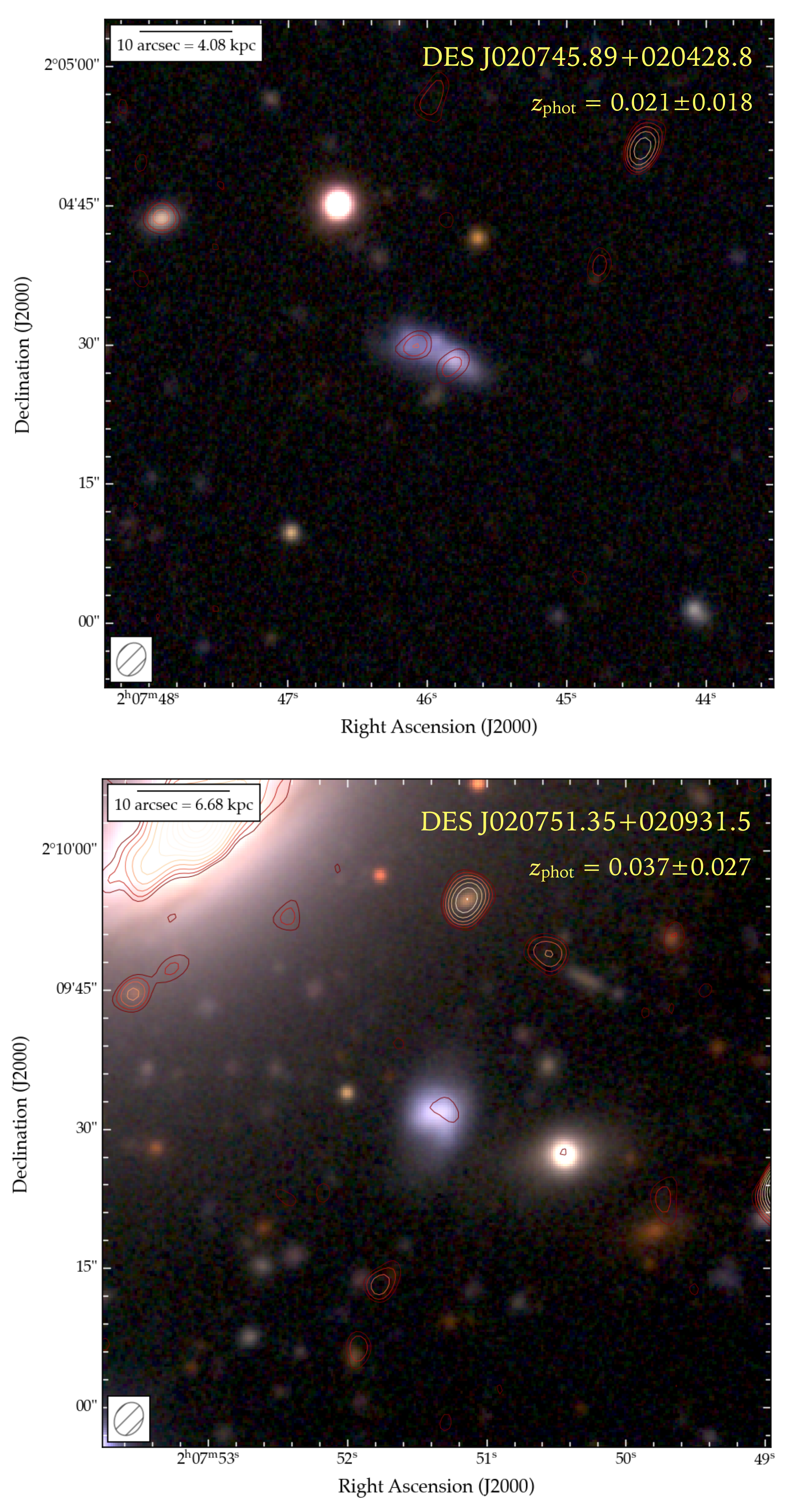}
\caption{Colour-composite images of candidate group members identified from the DESI Legacy Surveys DR8 \citep{Duncan2022_DESI_DR8}. Colour-composite and contours are as per Figure~\ref{fig:HCG15_Composite_HighRes}.}
\label{fig:HCG15_Candidates}
\end{figure}

As part of our investigation, we also searched for additional candidate group members from the Dark Energy Spectroscopic Instrument (DESI) Legacy Imaging Surveys Data Release 8 (DR8) catalogue \citep{Duncan2022_DESI_DR8}. We searched a sky area within 7~arcmin radius of HCG15, identifying four further candidate group members: DES~J020731.41+021050.8, DES~J020734.46+020841.0, DES~J020745.89+020428.8, and DES~J020751.35+020931.5 \citep{Abbott2021_DES-DR2}. All have photometric redshifts broadly consistent with the confirmed group members. We show colour-composite images of these galaxies in Figure~\ref{fig:HCG15_Candidates}. 
{\vspace{-1em}}

\paragraph{\textbf{DES~J020731.41+021050.8}} has a photometric redshift $z_{\rm phot} = 0.015 \pm 0.025$ in the DESI Legacy Surveys DR8. From the DES composite shown in Figure~\ref{fig:HCG15_Composite_HighRes}, this is likely a spiral galaxy. It is significantly fainter than the known group members, with ${\rm G} = 18.180 \pm 0.003$ mags \citep{Abbott2021_DES-DR2} although this is typical of the newly identified candidates.

This galaxy is located to the west of HCG15-D and -F, within the boundaries of the diffuse emission. As such, while the contours seen in Figure~\ref{fig:HCG15_Composite_HighRes} would appear to show radio emission associated with this galaxy, this is a chance overlay of diffuse emission that is shown at high resolution. We do not associate any significant radio emission with this galaxy.
{\vspace{-2em}}

\paragraph{\textbf{DES~J020734.46+020841.0}} lies close to the group centroid, around 1.3~arcmin from HCG15-C (around 43~kpc in projection). It is faint with ${\rm G} = 19.486 \pm 0.005$~mags \citep{Abbott2021_DES-DR2} and from Figure~\ref{fig:HCG15_Candidates} would appear to exhibit a somewhat barred-spiral morphology, with a largest linear size around 21~arcsec. At $z_{\rm phot} = 0.025 \pm 0.020$ \citep[DESI Legacy Surveys DR8;][]{Duncan2022_DESI_DR8} this would correspond to 10.6~kpc. We likewise note the presence of a single compact radio source in projection onto the western edge of this galaxy; at the same location there is a slight reddening of the optical colour-composite, which may hint at an association, although whether this might represent a background radio galaxy or emission associated with supernovae, for example, we cannot say conclusively.
{\vspace{-1em}}

\paragraph{\textbf{DES~J020745.89+020428.8}} is a faint (${\rm G} = 18.988 \pm 0.004$ mags; \citealt{Abbott2021_DES-DR2}) galaxy with a photometric redshift $z_{\rm phot} = 0.021 \pm 0.018$ in the DESI Legacy Surveys DR8. This galaxy lies around 4.7~arcmin (152~kpc projected) to the south of the group centre. From Figure~\ref{fig:HCG15_Candidates}, we see that in the optical this galaxy exhibits an irregular morphology, and we note two faint radio point sources (detected by MeerKAT at 2.41\,GHz) projected onto the optical emission. We do not have the data to determine whether these sources are physically associated with the galaxy or simply viewed in projection.

\begin{figure*}
 \centering
 \includegraphics[width=0.7\textwidth]{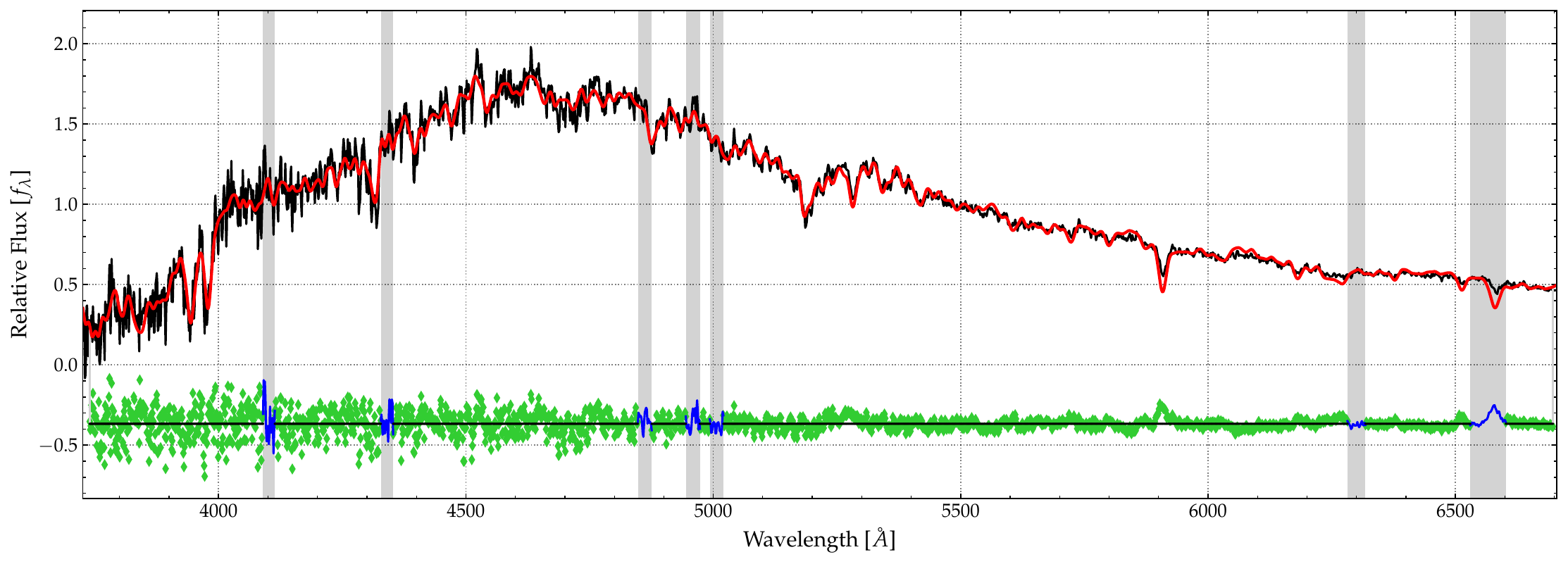}
 \includegraphics[width=0.45\textwidth]{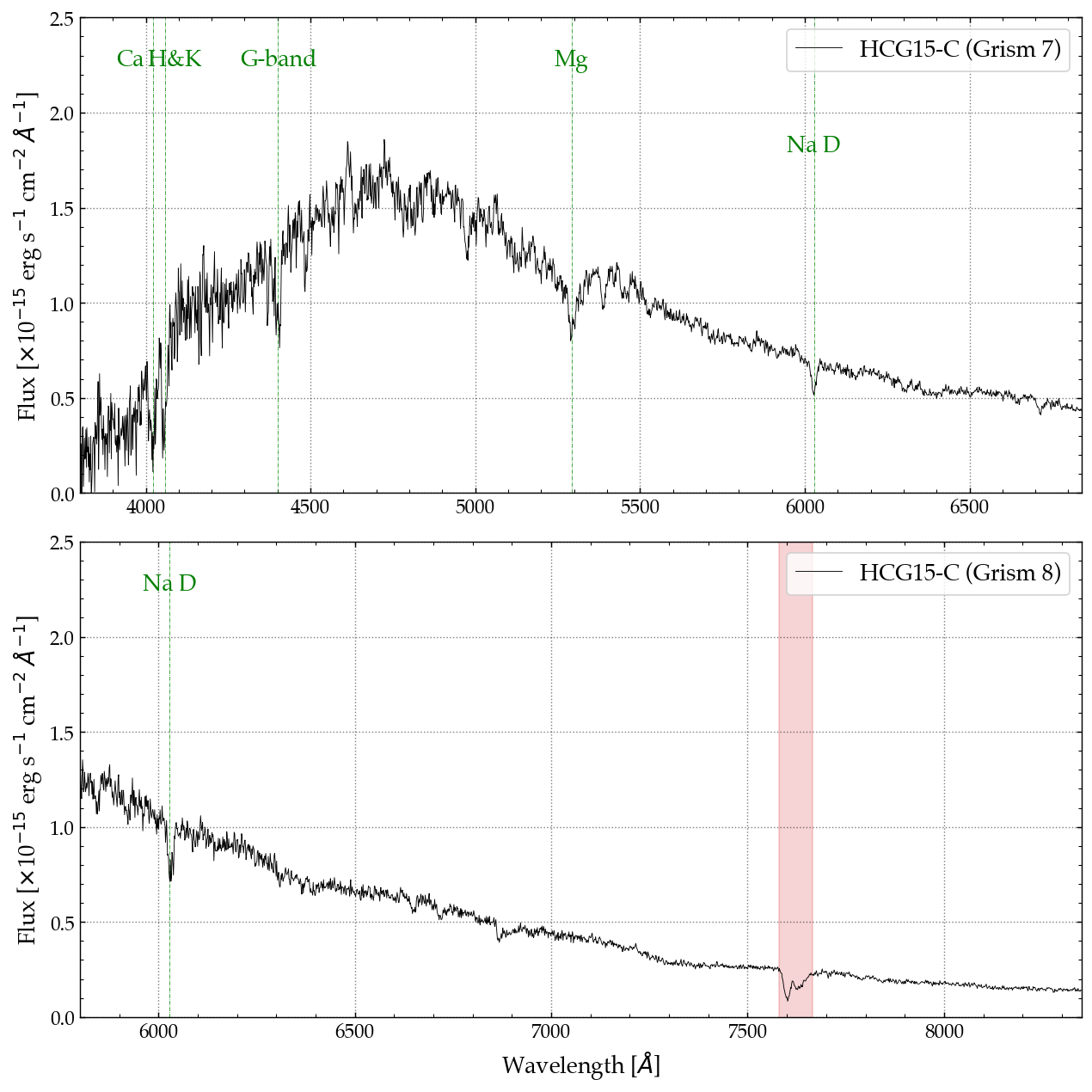}
 \includegraphics[width=0.45\textwidth]{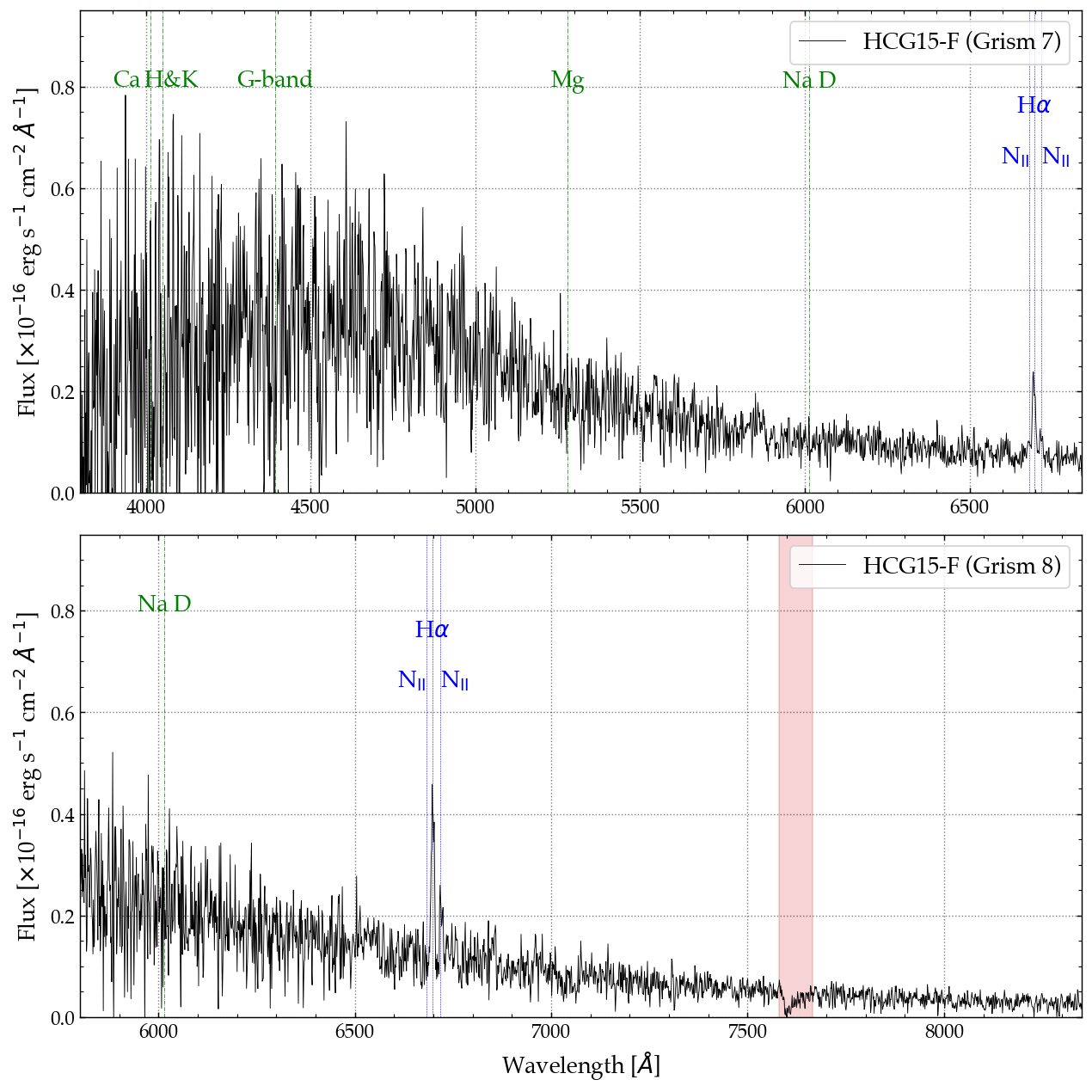}
\caption{Optical spectra from the HCT for HCG15-C (upper and left panels) and HCG15-F (right panels). The spectrum for each Grism is shown separately, with Grism-7 in the upper row and Grism-8 in the lower row. All wavelengths are quoted in the observer frame. Known emission and absorption features are marked at their observed wavelengths using the best-fit recession velocities of $6803 \pm 12$~km~s$^{-1}$ for HCG15-C and $6133 \pm 98$~km~s$^{-1}$ for HCG15-F, with green for absorption features and blue for emission lines. The shaded red region marks a known sky absorption feature. Note that in the spectrum of HCG15-C, no emission lines are detected, only the marked NaD, Mg, G-band and Ca H \& K absorption features. Conversely for HCG15-F the only detected feature is the H$\alpha$ emission line, with a tentative sign of the [N\textsc{ii}] doublet; no absorption features are detected.}
\label{fig:hcg15_hct_spectra}
\end{figure*}

This galaxy is small in angular size, around 12~arcsec in extent corresponding to a physical scale of 4.9~kpc at the catalogued $z_{\rm phot}$, fairly typical of dwarf irregular galaxies. However, we also note some discrepancy in this photometric redshift, as other surveys have derived different estimates. For example, from the DESI Peculiar Velocity Survey \citep{Saulder2023_DESI-PVS} the median value is $z_{\rm phot} = 0.048$, although the $2\sigma$ lower boundary estimate is consistent with that derived by \citep{Duncan2022_DESI_DR8}. Similarly, the Sloan Digital Sky Survey (SDSS) DR12 \citep{Alam2015_SDSS-DR12} derived a higher estimate of $z_{\rm phot} = 0.0556 \pm 0.0468$, although the large uncertainty again renders this estimate consistent. In general, we consider this galaxy as a less likely group member candidate than the others identified here due to its significant offset.
{\vspace{-1em}}

\paragraph{\textbf{DES~J020751.35+020931.5}} lies to the north-east of the group, at a projected distance of 38~arcsec (20~kpc) from HCG15-A. This galaxy has a G-band magnitude of ${\rm G} = 18.251 \pm 0.004$ mags \citep{Abbott2021_DES-DR2}. The photometric redshift of $z_{\rm phot} = 0.037 \pm 0.027$ is slightly higher than the representative redshift of the group, although given the large uncertainty it is consistent. 

The morphology of the galaxy is not easy to discern from Figure~\ref{fig:HCG15_Candidates}, although given the colour it is likely a spiral galaxy. In our high-resolution MeerKAT image at 2.41\,GHz, we recover a faint unresolved source co-located with this galaxy, although this emission is barely recovered at the $3\sigma$ level.\\

While the identification of these candidate group-member galaxies complicates the dynamical picture of HCG15 somewhat, and merits follow-up to confirm or refute their candidate status, our understanding of the nature of the diffuse emission is not significantly challenged. These galaxies all appear far fainter and more compact than the dominant six group-member galaxies; if they are indeed group members, then they are also far smaller. Further, none of them show evidence of significant AGN contribution via either radio or X-ray tracers, as none show compact components in our \textit{XMM-Newton} images.

\subsection{Optical Spectroscopy: HCG15-C and -F}\label{sec:spectra}
Originally identified as a group member by \cite{Hickson1982} with a radial velocity of $cz = 7222 \pm 30$\,km\,s$^{-1}$ \citep{Hickson1992}, later redshift survey data from the Updated Zwicky Catalogue \citep[UZC;][]{Falco1999} measured a significantly higher radial velocity of $cz = 9687$\,km\,s$^{-1}$. Our new HCT optical spectroscopic observations of HCG15-C and HCG15-F allow us to resolve the question of whether HCG15-C is a group member galaxy or an interloper at higher redshift. These spectra are presented in Figure~\ref{fig:hcg15_hct_spectra}.

\begin{figure*}
\begin{center}
\includegraphics[width=0.95\textwidth]{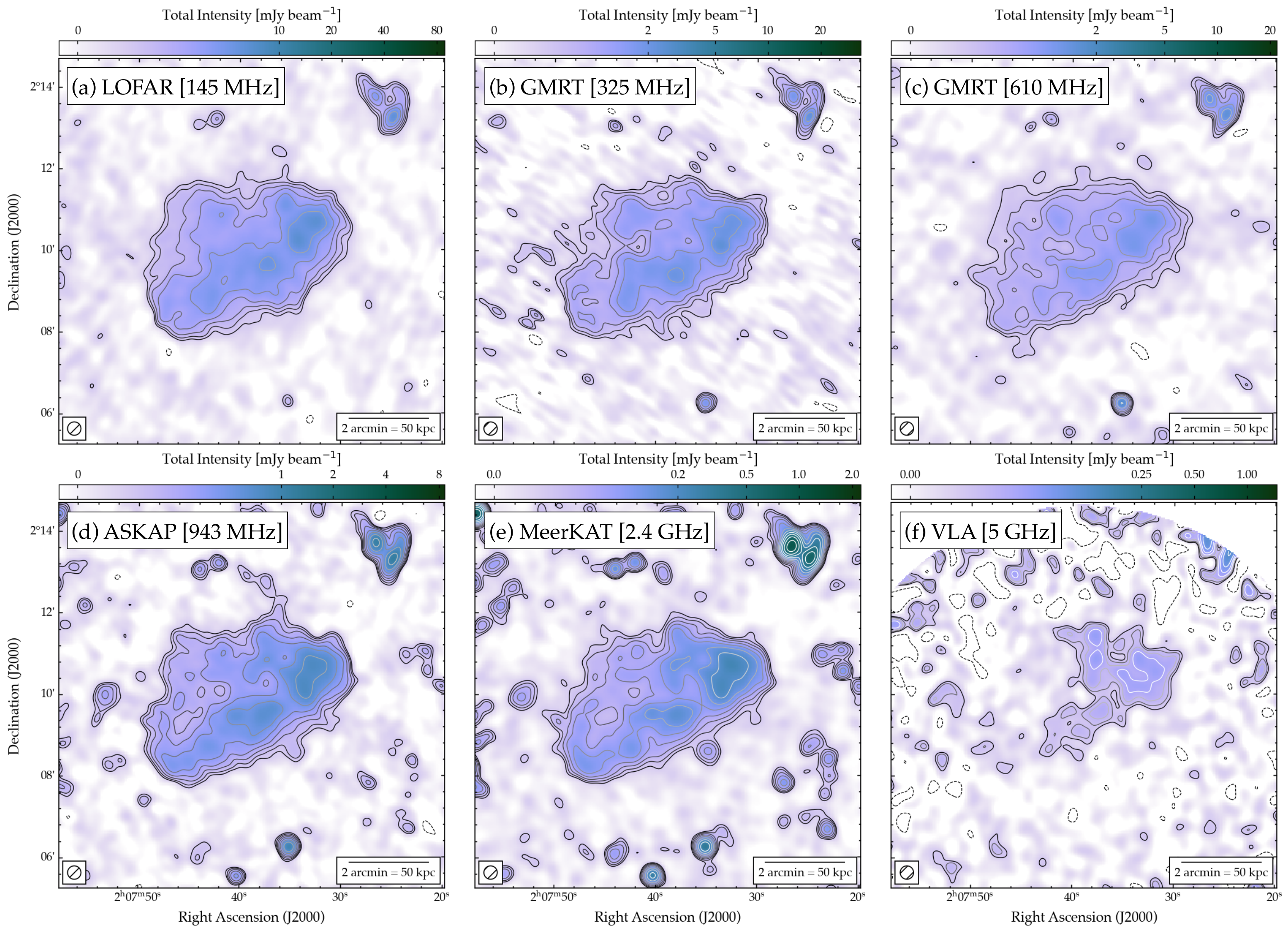}
\caption{Radio continuum maps of HCG15 from 145\,MHz to 5\,GHz, after subtraction of discrete radio sources embedded in the diffuse emission. From top-left, panels show (a) LOFAR at 145\,MHz, (b) GMRT at 325\,MHz, (c) GMRT at 610\,MHz, (d) ASKAP at 943\,MHz, (e) MeerKAT at 2.4\,GHz, and (f) VLA at 5\,GHz, produced from the combined VLA-C+D configuration dataset. All maps are shown at a resolution of 20\,arcsec, indicated by the hatched circle in the lower-left corner. Contours start at $3\sigma$ and scale by a factor of $\sqrt{2}$, where $\sigma$ values are reported in Tab.\,\ref{tab:img_summary}. Colourmaps range from $-1\sigma$ to $300\sigma$ on an arcsinh stretch to emphasise faint emission.}
\label{fig:HCG15_Continuum_Subtracted}
\vspace{-5mm}
\end{center}
\end{figure*}

We used the Python implementation of Penalized Pixel Fitting \citep[\texttt{PPxF}, e.g.][]{Cappellari2004_PPXF,Cappellari2017_PPXF,Cappellari2023_PPXF} to determine the best radial velocity for both HCG15-C and -F from our spectra, based on a linear combination of stellar templates that best describes the continuum. We performed these fits on our Grism-7 spectrum for HCG15-C as this spectrum showed the most features, and on our Grism-8 spectrum for HCG15-F as this spectrum shows greater signal-to-noise compared to our Grism-7 spectrum.

For HCG15-C, we find a best-fit radial velocity of $cz = 6803 \pm 12$~km~s$^{-1}$ (dimensionless redshift $z = 0.02269 \pm 0.0004$), confirming the original suggestion from \cite{Hickson1982} and \cite{Hickson1992} that HCG15-C \textit{is} a group-member galaxy. For HCG15-F we find a best-fit radial velocity of $cz = 6133 \pm 98$~km~s$^{-1}$ ($z = 0.02045$); while the uncertainties are large due to the faint nature of the galaxy ($\rm B = 16.73 \, mags$) this is consistent with the original measurement of $cz = 6242 \pm 102$~km~s$^{-1}$ from \cite{Hickson1992}.

From our optical spectra, we see that HCG15-C shows clear absorption features associated with the Sodium D ($\lambda\lambda = 5889 \, \& \, 5895$~\AA), Magnesium ($\lambda = 5175$~\AA), G-band ($\lambda = 4304$~\AA) and Calcium H \& K ($\lambda\lambda = 3968 \, \& \, 3933$~\AA) lines, but no emission lines which might indicate ongoing star formation or AGN activity. Conversely, while HCG15-F is significantly fainter and therefore our signal-to-noise is far lower, we see no absorption features but rather significant emission lines associated with H$\alpha$ ($\lambda = 6563$~\AA) and hints of the $[$N\textsc{ii}$]$ doublet ($\lambda\lambda = 6550 \, \& \, 6585$~\AA). The presence of these emission lines supports earlier evidence of ongoing or recent star formation activity.


\section{Results \& Discussion: Diffuse Emission}\label{sec:results:diffuse}
In order to isolate the diffuse emission associated with HCG15 for our analysis, we generated a model of the embedded sources by imaging our data with \texttt{WSclean}, applying a common inner \textit{uv}-cut of $5{\rm k}\lambda$ tuned to filter out the diffuse emission. This model was then subtracted from our data, which we re-imaged applying a combination of weighting, tapering, and image-plane convolution to achieve our target resolution of 20~arcsec.

In our analysis, we make use of two sets of source-subtracted radio images. For the first set, we do not employ a common inner \textit{uv}-cut, in order to estimate the total flux density of the source recovered at each frequency. For the second set, we adopt best practice and employ a common inner \textit{uv}-cut of 370$\lambda$, corresponding to the largest value of the shortest baseline in wavelengths (in this case the minimum baseline well sampled by the VLA in C-band). This corresponds to a largest recoverable angular scale of around 9~arcminutes, around twice the extent of the diffuse emission in HCG15, and so we do not anticipate a significant bias. Indeed, these images are qualitatively near-identical to those without a common inner \textit{uv}-cut, and so we do not show them separately. Quantitatively, the typical surface brightness differs by no more than seven per cent, less than the typical calibration uncertainty at each frequency, reinforcing the earlier suggestion that our analysis is not likely biased by scale size differences.

\begin{figure*}
\begin{center}
\includegraphics[width=0.75\textwidth]{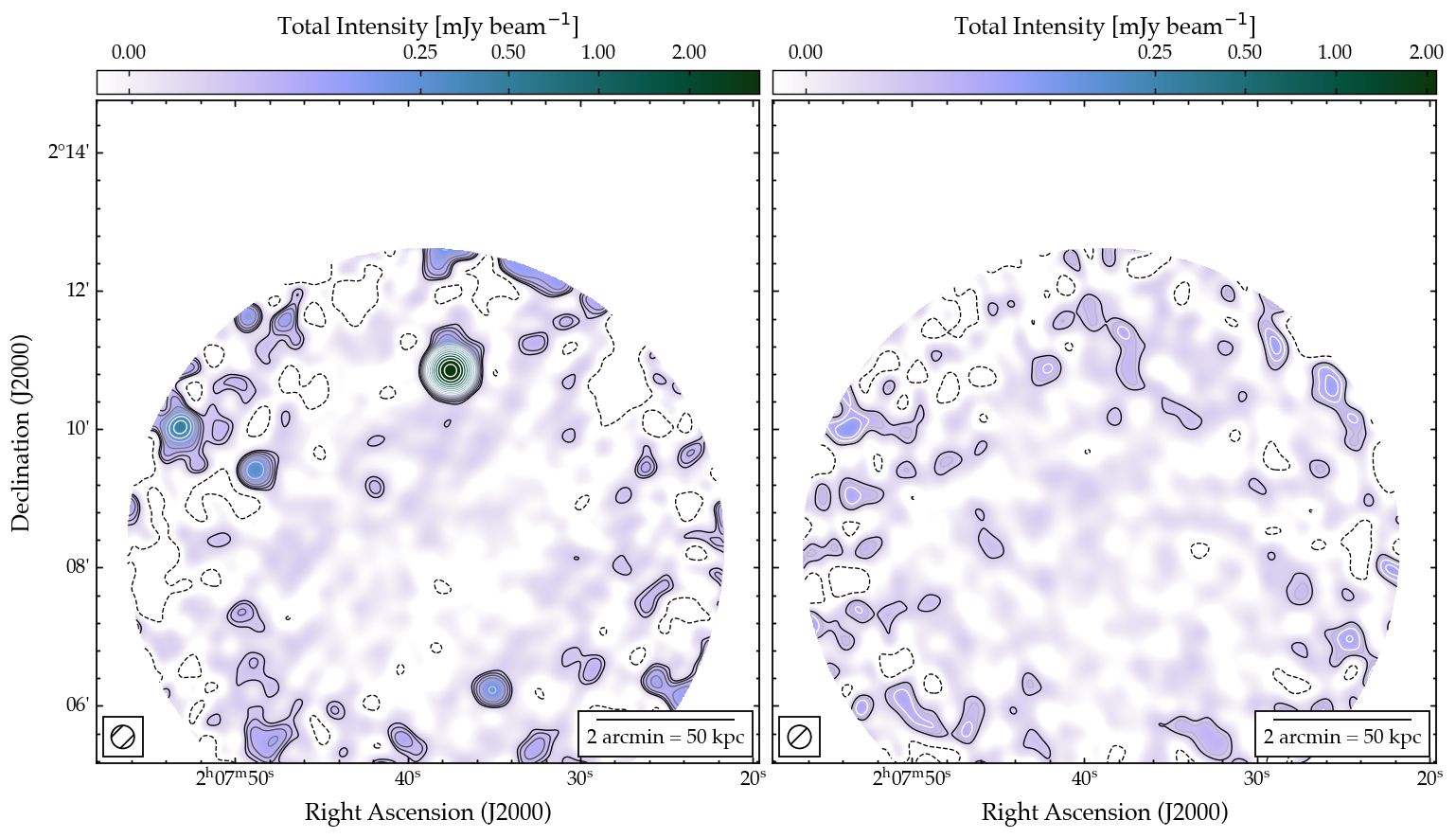}
\caption{VLA 7\,GHz maps of HCG15 at 20\,arcsec resolution, before (\textit{left}) and after (\textit{right}) subtracting the discrete radio sources associated with the group. The left-hand panel shows the VLA-D observations only, whereas the right-hand panel shows the combined VLA-C+D dataset. Contours start at $3\sigma$ and scale by a factor of $\sqrt{2}$, where $\sigma$ values are reported in Tab.\,\ref{tab:img_summary}. Colourmaps range from $-1\sigma$ to $300\sigma$ on an arcsinh stretch to emphasise faint emission, although we note that no significant diffuse emission is detected.}
\label{fig:HCG15_VLA7}
\vspace{-5mm}
\end{center}
\end{figure*}

\subsection{Radio continuum}
Figure~\ref{fig:HCG15_Continuum_Subtracted} presents our source-subtracted radio images at 20~arcsec resolution. Only panel (f) appears fundamentally different with respect to Figure~\ref{fig:HCG15_Continuum}, as we now perform joint deconvolution of the VLA-C and VLA-D configuration data. This joint deconvolution allows us to recover the diffuse emission to a far greater extent at 5\,GHz than with the VLA-D observations alone, although we note that no significant emission is seen at 7\,GHz. For clarity, we show the 7\,GHz VLA map in Figure~\ref{fig:HCG15_VLA7} both before and after subtraction.

Broadly-speaking, the diffuse emission shows a somewhat elliptical morphology. Tracing the extent using the $3\sigma$ contour, we measure a largest angular size of around 5.5\,arcmin with LOFAR at 145\,MHz, along a north-west/south-east axis, corresponding to a linear size of 144\,kpc. The minor axis of the emission, measured north-east/south-west, is around 4.4\,arcmin or 117\,kpc at 145\,MHz. However, the extent of the emission is not significantly smaller at higher frequencies: at 2.41\,GHz, MeerKAT recovers emission up to scales of 5.7\,arcmin by 4.0\,arcmin (major and minor axes) corresponding to 152\,kpc by 106\,kpc. We emphasise that the edge of the diffuse emission is very sharp, suggesting that deeper and/or lower-frequency observations (e.g. with LOFAR LBA) will not recover further diffuse emission.

As seen in Figure~\ref{fig:HCG15_Continuum_Subtracted}, while the extent changes very little moving from lower to higher frequencies, the morphology of the diffuse emission changes more significantly. At 145\,MHz, the surface brightness is far more uniform; transitioning to higher frequencies a `ridge' structure becomes more prominently defined --- the `ridge' first reported by \cite{Giacintucci2011_GalaxyGroups} --- following a south-east/north-west axis. The diffuse emission to the north of the ridge becomes increasingly faint toward higher frequencies and is largely undetected at 5\,GHz.

\subsubsection{Integrated spectrum}
\begin{figure}
\begin{center}
 \includegraphics[width=0.99\linewidth]{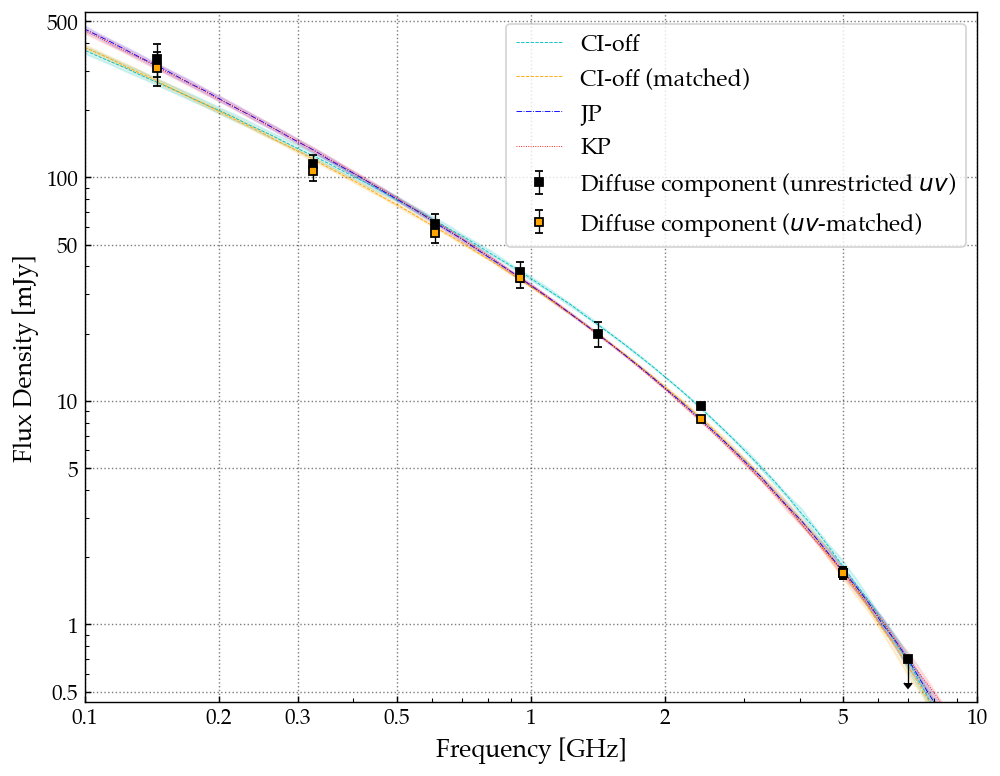}
\caption{Integrated SED for the diffuse radio emission associated with the IGrM of HCG15. The different curves trace the best-fit models for each injection and ageing model. We note that the best-fit `JP' and `KP' models are extremely similar.}
\label{fig:hcg15_sed_diffuse}
\end{center}
\end{figure}

Integrating over the volume of the diffuse radio emission seen in Figure~\ref{fig:HCG15_Continuum_Subtracted} we derive integrated flux density measurements, which are listed in Table~\ref{tab:diffuse_flux} and presented as an integrated SED in Figure~\ref{fig:hcg15_sed_diffuse}. At 7\,GHz we are only able to place an upper limit to the integrated flux density by integrating the $1\sigma$ noise over the volume of the emission.

\begin{table*}
\centering
\caption{Integrated flux density measurements for the diffuse emission associated with the IGrM of HCG15. \label{tab:diffuse_flux}}
\renewcommand{\arraystretch}{1.2}
\begin{tabular}{lp{3cm}p{3cm}p{6cm}}
\hline
Frequency $[$GHz$]$ & \multicolumn{2}{c}{Flux density $[$mJy$]$}        & Notes \& Comments    \\
          & Unrestricted $uv$-coverage & Common $uv$-min ($\geq 370\lambda$)  &        \\
\hline\hline
0.145     & $338.97 \pm 58.42$ & $309.78 \pm 53.40$   &        \\
0.325     & $114.77 \pm 11.55$ & $106.99 \pm 10.78$   &        \\
0.610     & $62.15  \pm 6.27$  & $56.59 \pm 5.72$   &        \\
0.943     & $37.83  \pm 3.80$  & $35.31 \pm 3.55$   &        \\
1.412     & $21.58  \pm 2.52$  & $-$  & Integrated over the NVSS map, subtracting compact flux from FIRST \\
2.412     & $9.51  \pm 0.30$   & $8.34 \pm 0.26$    &        \\
5.000     & $1.70  \pm 0.10$   & $1.70  \pm 0.10$   &        \\
7.000     & $\leq 0.70$        & $\leq 0.70$        & Upper limit   \\
\hline
\end{tabular}
\end{table*}

From Figure~\ref{fig:hcg15_sed_diffuse} we see that the diffuse emission follows approximately power-law behaviour at lower frequencies, becoming increasingly curved at higher frequencies. Below around 1.4\,GHz, where the SED follows a an approximate power-law trend, we find a steep spectral index of $\alpha = -1.28 \pm 0.04$.

We use the \textsc{SynchroFit} software package\footnote{Currently available at \url{https://github.com/synchrofit/synchrofit}} \citep[developed in][]{Quici2022} to fit physically-motivated models to this SED. The spectra of radio galaxies and remnants are often described using one of three common model flavours: the `JP' model \citep{JaffePerola1973}, the `KP' model \citep{Kardashev1962,Pacholczyk1970}, or a form of the `continuous injection' or `CI' model \citep{Komissarov1994}. Both the `JP' and `KP' models assume a single injection of electrons at $\tau = 0$ which subsequently undergoes radiative losses; the key difference between the two families of model is that the `JP' model assumes the electron population shows a uniform pitch angle whereas the `KP' model assumes a range of pitch angles. 

Conversely, the `CI' model describes a spectrum exhibited by an electron population that is injected over an extended time period from $t = 0$ to $t = \tau_{\rm total}$, where $\tau_{\rm total}$ is the source age. Two versions of the `CI' model exist: `CI-on', where energy injection is still occurring, and `CI-off', where the injection occurred for a given period of time $(\tau_{\rm on})$ and ceased some time ago ($\tau_{\rm off}$). In this latter scenario, the total source age is therefore given by $\tau = \tau_{\rm on} + \tau_{\rm off}$. Each of the aforementioned models predict a spectrum that becomes increasingly steep and curved as a source ages, with a spectral break that moves to increasingly lower frequencies. 

One of the key inputs required in spectral ageing modelling is the magnetic field strength in the source. Historically, various arguments have been used to estimate the magnetic field strength in group and cluster environments, such as minimum-energy arguments \citep[e.g.][]{Burbidge1956_M87,Pacholczyk1970} or the assumption of equipartition between magnetic fields and cosmic rays \citep[e.g.][]{Govoni_Feretti_2004}. These approaches rely on differing assumptions, which may or may not be valid depending on the nature of the target. More recent revised formalism for these methods exist --- see for example \cite{Pfrommer2004} and \cite{BeckKrause2005} --- which addressed some limitations of earlier formalisms.

However, given the dynamically-unrelaxed nature of many galaxy groups and clusters --- our target HCG15 among them --- there is no guarantee that either equipartition or minimum-energy assumptions are valid. We refer the reader to \cite{Ruszkowski_Pfrommer_2023} for an expansive review on the topic. As such, for our purposes, we assume a characteristic magnetic field strength of $B = 1.5 \upmu$G, although we note that while minimum-energy or equipartition arguments are problematic, applying such methods (e.g. following Eq.\,26 of \citealt{Govoni_Feretti_2004}) to our data suggest a magnetic field strength $B_{\rm eq} \simeq 1.3 - 1.5~\upmu$G. In such calculations we adopt a cylindrical geometry, take width equal to the width of the `ridge' structure (56~arcsec or 24.7~kpc) and a reference frequency of 943\,MHz, using the typical surface brightness measured by ASKAP.

Adopting this magnetic field value of $B = 1.5 \upmu$G for our modelling with \textsc{SynchroFit}, we derive the best-fit parameters for each model presented in Table~\ref{tab:synchrofit_fits}. Our models are parameterised in terms of the electron injection index $s$ and the break frequency $\nu_{\rm b}$, which represents the frequency above which the spectrum begins to steepen as energy loss mechanisms begin to dominate the spectral shape. The injection index describes the slope of the cosmic ray electron energy distribution at injection, and is related to the injection spectral index as $\alpha_{\rm inj.} = - ( s - 1 )/2$ at frequencies below the break frequency.

\begin{table}
\centering
\small
\caption{Results of our fitting to the integrated spectrum of HCG15 with \textsc{SynchroFit}. The best-fit parameters describing the `JP', `KP', and `CI-off' models are shown: the power law index of the injected electron distribution $s$, break frequency $\nu_b$ (in GHz), source age $\tau$ and remnant age $\tau_{\rm off}$ (both in Myr). \label{tab:synchrofit_fits}}
\renewcommand{\arraystretch}{1.2}
\begin{tabular}{lcccc}
\hline
Model   &   $s$                 &   $\nu_b$             &   $\tau_{\rm total}$          &   $\tau_{\rm off}$    \\
        &                       &   $[{\rm GHz}]$       &   $[\rm Myr]$     &   $[\rm Myr]$         \\
\hline\hline
\multicolumn{5}{c}{Unrestricted \textit{uv}-range}  \\
JP      &  $2.79 \pm 0.01$      &   $4.29 \pm 0.01$     &   $69.7 \pm 0.6$  &  $-$                  \\
KP      &  $2.73 \pm 0.01$      &   $2.35 \pm 0.01$     &   $41.7 \pm 0.4$  &  $-$                  \\
CI-off  &  $2.51 \pm 0.01$      &   $1.24 \pm 0.01$     &   $85.9 \pm 0.6$  &  $71.2 \pm 0.6$       \\
\hline
\hline
\multicolumn{5}{c}{Common \textit{uv}-min $\geq 370\lambda$}  \\
JP      &   $2.66 \pm 0.01$   & $3.84 \pm 0.01$  & $73.6 \pm 0.7$  &   $-$\\
KP      &   $2.51 \pm 0.01$   & $1.88 \pm 0.01$  & $46.8 \pm 0.5$  &   $-$\\
CI-off  &   $2.62 \pm 0.01$   & $3.21 \pm 0.01$ & $80.8 \pm 0.6$  &  $67.8 \pm 0.7$       \\
\hline
\end{tabular}
\end{table}

All ageing models predict a similar injection index $\alpha_{\rm inj.}$, and similarly the model SED derived using the best-fit parameters for each model flavour is almost indistinguishable. However, the break frequency and source age are remarkably different depending on whether the `JP', `KP' or `CI-off' model is chosen. In particular, the break frequency derived according to each model flavour lies within our observed frequency range, and so may suggest which model is the most appropriate. Both the `JP' and `KP' models suggest break frequencies above 2.3\,GHz, while the `CI-off' model suggests a lower break frequency of 1.2\,GHz; based on the spectrum in Figure~\ref{fig:hcg15_sed_diffuse}, it is possible to trace a reasonable power-law fit using our lower frequency data, but the 2.4~GHz datapoint from MeerKAT shows clear departure from power-law behaviour, and correspondingly we favour the `CI-off' model compared to the `JP' and `KP' models, although for clarity we show all models in Figure~\ref{fig:hcg15_sed_diffuse}.

For the \textit{uv}-matched images, the fit results are more mixed. Both the `JP' and `CI-off' models suggest break frequencies above 3\,GHz, whereas the `KP' model suggests the lowest break frequency at 1.9\,GHz. The total source age is also similar for the `JP' and `CI-off' models, with best-fit ages of around 75 to 80\,Myr; given that the source age is linked to the break frequency, this is unsurprising. We also show the `CI-off' model fit to the \textit{uv}-matched data in Figure~\ref{fig:hcg15_sed_diffuse}; the `JP' and `KP' models are not visually distinguishable.

\begin{figure*}
\begin{center}
 \includegraphics[width=0.8\textwidth]{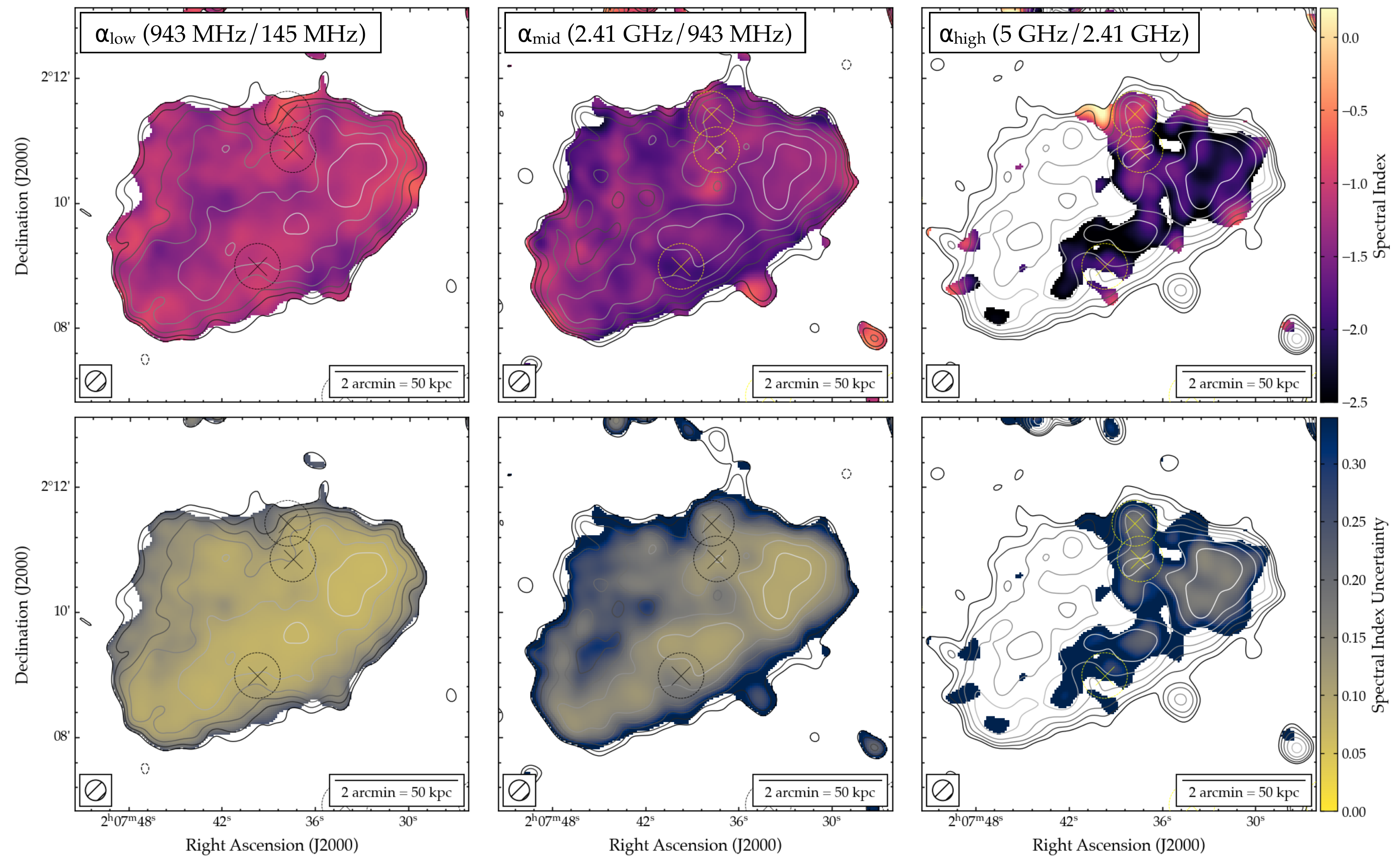}
\caption{Spectral index of the diffuse radio emission associated with the IGrM of HCG15. Upper panels show the spectral index, lower panels show the corresponding uncertainty. The spectral index is derived using maps from ASKAP/LOFAR (\textit{left}), MeerKAT/ASKAP (\textit{centre}) and VLA/MeerKAT (\textit{right}), all at 20~arcsec resolution. The frequencies involved are shown in the inset. Contours show the source-subtracted radio emission at the lower frequency, i.e. LOFAR, ASKAP and MeerKAT from left to right, with levels shown as per Figure~\ref{fig:HCG15_Continuum_Subtracted}. Markers show the positions of the group-member galaxies, whose radio counterparts have been subtracted from the data.}
\label{fig:hcg15_spectral_index_maps}
\end{center}
\end{figure*}

We must also consider what else is known about HCG15. While we know from our high resolution radio observations that none of the group-member galaxies show evidence of jets feeding into the diffuse emission, the overall morphology suggests that this source was once fed by one (or more) AGN in the group. This scenario also favours the `CI-off' model, whereby diffuse emission is seeded by an extended epoch of AGN activity that later ceases, over the `JP' or `KP' models, which each model a single epoch of electron injection.

Overall, from our best `CI-off' model fitted to the non-\textit{uv}-matched measurements, we use Equation~\ref{eq:radio_lum} to derive a $k$-corrected radio power of $P_{\rm 150 \, MHz} = (3.23 \pm 0.09) \times 10^{23}$~W~Hz$^{-1}$ and $P_{\rm 1.4 \, GHz} = (2.22 \pm 0.02) \times 10^{22}$~W~Hz$^{-1}$. Such radio power is consistent with the population of typical remnant radio sources \citep[e.g.][]{Quici2021_Remnants,Jurlin2021_Remnants,Riseley2022_Abell3266}.

\subsubsection{Resolved spectrum}
Figure~\ref{fig:hcg15_spectral_index_maps} shows the resolved spectral index maps (plus the associated uncertainty) derived using our source-subtracted maps at three pairs of frequencies. For the `low-frequency' spectral index $\alpha_{\rm low}$ we use ASKAP and LOFAR (respectively 943\,MHz and 145\,MHz), the `mid-frequency' spectral index $\alpha_{\rm mid}$ uses MeerKAT and ASKAP (respectively 2.41\,GHz and 943\,MHz), and we use the VLA and MeerKAT (respectively 5\,GHz and 2.41\,GHz) for the `high-frequency' spectral index $\alpha_{\rm high}$. In each case, we only consider pixels with signal at the $3\sigma$ level at both frequencies.

In a simple scenario where the diffuse emission in HCG15 is related to ongoing AGN activity from the brightest radio AGN in the group, HCG15-D, we might expect a flatter spectral index near to this source with a gradient along the `ridge', tracing an ageing electron population. The spectral index maps in Figure~\ref{fig:hcg15_spectral_index_maps} do not show this behaviour.

Instead, the spectral index is too steep for such a scenario: even at low frequencies, the median spectral index in the ridge is $\langle \alpha_{\rm low} \rangle = -1.13 \pm 0.08$. Moving towards higher frequencies we find that the spectral index steepens, consistent with the behaviour seen in the integrated spectrum, as we find average spectral index values of $\langle \alpha_{\rm mid} \rangle = -1.51 \pm 0.14$ and $\langle \alpha_{\rm high} \rangle = -2.14 \pm 0.30$.

\begin{figure}[ht]
\begin{center}
 \includegraphics[width=0.8\linewidth]{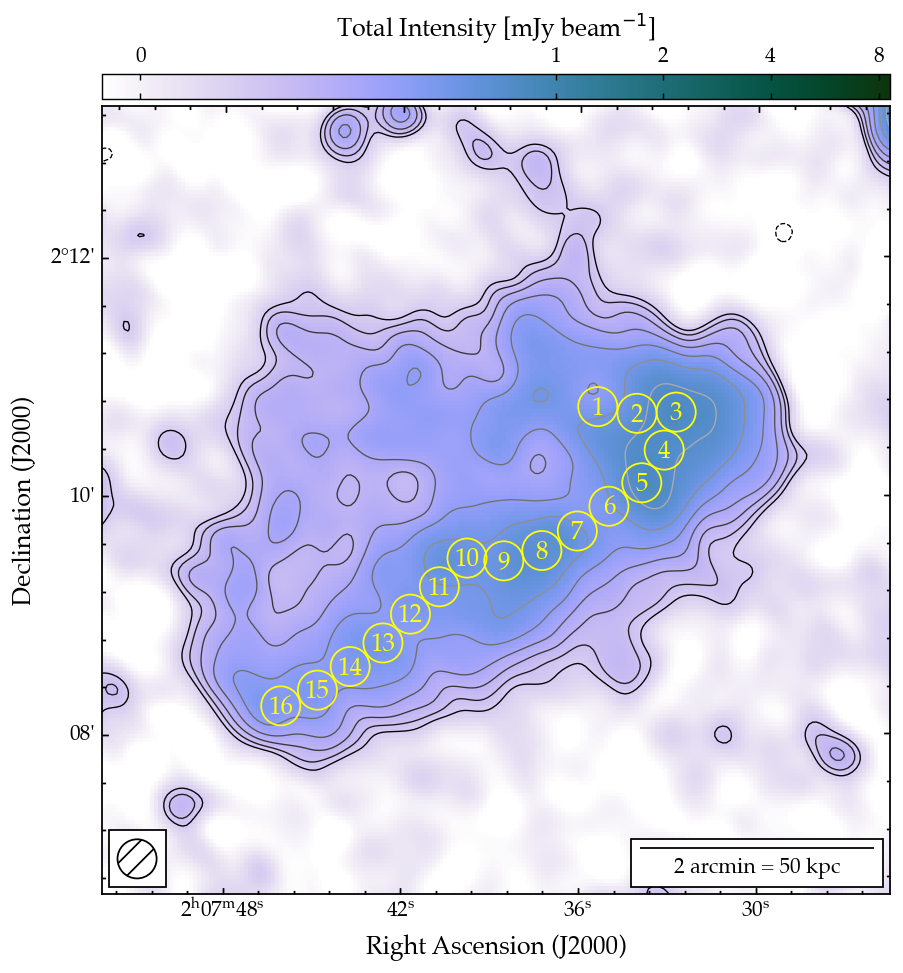}
 \includegraphics[width=0.99\linewidth]{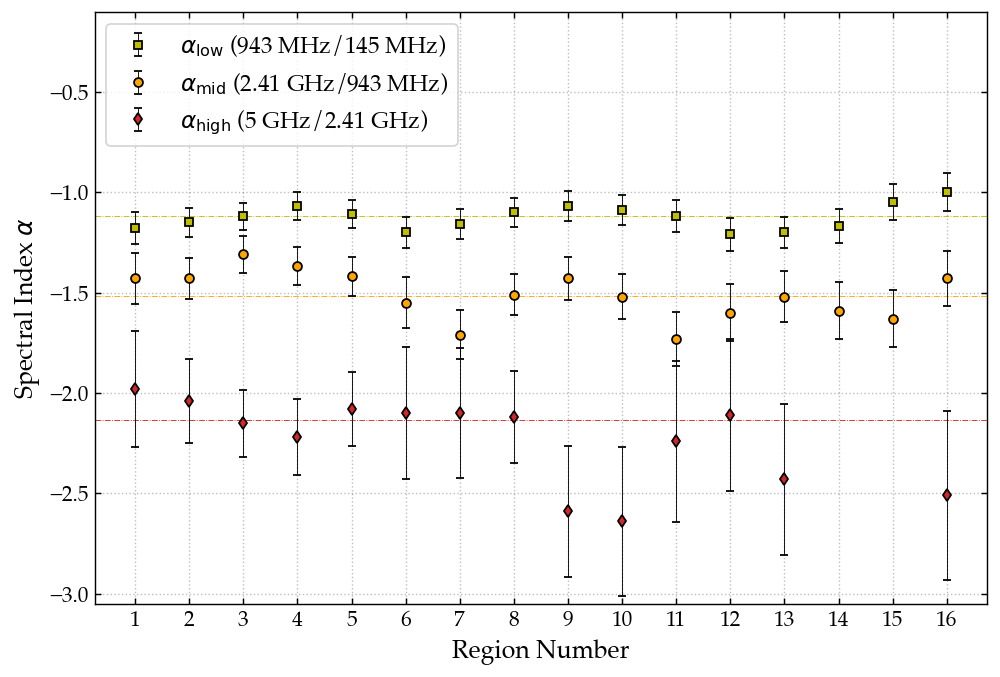}
\caption{Spectral index profile along the `ridge' of diffuse emission in HCG15. \textit{Top panel} shows the radio surface brightness at 943\,MHz as per Fig.~\ref{fig:HCG15_Continuum_Subtracted}, with circular regions of 20~arcsec diameter denoting those profiled. \textit{Bottom panel} shows the spectral index trend in the indicated regions. Horizontal lines show the median values for each of the $\alpha_{\rm low}$, $\alpha_{\rm mid}$ and $\alpha_{\rm high}$ profiles.}
\label{fig:hcg15_spectral_profile}
\end{center}
\end{figure}

We profiled the spectral index along the ridge using circular regions 20~arcsec in diameter, to further investigate whether there was evidence of any gradient, or whether the spectral index is more consistent with a bulk steepening. This profile is shown in Figure~\ref{fig:hcg15_spectral_profile} for each of the three spectral index maps, along with the regions profiled. There is clearly no overall trend, but rather there is a bulk shift to steeper spectra with increasing frequency.

Moreover, the spectral index is broadly uniform in each frequency interval considered here, but shows statistically-significant fluctuations in coherent patches up to several times the size of the restoring beam. This strongly suggests that the cosmic ray electron population is not fully homogeneous, but rather traces a distribution of electrons with different injection/acceleration times, and/or inhomogeneities in the ambient medium. Such a population cannot be modelled by the classical ageing models, which typically encapsulate a single injection (JP, KP) or continuous injection over a given time period (CI).

\subsubsection{Colour-colour analysis}
To further investigate the behaviour of the spectral index in different regions of the diffuse emission, we performed a colour-colour analysis \citep[e.g.][]{Katz-Stone1993,Katz-Stone_Rudnick_1997a,Katz-Stone_Rudnick_1997b} which studies the relation between the spectral index between two pairs of (ideally-independent) frequencies. Given the broad frequency coverage of our high-quality data, we used the same pairs of frequencies as used to derive our resolved spectral index maps, i.e. $\alpha_{\rm low}$ between 145\,MHz and 943\,MHz, $\alpha_{\rm mid}$ between 943\,MHz and 2.41\,GHz, and $\alpha_{\rm high}$ between 2.41\,GHz and 5\,GHz.

\begin{figure*}
\begin{center}
 \includegraphics[width=0.8\textwidth]{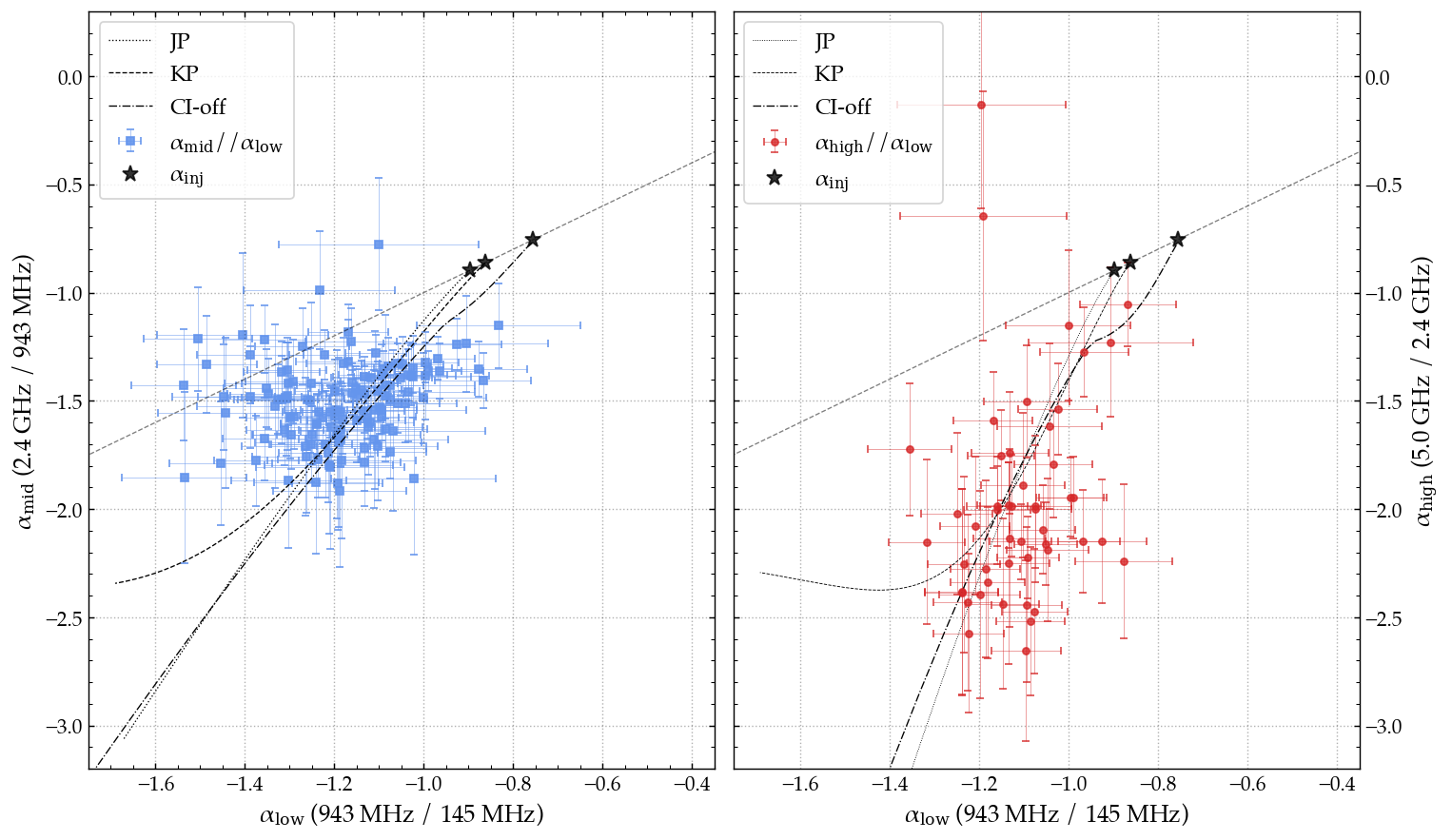}
\caption{Colour-colour diagrams for the diffuse emission associated with HCG15. Blue (red) datapoints indicate the position of regions in the $\alpha_{\rm low} // \alpha_{\rm mid}$ ($\alpha_{\rm low} // \alpha_{\rm high}$) plane; these datapoints are displayed in the left-hand (right-hand) panel. The stars indicate the injection spectral index derived from our modelling of the integrated spectrum, as discussed in the text. Different curves trace different evolutionary tracks of the spectrum following the JP, KP and CI-off models. The dashed line indicates the unity relation between the abscissa and ordinate in each case.}
\label{fig:hcg15_colourcolour}
\end{center}
\end{figure*}

We chose to derive region-wise spectral index values using our source-subtracted radio images. We placed square box regions of 20~arcsec on a side covering the full extent of the diffuse emission, and derived the region-wise spectral index using only those regions where the radio surface brightness was above the $3\sigma$ level at both frequencies. Our colour-colour diagrams are shown in Figure~\ref{fig:hcg15_colourcolour}.

Unlike in some active galaxies \citep[e.g.][]{Koribalski2024_Corkscrew} the colour-colour diagram for HCG15 shows only a very few regions where the spectral index is consistent with injection. Instead, the majority of regions lie below the unity line, indicating that the diffuse emission associated with HCG15 shows a steeper spectral index at higher frequencies than low frequencies, i.e. spectral curvature. This is particularly pronounced when comparing the left and right panels of Figure~\ref{fig:hcg15_colourcolour}: the bulk of the datapoints lie higher in the colour-colour plane for $\alpha_{\rm low} // \alpha_{\rm mid}$ and lower in the $\alpha_{\rm low} // \alpha_{\rm high}$ plane. Such behaviour is consistent with previous colour-colour analyses of steep-spectrum remnant sources \citep[e.g.][]{Candini2023_NGC6086,Shulevski2024_A1318_remnant} which are not experiencing active injection.

We note the presence of several regions which lie far above the unity line in the $\alpha_{\rm low} // \alpha_{\rm high}$ plane. While such behaviour can in some instances suggest the presence of overlapping structures (with different spectral indices) along the line of sight, as seen in some studies of radio relics in galaxy clusters \citep[e.g.][]{Rajpurohit2021_MACSJ0717_relic,Rajpurohit2022_Abell2256_relic}, the large uncertainty on the derived spectral index suggests that these regions may be spurious, and so these outliers should be treated with caution.

Using the best-fit parameters for our `JP', `KP', and `CI-off' models discussed earlier, we plot the evolutionary tracks of a cosmic ray electron population according to these scenarios. Cosmic ray electrons are injected with a spectrum that matches the injection index (the stars in Figure~\ref{fig:hcg15_colourcolour}) and then experience losses, falling away from the unity line following the respective curves. All three models describe broadly similar evolutionary tracks, although the `KP' model diverges from the `JP' and `CI-off' models outside the region where we have data. While it is curious to note that some regions show flatter spectra than predicted from either of the single-injection models, perhaps hinting at a slight preference for the `CI-off' model, none of the models provide a particularly good reproduction of the observed cosmic ray electron population.

Indeed, Figure~\ref{fig:hcg15_colourcolour} indicates that the cosmic ray electron population in HCG15 shows a broad distribution in colour-colour space. This is likely a consequence of the mechanism driving the spectral index fluctuations noted earlier: an inhomogeneous cosmic ray electron population. Inhomogeneities in injection/acceleration and/or the ambient medium (resulting in inhomogeneous ageing) will cause a superposition of different electron ages when viewed in projection. The result of this will be both a flatter and broader distribution in colour-colour space.

Recent theoretical work has begun to explore inhomogeneous cosmic ray electron populations in clusters, primarily from the point of view of explaining the colour-colour space behaviour of radio relics, which often show broad distributions \citep[e.g.][]{Rajpurohit2020_Toothbrush_wide} similar to that seen here for HCG15. Scenarios such as multiple shocks \citep{Inchingolo2022_MultiShockDSA} or fluctuations in density \citep{Whittingham2024_RadioRelicSims} in particular can replicate both the observed broadening in colour-colour space and the shallow curvature of integrated spectra, and may be applicable in the scenario considered here, although further theoretical exploration is beyond the scope of this work.

\subsubsection{Resolved spectral ageing}
We have extended our investigation of the spectral ageing properties of the diffuse emission in HCG15 by applying \textsc{SynchroFit}'s \texttt{spectral\_fitter} algorithm to the source-subtracted maps presented in Figure~\ref{fig:HCG15_Continuum_Subtracted} on a pixel-by-pixel basis. While this functionality is not currently present in \textsc{SynchroFit}, it is a reasonably trivial extension to the codebase, and reproduces one of the commonly-used features of the Broadband Radio Astronomy ToolS \citep[\textsc{BRATS}\footnote{\url{http://www.askanastronomer.co.uk/brats/}};][]{Harwood2013_FR2_ageing,Harwood2015_FR2_ageing,Harwood2018_BRATS_Software} software suite that is extensively employed in similar studies \citep[e.g.][]{Harwood2017_ageing,Harwood2017_FR2,Brienza2020_3C388,Biava2021_MS0735,Candini2023_NGC6086,Ubertosi2024_RBS797}.

\begin{figure*}
\begin{center}
\includegraphics[width=0.8\textwidth]{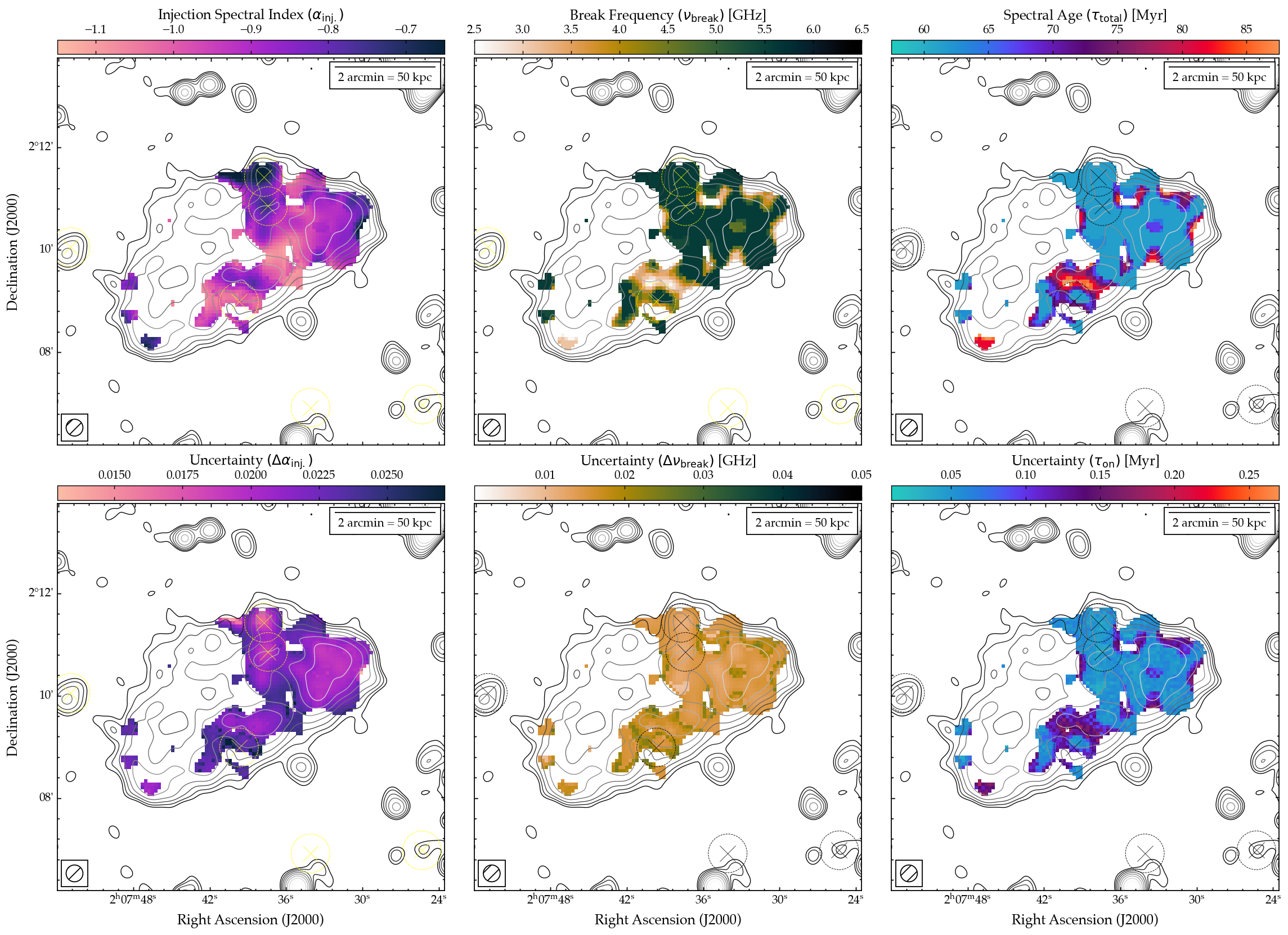}

\caption{Properties of the diffuse emission in HCG15 derived using the tools within the \textsc{SynchroFit} software suite. The upper row shows the injection index ($\alpha_{\rm inj.}$), break frequency ($\nu_{\rm break}$), and derived spectral age ($\tau_{\rm total}$); maps in the lower rows show the corresponding uncertainty. Contours show the MeerKAT continuum at 2.412\,GHz as per Figure~\ref{fig:HCG15_Continuum_Subtracted}. Markers show the positions of the group-member galaxies, whose radio counterparts have been subtracted from the data.}
\label{fig:HCG15_ageing_maps}
\vspace{-5mm}
\end{center}
\end{figure*}

Our analysis was performed using the maps from LOFAR (145\,MHz), ASKAP (943\,MHz), MeerKAT S-band (2.1\,GHz) and the JVLA (5\,GHz). We selected only those pixels within the extent of the diffuse emission where the surface brightness was above the $3\sigma$ level at all frequencies. As before, the GMRT maps at 325 and 610\,MHz were excluded from our analysis due to the comparatively poor \textit{uv}-coverage. We show the resulting maps of the source properties in Figure~\ref{fig:HCG15_ageing_maps}. We note that while our integrated analysis made use of the CI model, these models cannot be used for the resolved case as CI describes the spectrum of the \textit{entire} population of electrons injected by the source over its lifetime; in this resolved analysis we use the JP model instead, as in previous studies \citep[e.g.][]{Brienza2020_3C388,Candini2023_NGC6086,Shulevski2017_B0924+30,Shulevski2024_A1318_remnant}.

From this investigation, we find that the injection spectral index is steep throughout the diffuse emission with an average value of $\langle \alpha_{\rm inj., CI} \rangle = -0.91$, significantly steeper than the value derived from the `CI-off' model fit to the integrated spectrum: $\alpha_{\rm inj.} = - ( s - 1 )/2 = -0.75$, although it is consistent with the injection spectral index derived from the JP model fit $\alpha_{\rm inj., \, JP} = -0.90$ (see Table~\ref{tab:synchrofit_fits}). The injection spectral index is steeper in some regions, reaching values of $\alpha_{\rm inj.} = -1.1$, and the behaviour is somewhat inhomogeneous in that the structure of the injection index does not trace the structure of the continuum emission.

Conversely, the majority of the diffuse emission shows a largely uniform break frequency of around 5.5\,GHz. Similarly, the spectral age map in Figure~\ref{fig:HCG15_ageing_maps} shows a broadly uniform total age of $\tau_{\rm total} = 61.1 \pm 0.1$\,Myr, although given that the source age is derived using the break frequency, this is unsurprising. Overall, both our resolved analysis and our integrated analysis suggest an aged plasma, although there is a discrepancy between this resolved modelling and our CI fit to the integrated spectrum derived from the \textit{uv}-matched images, which yields a total age of $\tau_{\rm off} = 67.8 \pm 0.7$\,Myr. 

Such a discrepancy in total age derived using integrated and resolved models is not unexpected. As discussed by \cite{Harwood2017_ageing}, integrated modelling can yield significant \textit{underestimation} of the source age (by a factor up to three) or \textit{overestimation} (by up to a factor six) with respect to resolved modelling. The reasons for this discrepancy in remnant sources such as the diffuse emission seen in HCG15 remain unclear, although mechanisms such as adiabatic expansion/compression or magnetic field inhomogeneities which are not currently captured by models may go some way to resolving such disagreement.

\subsection{Polarimetry}
After performing RM-synthesis and RM-clean on our ASKAP and MeerKAT FDF cubes, we extracted several key maps for our analysis. These were (i) the peak polarised flux $P_{\rm max}$, (ii) the RM at the peak of the polarised flux, (iii) the polarisation angle at the RM of the peak of the polarised flux, derotated to reflect the polarisation angle \textit{at the point of emission}, $\chi_0$. These maps were extracted using the \texttt{rmtools\_peakfitcube} task within the \textsc{RM-tools} suite. For each of the maps, we extracted the values derived by fitting a parabola to the points around the peak of the FDF; see the \textsc{RM-tools} documentation for more details\footnote{Specifically: \url{https://github.com/CIRADA-Tools/RM-Tools/wiki/RMsynth1D}}.

\begin{figure*}
\begin{center}
\includegraphics[width=0.95\textwidth]{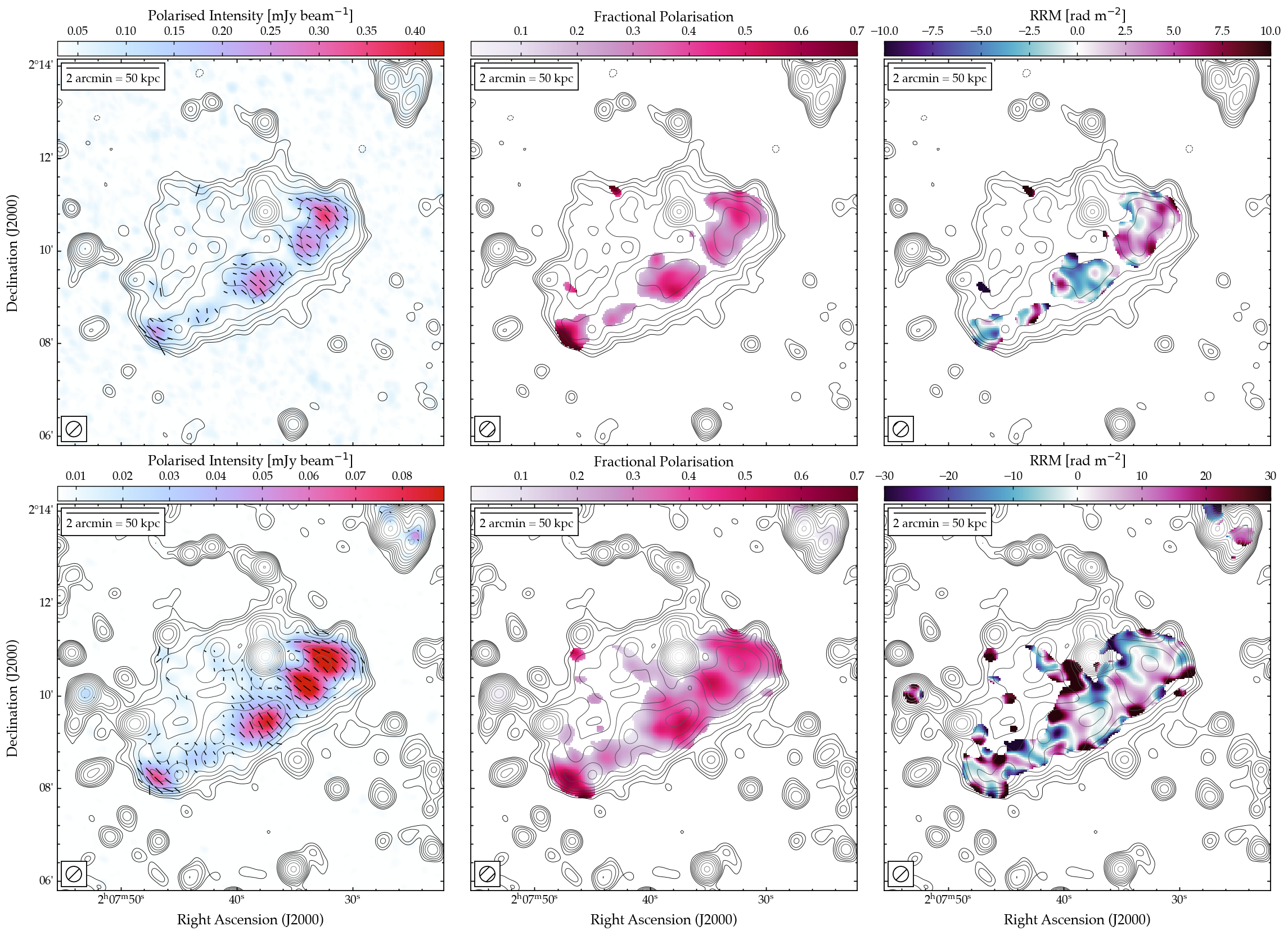}

\caption{Linear polarisation maps of HCG15 as seen by ASKAP at 943\,MHz (\textit{top}) and by MeerKAT at 2.41\,GHz (\textit{bottom}), both at a resolution of 20\,arcsec. Colourmaps show the linearly-polarised intensity (\textit{left}), the fractional polarisation (\textit{centre}) and the residual rotation measure (RRM; \textit{right}). Vectors denote the orientation of the magnetic field, scaled according to the magnitude of the fractional polarisation. Contours show radio continuum surface brightness as per Figure~\ref{fig:HCG15_Continuum}. The fractional polarisation, polarisation angle and RRM are blanked below a threshold of $4\sigma_{\rm P}$ where $\sigma_{\rm P} = 28~\upmu$Jy~beam$^{-1}$ for ASKAP and $3~\upmu$Jy~beam$^{-1}$ for MeerKAT.}
\label{fig:HCG15_Polarisation}
\vspace{-5mm}
\end{center}
\end{figure*}

Figure~\ref{fig:HCG15_Polarisation} presents the maps derived from our FDF cubes, showing the peak polarised flux ($P_{\rm max}$) corrected for Ricean bias, overlaid with the intrinsic polarisation angle $\chi_0$ (rotated by $90\degree$ to reflect the magnetic field orientation) as well as the fractional polarisation and residual rotation measure (RRM). The RRM describes the observed rotation measure corrected for Faraday rotation by the Galactic foreground; we used the most recent version of the Galactic RM map reconstructed by \cite{Hutschenreuter2022_GalacticRM} to estimate the Galactic contribution, which we find to be ${\rm RM_{Gal}} = 0.99 \pm 4.28$\,rad\,m$^{-2}$. The maps of $\chi_0$ presented herein have also been corrected for this Galactic Faraday rotation.

Given our ASKAP frequency coverage ($800 - 1088$~MHz) ionospheric Faraday Rotation is also a consideration. While $\lambda_{\rm ref} \simeq 0.318~{\rm m}$ (and so $\lambda^2 \ll 1~{\rm m}^2$), sufficiently large ionospheric RMs $({\rm RM_{ion}})$ could impact our analysis. We used the \textsc{RMextract} \citep{Mevius2018_RMextract} to estimate the total electron content (TEC) along the LOS to HCG15, and then derive ${\rm RM_{ion}}$. Over the course of each ASKAP observing run we find $-1.6 \lesssim {\rm RM_{ion}} ~ [\rm rad~m^{-2}]\lesssim -1.1$ for the first epoch, and $-1.0 \lesssim {\rm RM_{ion}} ~ [\rm rad~m^{-2}]\lesssim -0.8$ for the second epoch. These values of ${\rm RM_{ion}}$ would lead to a polarisation angle rotation of between 4~degree and 9~degree, which is likely small compared to other uncertainties, and so we do not apply a correction for this rotation, although we adopt an upper limit of 9~degree to the polarisation angle uncertainty. Similarly, we incorporate an additional factor of the differential ${\rm RM_{ion}}$ into the RMs of $\rm 0.35~rad~m^{-2}$ quoted from our ASKAP data.

We note that the tools used to extract these maps are only capable of fitting to the \textit{peak} of the FDF, and will thus not encapsulate any more complex behaviour than a single Faraday-thin component. However, inspection of our FDF cubes revealed no evidence of the presence of additional polarisation components with different RMs; the emission is well-described by a single Faraday-thin component with a single value of RM. 

\begin{figure}[!ht]
\begin{center}
 \includegraphics[width=0.9\linewidth]{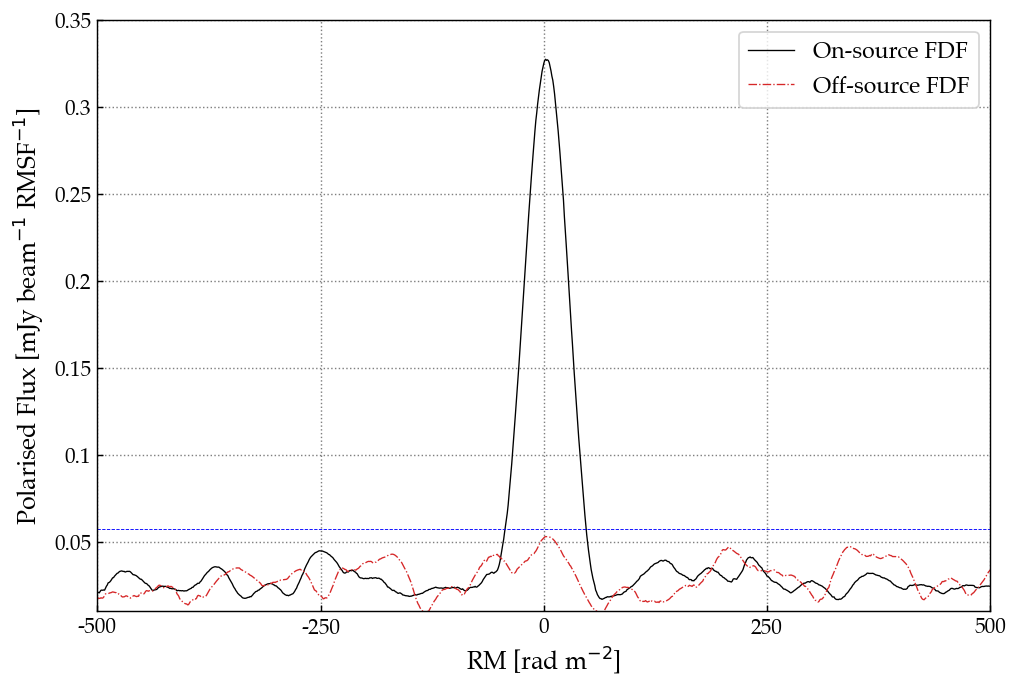}
 \includegraphics[width=0.9\linewidth]{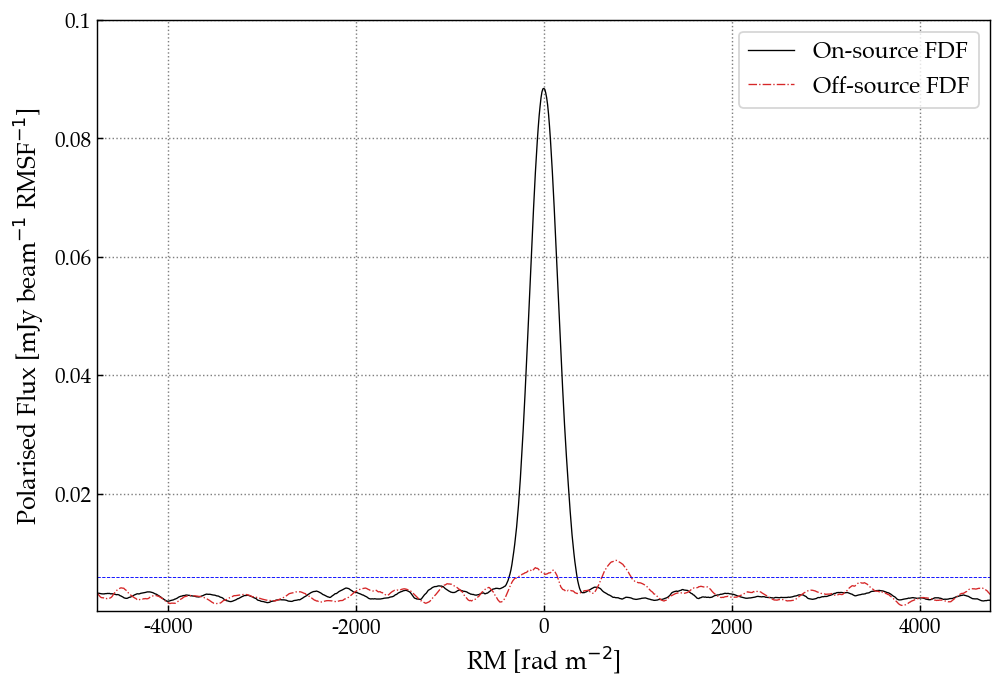}
\caption{Example FDF spectra from our ASKAP (943\,MHz, \textit{top}) and MeerKAT (2.41\,GHz, \textit{bottom}) polarisation cubes. The `on-source' spectrum was measured within a 20\,arcsec radius of the peak of diffuse polarised emission $({\rm RA, Dec.}) = (31.8849062\degree, 2.1790544\degree)$; the `off-source' spectrum within 20\,arcsec of HCG15-D, the brightest continuum source embedded in the diffuse emission. The horizontal line indicates the $2\sigma_{\rm P}$ level, where $\sigma_{\rm P} = 28.6 \, \upmu$Jy beam$^{-1}$ for ASKAP and $2.96 \, \upmu$Jy beam$^{-1}$ for MeerKAT.}
\label{fig:hcg15_fdfs}
\end{center}
\end{figure}

Figure~\ref{fig:hcg15_fdfs} shows example FDF spectra for an `on-source' region extracted at the peak of the polarised emission, using a circular region of 20\,arcsec radius centred on $({\rm RA, Dec.}) = (31.885\degree, 2.179\degree)$ as well as a nearby `off-source' region of 10\,arcsec radius centred on HCG15-D, the embedded bright radio source. These spectra serve two verification purposes: firstly, they demonstrate that at our Faraday-space resolution, the FDF is dominated by a single component at low $|{\rm RM}|$; secondly, they demonstrate that residual instrumental leakage after correction is negligible. The most significant component of instrumental leakage would appear at zero RM in these spectra --- which have not been corrected for Galactic Faraday rotation.

From the `off-source' FDF spectra, we can estimate an upper limit to any residual leakage. The maximum value at zero RM measured from our MeerKAT FDF is $P_{\rm max} = 6.5 \, \upmu$Jy beam$^{-1}$ RMSF$^{-1}$, and for ASKAP we find $P_{\rm max} = 19 \, \upmu$Jy beam$^{-1}$ RMSF$^{-1}$; given the Stokes I flux density measurements for HCG15-D in Table~\ref{tab:hcg15_galaxies}, we hence find an upper limit to any residual leakage of $1.9\%$ for ASKAP at 943\,MHz and $0.2\%$ for MeerKAT at 2.41\,GHz. While the former value is higher than target polarisation purity of $\lesssim 1\%$ for ASKAP's POlarisation Sky Survey of the Universe's Magnetism \citep[POSSUM; e.g.][]{Gaensler2010_POSSUM} it is in line with expectations from POSSUM early science \citep[e.g.][]{Anderson2021_Fornax}. We additionally emphasise that these values are \textit{upper limits}: neither off-source FDF shows a significant peak at zero RM, but is rather consistent with the typical noise level in the rest of the FDF spectrum away from zero RM. Thus, to emphasise, we are confident that the diffuse polarised emission seen in HCG15 is of true astrophysical origin.

\subsubsection{Polarimetric profiles}
From Figure~\ref{fig:HCG15_Polarisation}, we see that the radio ridge exhibits significant linearly-polarised emission in both the ASKAP and MeerKAT maps, whereas the diffuse emission extending to the north-east is unpolarised at 943\,MHz and only shows faint polarised signal at 2.41\,GHz in some regions.

Within the ridge, we see five primary concentrations of linearly-polarised flux, which have a uniformly high linear polarisation fraction ($\Pi = P/I$), with a median value of $\langle \Pi \rangle = (31.3 \pm 3.2)\%$ at 943\,MHz and $(30.2 \pm 2.2)\%$ at 2.41\,GHz. However, in some regions the observed polarisation fraction is higher, reaching up to $\Pi = 78.2\%$ at 943\,MHz and $62.7\%$ at 2.41\,GHz. The former value is of the order of the theoretical maximum for synchrotron emission \citep[$\simeq 70 - 80\%$; e.g.][]{Feretti2012_review}, although there are only a handful of pixels which show values this high. The highest values of the polarisation fraction are concentrated toward the far south-east extremity of the diffuse emission.

Similarly, the intrinsic polarisation angle (and thus magnetic field) is extremely well-ordered. The orientation of the magnetic field is \textit{transverse} to the long axis of the ridge, however, which provides crucial evidence in assessing different hypotheses for the underlying mechanism (see \S\ref{sec:analysis}).

From the RRM maps presented in the right hand panels of Figure~\ref{fig:HCG15_Polarisation}, we see that the typical RM is low, with $\langle |{\rm RM}| \rangle = (2.5 \pm 2.3)$~rad~m$^{-2}$ at 943\,MHz. As the RM precision is inversely-proportional to the observing wavelength --- from Equation~\ref{eq:rm_drm} our theoretical Faraday-space resolution is 61~rad~m$^{-2}$ --- and this nominal uncertainty decreases with signal-to-noise, the typical uncertainty seen in the map is $\Delta{\rm RM} \lesssim 1$~rad~m$^{-2}$ at 943\,MHz.

For completeness, we wish to clarify that the maximum absolute value of RM recovered in our ASKAP map is $|{\rm RM_{max}}| = 17$~rad~m$^{-2}$, emphasising that all observed RM values are low. We also note that there is a slight tendency for the north-western component of the diffuse polarised emission to show broadly positive RMs with $\langle {\rm RM} \rangle = (+2.1 \pm 2.1)$~rad~m$^{-2}$, whereas the central component shows broadly negative RMs, with $\langle {\rm RM} \rangle = (-1.9 \pm 2.2)$~rad~m$^{-2}$, although these values are consistent and so we caution against interpreting this too strongly.

Conversely, at 2.41\,GHz, the picture is somewhat different. The RMs are still generally relatively low, as we find a median value of $\langle |{\rm RM}| \rangle = (7.7 \pm 8.2)$~rad~m$^{-2}$. However, while our signal-to-noise ratio is greater than for our ASKAP RM cube, and the observing bandwidth is broader, due to the higher frequencies, from Equation~\ref{eq:rm_drm} our theoretical Faraday-space resolution is much broader at $\simeq360$~rad~m$^{-2}$. Hence, it is expected that the median uncertainty quoted above is larger. 

\section{Analysis and Interpretation}\label{sec:analysis}
One of the key goals of this study is to determine the origins of the diffuse emission in HCG15. Previous studies have proposed a range of hypotheses, from turbulence-driven (mini-)halo-type emission \citep{NikielWroczynski2017_GalaxyGroups} to shock-driven relic-type emission \citep{Giacintucci2011_GalaxyGroups} to fossil/remnant emission associated with historic AGN activity \citep{Giacintucci2011_GalaxyGroups,NikielWroczynski2017_GalaxyGroups,NikielWroczynski2021_GalaxyGroups}.

While it is clear that the group is dynamically unrelaxed, as the group-member galaxy velocity dispersion is of the order of $\sigma_v \simeq 1000 ~ {\rm km ~ s}^{-1}$ and the X-ray from the hot plasma of the IGrM appears to show two primary concentrations (although see the following sub-section) the presence of significant polarised emission provides strong evidence against a turbulence-driven hypothesis. Turbulence is an inherently chaotic process typically invoked in the generation of radio haloes in galaxy clusters \citep[e.g.][]{Brunetti2001_Coma,Brunetti2014,vanWeeren2019_review}; these objects are known to be unpolarised due to disruption of the large-scale magnetic field coherence required for significant linearly polarised emission. 

While the young AGN associated with the somewhat central galaxy HCG15-D dominates the contribution from group-member galaxies, we can also strongly reject the null hypothesis that the diffuse emission is simply associated with a current epoch of AGN activity. We see no evidence of either jets feeding the diffuse radio emission (see Figure~\ref{fig:HCG15_Composite_HighRes}) nor hotspots in which might represent the termination point of jets, at any of the observing frequencies among our radio data. Furthermore, the observed spectral index (e.g. $\langle \alpha_{\rm low} \rangle = -1.13 \pm 0.08$) is steeper than is typically observed in the diffuse component of radio galaxies which are being actively fed ($\alpha \simeq -0.8$), and the spectral index profile does not show the behaviour expected of a simple active radio galaxy.

One further possibility is that we are viewing an extended radio galaxy close to the line of sight. In such a scenario, HCG15-D would be powering the extended radio emission, with the diffuse component being the radio lobe seen in projection, and the polarised `ridge' potentially corresponding to the edge of the foreground lobe. While this scenario would explain the absence of jets seen at $\sim$arcsec resolution, as well as the flat spectrum and variability---particularly toward higher frequencies---seen for the radio counterpart to HCG15-D \citep[e.g.][]{Rani2013_Blazar_Variability}, this scenario struggles to explain some of the other characteristics.

Specifically, in this scenario we would expect to observe a more `roundish' morphology, reminiscent of Odd Radio Circles \citep[ORCs;][]{Norris2021_ORCs,Koribalski2021_ASKAP_ORC,Norris2022_MeerKAT_ORC,Norris2025_MIGHTEE_ORC}. While the mechanism by which ORCs are generated is still discussed, simulations show that ORCs can be generated directly by AGN jet-inflated bubbles viewed in projection \citep[e.g.][]{Lin_Yang_2024_ORC_AGN_Simulations} or the interactions of fossil AGN lobes with propagating shocks \citep[e.g.][]{Shabala2024_ORC_Simulations}, and as such there is likely a strong AGN connection.

Both simulations and observations show that such sources are edge-brightened, with a flatter spectral index in the `ring' (corresponding to the `ridge' in HCG15) and a steeper spectrum internally. Similarly, the polarisation fraction is enhanced in the `ring', although the magnetic field orientation follows the orientation of the ring \citep{Ensslin2002}.

In the case of HCG15, we do not see a `roundish' morphology, but rather a relatively sharp edge to the south-west of the `ridge', and extended morphology to the north-east, and the spectral index shows no flattening in the ridge. Similarly, while the polarisation fraction in the ridge is high, the magnetic field orientation is transverse to the ridge rather than following its long axis. As such, we disfavour such a scenario for the origins of the diffuse emission. However, we do not rule out a viewing angle close to the line of sight for the AGN in HCG15-D, as it is consistent with our findings. Such scenarios involving AGN realignment in new active phases has been observed in several giant radio galaxies to-date \citep[e.g.][]{HernandezGarcia2017_AGN_realignment,Bruni2020_Xray_GRGs_Paper2,Bruni2025_GRACE_Project}. New optical spectroscopy and sub-arcsecond resolution VLBI observations would allow us to test this scenario; this will be explored in future work using our eMERLIN observations (time awarded).

\subsection{On the nature of the diffuse X-ray components}
While the X-ray surface brightness maps presented in Figures~\ref{fig:HCG15_Composite} and \ref{fig:HCG15_Xray} show two principal concentrations of diffuse thermal emission, their nature is not immediately obvious. The primary component is centred on the galaxy HCG15-D, whereas the secondary component is offset from any group-member galaxies and lies just to the west of HCG15-A, within around 50~arcsec radius of $({\rm J2000 \, RA, \, Dec}) = (02:07:49.60, \, +02:09:21.8)$. The initial hypothesis would be that both components are associated with the IGrM of HCG15. Such a scenario would be highly unusual however, particularly given the offset secondary component with respect to all group-member galaxies if this emission represents a diffuse thermal component of the IGrM.

This diffuse component could represent a stripped galaxy halo instead of a separate IGrM component. While such processes are more likely to occur in massive galaxy clusters than in low-mass galaxy groups, stripping of hot galactic haloes can occur on Gyr timescales even in group environments \citep[e.g.][]{Hester2006_HaloStripping,Kawata_Mulchaey_2008}. However, the likely candidate for a halo-stripping origin --- HCG15-A --- still shows significant extended X-ray emission (Figures~\ref{fig:HCG15_Composite},~\ref{fig:HCG15_Xray}). Both the extended thermal component co-located with HCG15-A and the secondary diffuse component show broadly relaxed morphologies, and it is difficult to envision a scenario where an entire galaxy halo could become stripped without being significantly disturbed in the process.

\begin{figure}
\begin{center}
 \includegraphics[width=0.93\linewidth]{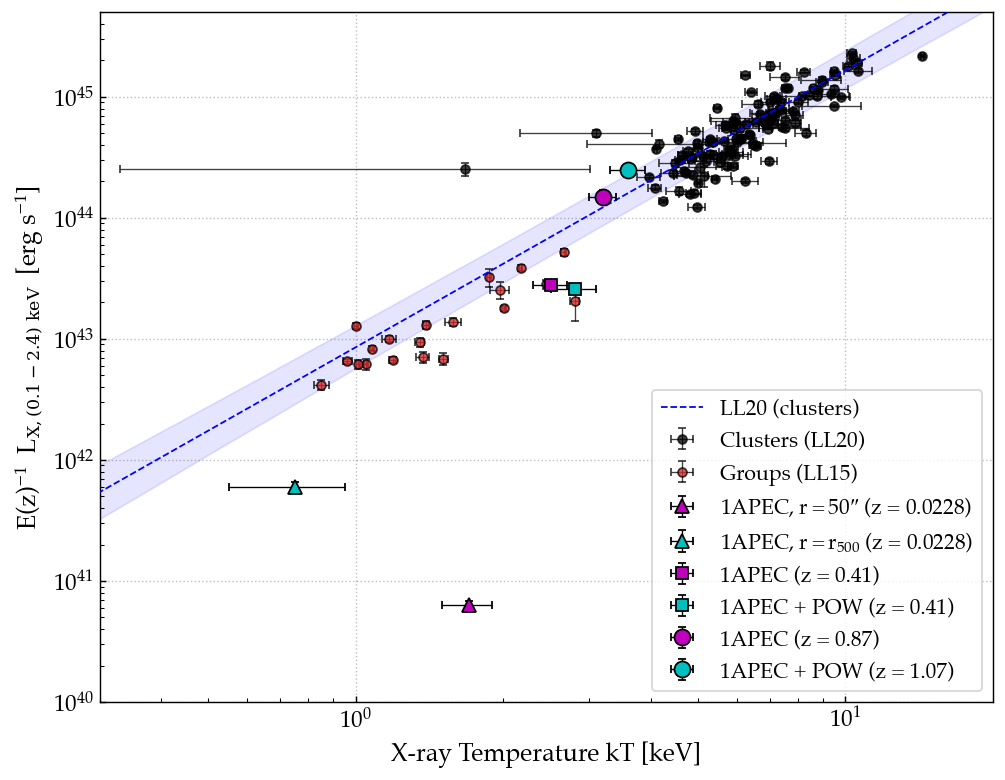}
\caption{Scaling relation between X-ray temperature (kT) and luminosity in the $0.1-2.4$~keV band for samples of galaxy groups \citep[LL15;][]{Lovisari2015_Xray_groups} and clusters \citep[LL20;][]{Lovisari2020_PSZXMM}. Blue line and shaded region denotes the best-fit relation from LL20 (their Table~4, $L_X - T$, third row). Cyan and pink markers show fits to the secondary diffuse X-ray component viewed in projection on HCG15, as described in the text.}
\label{fig:hcg15_LTrelation}
\end{center}
\end{figure}

To understand the nature of this component, we extracted a spectrum from a circular region of 50~arcsec radius and fit one APEC thermal plasma model with, first the redshift fixed at $z=0.0228$, and then free to vary. This radius was chosen to maximise the S/N for the spectral fit. In the former case we obtained a temperature of $1.7\pm0.2$ keV, a very low abundance (i.e. ${\rm Z} < 0.06$) and an estimated luminosity (in the 0.1-2.4 keV band) $L_{\rm X} \approx 6.3 \times 10^{40}$ erg~s$^{-1}$ which is significantly off from observed $L-T$ relation (see Figure~\ref{fig:hcg15_LTrelation}). For this fit, \texttt{cstat/dof} 408/406.

At the representative redshift of HCG15 $(z = 0.0228)$, a radius of 50~arcsec corresponds to a far smaller region than radius of $r_{500}$ used by \cite{Lovisari2020_PSZXMM}. For HCG15, $r_{500} \sim 11$~arcmin, a radius within which the primary concentration of X-ray emission in the vicinity of HCG15-D/F also lies. We repeated the spectral extraction and fitting process using this radius, deriving a best-fit temperature of $T \approx 0.75$ keV and luminosity of $L_{\rm X} \approx 6 \times 10^{41}$ erg~s$^{-1}$. While this point lies closer to the relation, it would still place HCG15 far below both the group sample of \cite{Lovisari2015_Xray_groups} and the $L-T$ relation of \cite{Lovisari2020_PSZXMM}. Additionally, this extraction region also includes emission associated with HCG15, meaning that this datapoint is less informative regarding the origin of the secondary component. 

When the redshift is left free to vary the results depend on the initial fitting values, indicating that there are several local minima. If the initial redshift is set to the HCG15 value the fit converges to a temperature of ${\rm kT} \sim 2.5 \pm 0.2$ keV, ${\rm Z} \sim 0.2 \pm 0.1$, $z \approx 0.41$, and  $L_{\rm X} \approx 2.8 \times 10^{43}$ erg/s. Conversely, if the initial redshift is set to 1 then the fit converges to a temperature of $\sim 3.2 \pm 0.2$~keV, ${\rm Z} \sim 0.8 \pm 0.2$, $z \approx 0.87$, and $L_{\rm X} \approx 1.5 \times 10^{44}$~erg/s. The latter result fits perfectly to the L-T relation but also the former is consistent at $\sim$2$\sigma$ level. For both of these fits, we find \texttt{cstat/dof} 408/405.

We also attempted to fit a two-component APEC model. However, the F-test functionality of XSPEC allowed us to exclude that a second thermal component is needed (significance was lower than 0.05).

Finally, we attempted to fit an APEC+POW model with the power-law index set to 2. Again, the results depend on the initial fitting values. When starting from the HCG15 redshift the fit converges to $T \sim 2.8 \pm 0.3$ keV, ${\rm Z} \sim 0.4 \pm 0.2$, $z \approx 0.41$, and $L_{\rm X} \approx 2.6 \times 10^{43}$~erg/s while when starting from redshift 1 the fit converges to $T \sim 3.6 \pm 0.3$ keV, ${\rm Z} \sim 0.3 \pm 0.2$, $z \approx 1.07$, and  $L_{\rm X} \approx 2.5 \times 10^{44}$~erg/s. Again, the latter result fit better the $L-T$ relation by \cite{Lovisari2020_PSZXMM}. For both of these fits, \texttt{cstat/dof} 415/402.

As such, while it is likely that the diffuse thermal plasma of the IGrM contributes to some extent, we suggest that the majority fraction of this secondary diffuse X-ray component seen within 50~arcsec radius of $(02:07:49.60, \, +02:09:21.8)$ is not associated with the IGrM of HCG15, but rather represents a background object. With the current data we cannot draw firm conclusions on the nature of this background object, but based on both the morphology of the emission and the L-T relation in Figure~\ref{fig:hcg15_LTrelation} it would appear that a background cluster at $z \approx 0.87 - 1.07$ is most consistent with both our observations and established scaling relations from \cite{Lovisari2020_PSZXMM}.

This hypothesis is reinforced by the recent cluster catalogue derived by \cite{Klein2024_ACT-DR5_MCMF} from Data Release 5 of the Atacama Cosmology Telescope (ACT) thermal Sunyaev–Zel`dovich Effect (tSZE) survey \citep[ACT-DR5;][]{Naess2020_ACT-DR5}. \citeauthor{Klein2024_ACT-DR5_MCMF} report a strong $(\rm{S/N} = 5.91)$ detection of a cluster\footnote{Identified as ACT-CL~J0207.8+0209.} at the location of this X-ray component, with a spectroscopic redshift of $z_{\rm spec} = 0.88$ for the best optical counterpart. This cluster counterpart has a mass of $M_{500} \simeq 2 \times 10^{14} ~ {\rm M_{\odot}}$. We indicate the location of this cluster in Figure~\ref{fig:HCG15_Composite} and Figure~\ref{fig:HCG15_Xray} for reference.

\subsection{Origins of the polarised emission: shocks in the IGrM}
\subsubsection{Viewing angle}
In a scenario where the polarised emission generated by shocks in the IGrM, it is possible to estimate the viewing angle of the propagating shock front (from the observer's point of view) based on the polarisation information. The mathematical framework was first discussed by \cite{Ensslin1998} in the context of understanding the viewing angle of radio relics in merging clusters, and has been invoked in several studies since \citep[e.g.][]{Hoang2018_Abell1240,Rajpurohit2022_MACSJ0717_pol}.

Following previous studies, we take the weak-field approximation from Section~3.2 of \cite{Ensslin1998}. In this situation, the average fractional polarisation $\langle \Pi \rangle$ is related to the viewing angle $\Theta$ via Eq.~22 of \cite{Ensslin1998}:
\begin{equation}\label{eq:viewing_angle}
   \langle \Pi \rangle = \frac{ \gamma +1 }{ \gamma + \frac{7}{3} } \frac{ {\rm sin}^2 \Theta }{ \left( \frac{2 R^2 }{ R^2 - 1 } \right) - {\rm sin}^2 \Theta }
\end{equation}
where $\gamma$ relates to the observed spectral index via $\gamma = 1 - 2 \alpha$ and $R$ is the shock compression ratio defined as $R = (1 -\alpha )/(-\alpha - 0.5)$, adjusted for our spectral index convention. See \cite{Ensslin1998}. Note that in Equation~\ref{eq:viewing_angle}, $\Theta = 90\degree$ would indicate a shock propagating in the plane of the sky. Following previous studies, we take the value of the integrated spectrum $\alpha = -1.28 \pm 0.04$ from the lower-frequency regime where it follows a single power-law trend.

\begin{figure}
\begin{center}
 \includegraphics[width=0.95\linewidth]{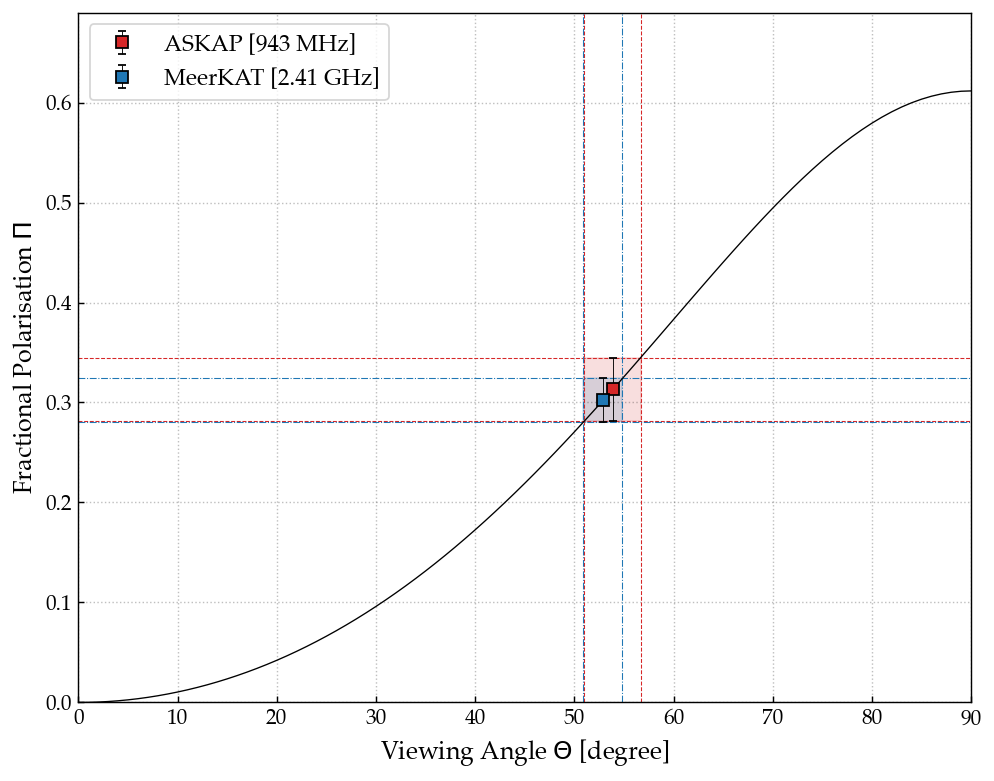}
\caption{Viewing angle of HCG15. Black curve denotes the prediction of Equation~\ref{eq:viewing_angle} evaluated for viewing angles in the range $0 \leq \Theta \, \left[ {\rm deg} \right] \leq 90$ given our observed integrated spectral index. Datapoints show our polarisation results as measured by ASKAP at 943\,MHz and MeerKAT at 2.41\,GHz.}
\label{fig:hcg15_viewangle}
\end{center}
\end{figure}

We then evaluated Equation~\ref{eq:viewing_angle} for viewing angles between $0 \leq \Theta \, \left[ {\rm deg} \right] \leq 90$ to predict the observed fractional polarisation. We plot the results of this model in Figure~\ref{fig:hcg15_viewangle}, overlaid with our observational results at 943\,MHz and 2.41\,GHz.

Given our observed fractional polarisation of $\langle \Pi \rangle = (31.3 \pm 3.2)\%$ at 943\,MHz and $(30.2 \pm 2.2)\%$ at 2.41\,GHz, our modelling suggests as a viewing angle of $\Theta = (53.9^{+2.8}_{-2.9})\degree$ at 943\,MHz or $\Theta = (52.9^{+2.0}_{-2.0})\degree$ at 2.41\,GHz. These results are consistent, suggesting an inclined viewing angle.

\subsubsection{Searching for shocks in the IGrM}
\cite{Giacintucci2011_GalaxyGroups} first suggested that HCG15 may represent an analogue of Stephan's Quintet \citep[e.g.][]{OSullivan2009_StephansQuintet} where a high-velocity interloper galaxy is driving shocks in the IGrM. Such an interloper would need to be moving with a relative velocity greater than the speed of sound in the IGrM, which can be estimated as $c_{\rm s} \, [{\rm km \, s^{-1}}] = 1480 \, \sqrt{ {T_{\rm X}}/{10^8 {\rm K}}}$.

From our \textit{XMM-Newton} thermodynamic maps (Fig.~\ref{fig:HCG15_Xray}) we find an average temperature of $\langle T_{\rm X} \rangle = (0.72 \pm 0.13) \, {\rm keV}$, suggesting a sound speed $c_{\rm s} \simeq (427 \pm 77)$~km~s$^{-1}$. Given on the relative positions of the group-member galaxies and the diffuse polarised emission, HCG15-C could represent a candidate for such an interloper. As can be seen in Table~\ref{tab:hcg15_galaxies}, this galaxy is the most massive group member with a stellar mass of $M_* = 1.07 \times 10^{11} ~ M_{\odot}$ \citep{Bitsakis2011}. Previous velocity measurements would appear to support the hypothesis that this galaxy is moving through the IGrM at high relative velocity, however given our new velocity measurements (Section~\ref{sec:spectra}) it is prudent to examine this specifically.

From earlier, our revised radial velocity measurement for HCG15-C is $cz = (6803 \pm 12)$~km~s$^{-1}$. The average radial velocity for the other five group-member galaxies is $\langle cz \rangle = (6753 \pm 129)$~km~s$^{-1}$, and as such the \textit{relative} velocity of HCG15-C (some $\sim 50$~km~s$^{-1}$) is highly subsonic. However, this approach is only sensitive to motions with a significant velocity component along the line of sight, whereas there could be a significant velocity component in the plane of the sky.

As seen in Figure~\ref{fig:HCG15_Polarisation}, the magnetic field is oriented perpendicular to the long axis of the bar, and so in a scenario where shocks are driven into the IGrM by HCG15-C, we might expect a direction of motion largely along a north-west/south-east direction. To examine this hypothesis further, we searched our \textit{XMM-Newton} maps for evidence of the presence of shocks in the IGrM, which might be discontinuities in temperature, pressure, and/or surface brightness for example.

While the tessels used to reconstruct the thermodynamic maps presented in Figure~\ref{fig:HCG15_Xray} are relatively large due to the signal-to-noise requirement, we do not see clear evidence of a temperature enhancement at the location of HCG15-C $(T_{\rm X} = 1.10 \pm 0.23 ~ {\rm keV})$ compared to the average in the IGrM (from earlier $\langle T_{\rm X} \rangle = 0.72 \pm 0.13$~keV). The only regions of enhanced temperature are those associated with the background cluster discussed in the previous section, and are unrelated to the IGrM of HCG15.

We also applied the Gaussian Gradient Magnitude (GGM) filtering technique \citep{Sanders2016b,Sanders2016a} as employed in the \texttt{ggm} package described by \citep{Sanders2021_GGM}\footnote{Currently accessible via \url{https://github.com/jeremysanders/ggm}} to the \textit{XMM-Newton} surface brightness map in an attempt to highlight any discontinuities in the IGrM. This technique is often used to highlight shocks and cold fronts in groups and clusters \citep[e.g.][]{Sanders2022_A3266_eROSITA,Biava2024_CoolCoreSample,Mirakhor2023_NGC4839_Coma,Rajpurohit2024_NGC741}. In the case of HCG15 however, GGM filtering did not reveal significant evidence of credible discontinuities. Similarly, we examined surface brightness profiles along the direction of the bar, finding no evidence of discontinuities.

\subsection{Origins of the polarised emission: magnetic draping}
\subsubsection{Physical conditions}
Let us consider a scenario where the diffuse emission represents fossil emission expanding into the IGrM of HCG15. Following earlier works by \cite{Guidetti2012_B20755} and \cite{Adebahr2019_B20258} we evaluate the potential for magnetic draping to be responsible for the strongly linearly-polarised emission seen in the IGrM.

For magnetic draping to be plausible, two conditions must be met. Firstly, we require super-Alfv\'{e}nic expansion in the IGrM. Let us take the standard textbook Equation for Alfv\'{e}n velocity, modified to reflect the (primarily H and He) composition of the IGrM:
\begin{equation}\label{eq:alfven_velocity}
   v_{\rm A} = 2.05 \, \left( n_{\rm e} \, \, [{\rm cm}^{-3}] \right)^{- \frac{1}{2}}  \, \left( B \, \, [\upmu{\rm G}] \right) \, \, \, {\rm km \, s}^{-1}
\end{equation}
where $n_{\rm e}$ is the electron density and $B$ is the magnetic field strength. Continuing from earlier, we take $B$ to be the magnetic field strength assumed earlier, $B_{\rm eq} = 1.5 \, \upmu$G. For the electron density we use the median value derived from our tessellated \textit{XMM-Newton} map, $n_{\rm e} = 2.57 \times 10^{-4}$~cm$^{-3}$, and assume this value holds constant across the extent of the source. Using these values, Equation~\ref{eq:alfven_velocity} yields an Alfv\'{e}n velocity of $v_{\rm A} = 191.4$~km~s$^{-1}$.

We can use the projected distance in conjunction with the source age to estimate the expansion velocity $v_{\rm exp}$ of the plasma in the IGrM. For the source age, we can use the values derived earlier from our spectral fitting, where we found values in the range $\tau \simeq 70 - 86$~Myr; in this case we adopt the more conservative age of $\tau = 80.8$~Myr derived from our fit to the \textit{uv}-matched integrated spectrum. For the projected distance, we continue with our hypothesis that the diffuse emission seen in HCG15 was seeded by an historic epoch of activity from the roughly central galaxy HCG15-D. Hence we find a largest (projected) distance of 240~arcsec (106~kpc), although we note that this is measured in a straight line from HCG15-D to the outer extent of the emission to the south-east. This distance would increase to some 350~arcsec (155~kpc) if we followed a ridge-line along the spine of the diffuse plasma from HCG15-D to the outer edge. Given the projected distance and source age, we find an expansion velocity of $v_{\rm exp} \simeq 1280$~km~s$^{-1}$, indicating a super-Alfv\'{e}nic expansion with a Mach number $\mathcal{M}_{\rm A} = 6.7$.

Secondly, the coherence scale of the magnetic field $\lambda_{B}$ must be sufficiently large for the magnetic field to remain coherent while draping around the expanding plasma. From simulations \citep[e.g.,][]{Ruszkowski2007,Dursi_Pfrommer_2008,Pfrommer_Dursi_2010_Nature,Ruszkowski_Pfrommer_2023} we know that magnetic draping can occur when
\begin{equation}\label{eq:magnetic_coherence}
   \lambda_{B} \gtrsim R_c \, / \, \mathcal{M}_{\rm A}
\end{equation}
where $R_c$ is the radius of curvature around the stagnation point of the draping layer and $\mathcal{M}_{\rm A}$ is the Alfv\'{e}nic Mach number derived above \citep[see discussion in][for example]{Dursi_Pfrommer_2008,Adebahr2019_B20258}. Assuming an approximately cylindrical geometry for the polarised bar, the curvature radius $R_c$ can be approximated as half the minor axis of the polarised bar, i.e. $R_c \sim 81 / 2 = 40.5$~arcsec (17.9~kpc), and so Equation~\ref{eq:magnetic_coherence} suggests that for magnetic draping to occur, $\lambda_{B} \gtrsim 3$~kpc. The extent of the polarised emission and the uniformity of the magnetic field vectors suggest a coherent magnetic field on scales up to at least $\sim 35$~kpc, and as such, both conditions --- super-Alfv\'{e}nic expansion and a sufficiently large magnetic field coherence length --- appear to be met. This would suggest that a magnetic draping layer could develop between the expanding remnant plasma and the IGrM of HCG15.


\subsubsection{Numerical simulations}
Detailed exploration of magnetic draping through bespoke simulations is beyond the scope of this work, although we can compare with previously published simulation sets. In particular, \cite{Pfrommer_Dursi_2010_Nature} published an extensive suite of magnetic draping simulations as observed from a range of viewing angles. We draw the reader's attention to their Supplementary Figure 13, which shows the polarised synchrotron emission observed at a variety of viewing angles. Those simulations of a magnetic draping layer viewed at an inclination of $\Theta \simeq 45\degree ~ {\rm to } ~ 60\degree$ in particular bear a strong resemblance to the observed linearly-polarised emission in Figure~\ref{fig:HCG15_Polarisation}. It is also curious to note that this viewing angle range broadly agrees with the viewing angle of $\Theta \simeq 53\degree$ estimated earlier from the observed polarised emission.

\section{Summary and Conclusions}\label{sec:conclusions}
In this paper, we have presented new multi-frequency observations of Hickson Compact Group (HCG) 15. This source was visually identified in radio continuum data from ASKAP's Evolutionary Map of the Universe (EMU) survey, and full polarisation processing with the POSSUM pipeline revealed significant strongly linearly-polarised emission and a highly-ordered magnetic field that sparked a follow-up campaign.

We obtained new and/or archival data from MeerKAT, LOFAR, the VLA and the GMRT at radio wavelengths, the Himalayan Chandra Telescope (HCT) at optical wavelengths, and with re-analysed X-ray data from \textit{XMM-Newton}, with the aim of understanding the nature of the diffuse radio continuum and linear polarisation emission associated with this compact group. Our key findings are as follows:

\begin{itemize}
    \setlength\itemsep{0.5em}
    \item We detect radio counterparts to all galaxies in the group at one or more radio frequency, with the exception of HCG15-B. Our observations suggest that \textbf{(i)} the synchrotron emission associated with HCG15-A and HCG15-E is consistent with star formation, including the nuclear region; (ii) HCG15-C likely hosts both an AGN and star formation, although separating these components is not possible with the available data; (iii) HCG15-D is clearly AGN-dominated; (iv) HCG15-F appears to be star-formation dominated, with extended synchrotron emission recovered associated with the disk. \textbf{HCG15-D hosts an inverted-spectrum radio AGN} with a spectral index of $\alpha = +0.26 \pm 0.02$. This AGN is also highly variable at frequencies above $\nu \approx 3$~GHz, with hints of variability extending down to $\sim 1$~GHz. As such, we hypothesise that the AGN associated with HCG15-D is young, having restarted relatively recently, although we cannot determine an age as all our observations lie below any spectral break.

    \item From our \textbf{optical spectroscopy} we find a best-fit radial velocity of $cz = 6803 \pm 12$~km~s$^{-1}$ for HCG15-C and $cz = 6133 \pm 98$~km~s$^{-1}$ for HCG15-F. The latter is consistent with published measurements, albeit with improved uncertainties, but the former value allows us to resolve an open question from historic large optical surveys: HCG15-C is a part of the group, as originally suggested by \cite{Hickson1982}.

    \item The \textbf{diffuse emission} associated with HCG15 shows a steep and curved integrated spectrum consistent with aged plasma. Integrated spectral modelling suggests an age $\tau_{\rm total} \simeq 80 - 86$~Myr.

    \item The \textbf{resolved spectral index} profiles show no gradients that might be expected in the case of active cosmic ray electron injection from an AGN, but rather broadly uniform spectral index across the whole extent. The median spectral index is steep: $\langle \alpha_{\rm low} \rangle = -1.13 \pm 0.08$ between 943\,MHz and 143\,MHz, $\langle \alpha_{\rm mid} \rangle  = -1.51 \pm 0.14$ between 2.41\,GHz and 943\,MHz, and $\langle \alpha_{\rm high} \rangle = -2.14 \pm 0.30$ between 5.0\,GHz and 2.41\,GHz. 
    \vspace{0.5em}
    
    The spectral index profiles also show statistically-significant fluctuations between all pairs of frequencies, suggesting an inhomogeneous cosmic ray electron population that traces a distribution of different injection and/or acceleration timescales, inhomogeneities in the environment, or both.

    \item The \textbf{colour-colour diagrams} show clear evidence of spectral curvature, with a broad distribution of datapoints in colour-colour space. Standard injection and ageing models provide a fairly poor description, failing to capture the broad distribution; instead, an inhomogeneous cosmic ray electron population may be able to replicate our findings.

    \item Both ASKAP (at 943\,MHz) and MeerKAT (at 2.41\,GHz) recover \textbf{significant strongly linearly-polarised emission} associated with the main `ridge' structure of HCG15. The typical polarisation fraction is high, with a median value of $\langle \Pi \rangle = (31.3 \pm 3.2)\%$ at 943\,MHz and $(30.2 \pm 2.2)\%$ at 2.41\,GHz, although in some regions the polarisation fraction reaches toward the theoretical maximum for synchrotron emission.

    \item Additionally, both ASKAP and MeerKAT show a \textbf{highly-ordered magnetic field} across the extent of the linearly-polarised emission, with an orientation that is transverse to the major axis of the `ridge'. The typical \textbf{RM} recovered for the linearly-polarised emission is low, suggesting that the line of sight to the emission does not pass through a significant depth of the IGrM --- i.e. the linearly-polarised emission originates on the near side of the IGrM.

    \item The \textbf{X-ray emission} associated with the group is complex: not only is there significant diffuse thermal emission, but five of the six galaxies show associated X-ray counterparts. The diffuse emission is broadly characterised by two principal components: the primary centred on galaxy HCG15-D, and the secondary offset from any group members, centred around $({\rm J2000 \, RA, \, Dec}) = (02:07:49.60, \, +02:09:21.8)$. This secondary component also shows a high temperature ($T_{\rm X} \approx 3.5$~keV) and a divergent pseudo-entropy ($\mathcal{K} \approx 13$~keV~cm$^2$) compared to the group average $(\langle T_{\rm X} \rangle = (0.72 \pm 0.13) \, {\rm keV})$.
    \vspace{0.5em}
    
    Spectral analysis suggests that this component is in fact a background cluster at redshift $z \approx 0.87$; consistent with the position and redshift of the SZ-detected cluster ACT-CL~J0207.8+0209 reported by \cite{Klein2024_ACT-DR5_MCMF} during preparation of this manuscript. Thus, we conclude that this second thermal component is only viewed in projection onto the IGrM, and is not part of HCG15.
\end{itemize}

We have systematically investigated the previous hypotheses for the nature of the diffuse emission. Previous studies, which did not focus on this group in such detail, hypothesised that the diffuse emission may represent either a remnant AGN, turbulence-driven radio-(mini-)halo type emission, or shock-driven radio-relic-type emission.

The detection of significant linearly-polarised emission strongly disfavours a turbulence-driven origin. In a shock-driven scenario, analogous to Stephan's Quintet, the most likely candidate for a high-velocity interloper would be HCG15-C; however, our optical spectroscopy reveals a highly sub-sonic relative velocity along the line of sight. While we cannot conclusively rule out a significant velocity component in the plane of the sky, we do not detect any suggestion of discontinuities in the X-ray surface brightness or temperature that would hint at shocks in the IGrM; Gaussian gradient magnitude (GGM) filtering similarly reveals no significant edges.

In the broader context of AGN life-cycles and evolution, our evidence strongly favours the interpretation that the diffuse emission associated with the IGrM of HCG15 represents remnant emission from an historic episode of AGN activity from the group-member galaxies. Based on the relative galaxy positions and the morphology of the diffuse emission, as well as the distribution of the thermal plasma of the IGrM, we suggest that HCG15-D was the original AGN responsible for seeding this diffuse emission.

In such a scenario, the diffuse emission may constitute some form of bent remnant radio galaxy viewed in projection. The highly-polarised `ridge' structure would be viewed in the foreground of HCG15-D, toward the nearer edge of the group, and the faint unpolarised emission toward the north-east in the background of HCG15-D, viewed toward the farther edge of the group. Such an interpretation is broadly supported by our polarisation analysis, from which we derive an inclination angle of around $\Theta \approx 57\degree$ to the line of sight. The uniformly low values of the $|{\rm RRM}|$ support this, as they suggest the polarised emission arises on the near side of the IGrM.

While the polarisation properties of remnant radio galaxies are poorly explored, in the standard picture of fossil cosmic rays evolving passively in the IGrM, we might expect that the dynamic environment would act to homogenise the magnetic field topography. Thus we would expect a reduced linearly-polarised signal and a more turbulent magnetic field. Conversely, both ASKAP and MeerKAT see a strongly linearly-polarised `ridge' with a highly-ordered magnetic field structure, which is challenging to explain within the standard picture.

While large-scale shocks are known to compress and amplify magnetic fields, leading to significant linearly-polarised emission, shocks also act to flatten the radio spectrum toward the leading edge, which is not observed in HCG15. As such, and given the lack of evidence of shocks from our \textit{XMM-Newton} analysis, we disfavour a shock-driven origin for the polarised emission. 

Instead, we favour an explanation where the fossil plasma in the IGrM is expanding sub-sonically but super-Alfv\'{e}nically. In such a scenario, a magnetic draping layer could form between the expanding fossil cosmic ray electron population and the thermal IGrM of HCG15. Our analysis suggests that the key conditions for magnetic draping in the IGrM are met, and our observational evidence bears strong resemblance to simulations of magnetic draping, although further detailed work is highly motivated.

Looking ahead to the near future, the SKA will become operational within the coming years. We fully expect the SKA to deliver transformational science in the under-explored area of galaxy groups, due to its excellence in mapping extremely low surface brightness emission at high resolution. With the wide frequency coverage provided by combining SKA-Low and SKA-Mid, full mapping of the synchrotron spectrum and precision polarimetric studies on galaxy groups, of the kind performed in this paper, will become routinely possible.

\begin{acknowledgement}
We thank the referee for their constructive and informative comments and suggestions, which both strengthened our manuscript and improved the clarity of the text. CJR acknowledges financial support from the German Science Foundation DFG, via the Collaborative Research Center SFB1491 `Cosmic Interacting Matters – From Source to Signal', and from the ERC Starting Grant `DRANOEL', number 714245. TV acknowledges funding support from the University of Western Australia, RCA 2023/GR001284. LL acknowledges support from INAF grant 1.05.12.04.01. MB acknowledges financial support from Next Generation EU funds within the National Recovery and Resilience Plan (PNRR), Mission 4 - Education and Research, Component 2- From Research to Business (M4C2), Investment Line 3.1 - Strengthening andcreation of Research Infrastructures, Project IR0000034 – ``STILES - Strengthening the Italian Leadership in ELT and SKA''. PKN acknowledges support from the Centro de Astrofisica y Tecnologias Afines (CATA) fellowship via grant Agencia Nacional de Investigacion y Desarrollo (ANID), BASAL FB210003. AB acknowledges financial support from the ERC Starting Grant `DRANOEL', number 714245. SG acknowledges funding support from the US Naval Research Laboratory. Basic research in radio astronomy at the Naval Research Laboratory is supported by 6.1 Base funding. CP acknowledges support by the European Research Council under ERC-AdG grant PICOGAL-101019746. MR acknowledges support from the National Aeronautics and Space Administration grant ATP 80NSSC23K0014 and the National Science Foundation Collaborative Research grant NSF AST-2009227. DJB. acknowledges funding from the German Science Foundation DFG, via the Collaborative Research Center SFB1491 `Cosmic Interacting Matters – From Source to Signal'. AD acknowledges support by the BMBF Verbundforschung under the grant 05A20STA. R.S. acknowledges the support of the Department of Atomic Energy, Government of India, under project no. 12-R\&D-TFR-5.02-0700. CJR wishes to thank K.~Rajpurohit for useful discussions around aspects of this paper. 

This scientific work uses data obtained from Inyarrimanha Ilgari Bundara / the Murchison Radio-astronomy Observatory. We acknowledge the Wajarri Yamaji People as the Traditional Owners and native title holders of the Observatory site. The Australian SKA Pathfinder is part of the Australia Telescope National Facility (\url{https://ror.org/05qajvd42}) which is managed by CSIRO. Operation of ASKAP is funded by the Australian Government with support from the National Collaborative Research Infrastructure Strategy. ASKAP uses the resources of the Pawsey Supercomputing Centre. Establishment of ASKAP, the Murchison Radio-astronomy Observatory and the Pawsey Supercomputing Centre are initiatives of the Australian Government, with support from the Government of Western Australia and the Science and Industry Endowment Fund. The POSSUM project (https://possum-survey.org) has been made possible through funding from the Australian Research Council, the Natural Sciences and Engineering Research Council of Canada, the Canada Research Chairs Program, and the Canada Foundation for Innovation.

The MeerKAT telescope is operated by the South African Radio Astronomy Observatory, which is a facility of the National Research Foundation, an agency of the Department of Science and Innovation. We wish to acknowledge the assistance of the MeerKAT science operations team in both preparing for and executing the observations used in this paper.

This work uses data from the Himalayan Chandra Telescope based at Hanle Observatory, India. We thank the staff of IAO, Hanle and CREST, Hosakote, that made these observations possible. The facilities at IAO and CREST are operated by the Indian Institute of Astrophysics, Bangalore.

This work uses data from the NRAO Karl G. Jansky Very Large Array. The National Radio Astronomy Observatory is a facility of the National Science Foundation operated under cooperative agreement by Associated Universities, Inc. The NVAS images used in this work were produced as part of the NRAO VLA Archive Survey, (c) AUI/NRAO.

LOFAR is the Low Frequency Array designed and constructed by ASTRON. It has observing, data processing, and data storage facilities in several countries, which are owned by various parties (each with their own funding sources), and which are collectively operated by the ILT foundation under a joint scientific policy. The ILT resources have benefited from the following recent major funding sources: CNRS-INSU, Observatoire de Paris and Universit\'e d'Orl\'eans, France; BMBF, MIWF-NRW, MPG, Germany; Science Foundation Ireland (SFI), Department of Business, Enterprise and Innovation (DBEI), Ireland; NWO, The Netherlands; The Science and Technology Facilities Council, UK; Ministry of Science and Higher Education, Poland; The Istituto Nazionale di Astrofisica (INAF), Italy.

This research made use of the Dutch national e-infrastructure with support of the SURF Cooperative (e-infra 180169) and the LOFAR e-infra group. The J\"{u}lich LOFAR Long Term Archive and the German LOFAR network are both coordinated and operated by the J\"{u}lich Supercomputing Centre (JSC), and computing resources on the supercomputer JUWELS at JSC were provided by the Gauss Centre for Supercomputing e.V. (grant CHTB00) through the John von Neumann Institute for Computing (NIC).

This research made use of the University of Hertfordshire high-performance computing facility and the LOFAR-UK computing facility located at the University of Hertfordshire and supported by STFC [ST/P000096/1], and of the Italian LOFAR IT computing infrastructure supported and operated by INAF, and by the Physics Department of Turin university (under an agreement with Consorzio Interuniversitario per la Fisica Spaziale) at the C3S Supercomputing Centre, Italy.

The Australia Telescope Compact Array is part of the Australia Telescope National Facility (grid.421683.a) which is funded by the Australian Government for operation as a National Facility managed by CSIRO. We acknowledge the Gomeroi people as the traditional owners of the Observatory site.

This research has made use of the CIRADA cutout service at URL \url{cutouts.cirada.ca}, operated by the Canadian Initiative for Radio Astronomy Data Analysis (CIRADA). CIRADA is funded by a grant from the Canada Foundation for Innovation 2017 Innovation Fund (Project 35999), as well as by the Provinces of Ontario, British Columbia, Alberta, Manitoba and Quebec, in collaboration with the National Research Council of Canada, the US National Radio Astronomy Observatory and Australia’s Commonwealth Scientific and Industrial Research Organisation.

This project used public archival data from the Dark Energy Survey (DES). Funding for the DES Projects has been provided by the U.S. Department of Energy, the U.S. National Science Foundation, the Ministry of Science and Education of Spain, the Science and Technology Facilities Council of the United Kingdom, the Higher Education Funding Council for England, the National Center for Supercomputing Applications at the University of Illinois at Urbana-Champaign, the Kavli Institute of Cosmological Physics at the University of Chicago, the Center for Cosmology and Astro-Particle Physics at the Ohio State University, the Mitchell Institute for Fundamental Physics and Astronomy at Texas A\&M University, Financiadora de Estudos e Projetos, Funda{\c c}{\~a}o Carlos Chagas Filho de Amparo {\`a} Pesquisa do Estado do Rio de Janeiro, Conselho Nacional de Desenvolvimento Cient{\'i}fico e Tecnol{\'o}gico and the Minist{\'e}rio da Ci{\^e}ncia, Tecnologia e Inova{\c c}{\~a}o, the Deutsche Forschungsgemeinschaft, and the Collaborating Institutions in the Dark Energy Survey.

The Collaborating Institutions are Argonne National Laboratory, the University of California at Santa Cruz, the University of Cambridge, Centro de Investigaciones Energ{\'e}ticas, Medioambientales y Tecnol{\'o}gicas-Madrid, the University of Chicago, University College London, the DES-Brazil Consortium, the University of Edinburgh, the Eidgen{\"o}ssische Technische Hochschule (ETH) Z{\"u}rich,  Fermi National Accelerator Laboratory, the University of Illinois at Urbana-Champaign, the Institut de Ci{\`e}ncies de l'Espai (IEEC/CSIC), the Institut de F{\'i}sica d'Altes Energies, Lawrence Berkeley National Laboratory, the Ludwig-Maximilians Universit{\"a}t M{\"u}nchen and the associated Excellence Cluster Universe, the University of Michigan, the National Optical Astronomy Observatory, the University of Nottingham, The Ohio State University, the OzDES Membership Consortium, the University of Pennsylvania, the University of Portsmouth, SLAC National Accelerator Laboratory, Stanford University, the University of Sussex, and Texas A\&M University.

Based in part on observations at Cerro Tololo Inter-American Observatory, National Optical Astronomy Observatory, which is operated by the Association of Universities for Research in Astronomy (AURA) under a cooperative agreement with the National Science Foundation.

This research has made use of numerous python packages not cited explicitly elsewhere in the text: \textsc{astropy}, a community-developed core Python package for Astronomy \citep{Astropy_2013,Astropy_2018,Astropy_2022}, \textsc{aplpy} \citep{Robitaille2012}, \textsc{cmasher} \citep{vanderVelden2020}, \textsc{colorcet} \citep{Kovesi2015}, \textsc{matplotlib} \citep{Hunter2007}, \textsc{numpy} \citep{Numpy2011,Harris_2020_NumPy} and \textsc{scipy} \citep{Jones2001}. This research made extensive use of the Astrophysics Data System (ADS), funded by NASA under Cooperative Agreement 80NSSC21M00561. This work also made use of the Cube Analysis and Rendering Tool for Astronomy \citep[\textsc{CARTA};][]{Comrie2021_CARTA}.
\end{acknowledgement}

\paragraph{Data Availability}
The images underlying this article will be shared on reasonable request to the corresponding author. Raw MeerKAT visibilities for DDT-20230705-CR-01 can be accessed via the SARAO archive (\url{https://apps.sarao.ac.za/katpaws/archive-search}). Raw LOFAR visibilities can be accessed via the LOFAR Long-Term Archive (LTA; \url{https://lta.lofar.eu}). Visibilities and pipeline-processed ASKAP images are available from CASDA (\url{https://research.csiro.au/casda/}). GMRT data can be sourced via the GMRT Online Archive (GOA; \url{https://naps.ncra.tifr.res.in/goa/data/search}). Raw visibilities from the ATCA are available via the Australia Telescope Online Archive (ATOA; \url{https://atoa.atnf.csiro.au}). The \XMM{} data used in this paper are available through the Science Data Archive (\url{https://www.cosmos.esa.int/web/xmm-newton/xsa}).

\bibliographystyle{aa}
\bibliography{HCG15}

\begin{appendix}
\section{MeerKAT S-band polarimetry verification}\label{appendix_meerkat_pol}
As mentioned in the main body of the paper, to the best of our knowledge this is the first publication reporting linear polarisation data from MeerKAT's S-band receivers. We followed the calibration procedure described in the text, which makes use of standard techniques in \textsc{CASA} \texttt{v5.5.0-149.el7}.

\begin{figure}
  \centering
 \includegraphics[width=0.9\linewidth]{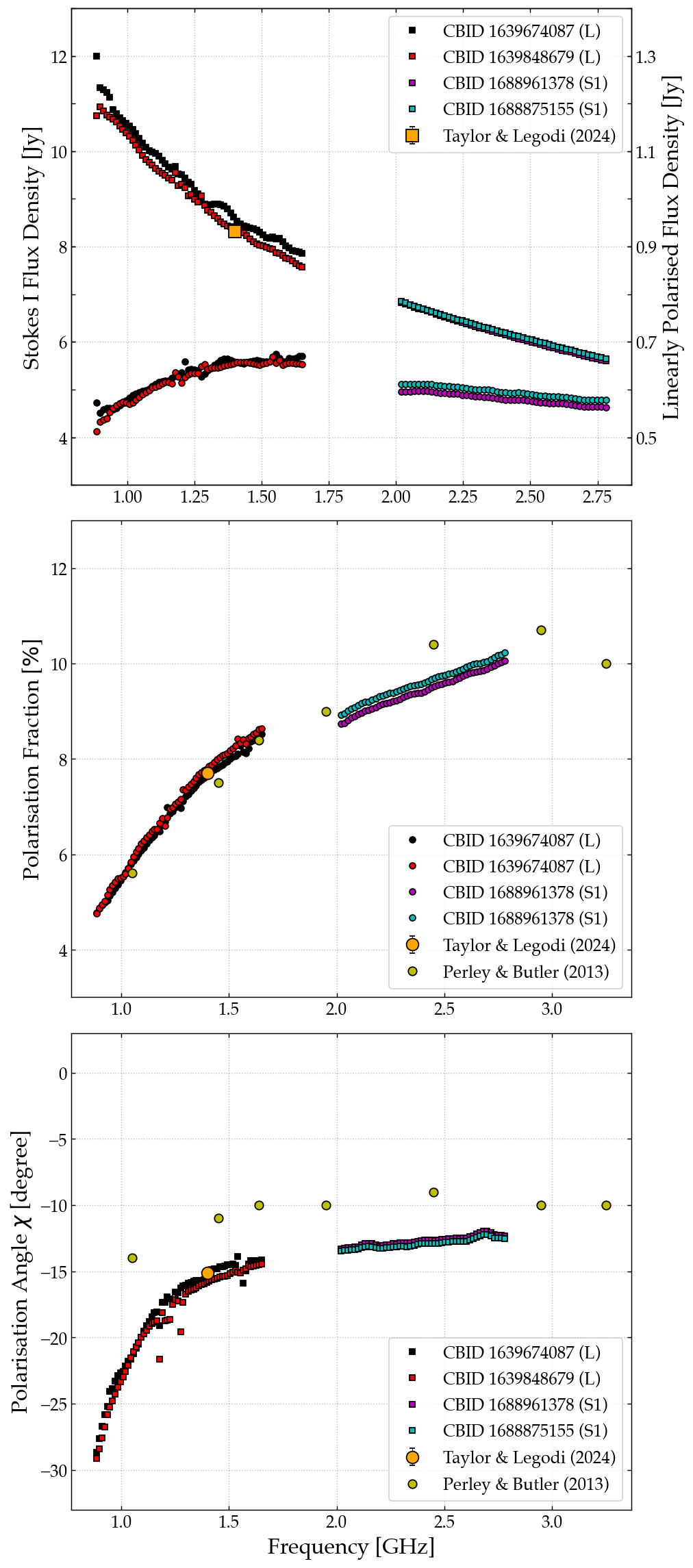}
\caption{Polarisation properties of 3C\,138 measured from MeerKAT data at both L- and S-band. The S-band data are those presented in this paper, whereas the L-band data are from published CBIDs 1639674087 and 1639848679 \citep{Riseley2024_A2142}. Also shown are reference measurements with the VLA from \cite{PerleyButler2013} and with MeerKAT from \cite{Taylor2024_MeerKAT_polcals}.}
\label{fig:3c138_measurements}
\end{figure}

To check our calibration, we imaged our polarisation calibrator 3C\,138 using \textsc{WSclean} \texttt{v3.3} with standard polarisation imaging options enabled (\texttt{-join-channels}, \texttt{-join-polarizations}, and \texttt{-squared-channel-joining}) producing images in all four Stokes parameters (\textit{I,Q,U,V}) before generating linearly-polarised intensity and polarisation angle maps. The results of the calibration  derived from multi-frequency synthesis (MFS) images are shown in Table~\ref{tab:polcal_results}, and we present the full spectral measurements in Figure~\ref{fig:3c138_measurements}.

\begin{table}[h!]
\centering
\caption{Polarisation properties of 3C\,138 as measured from our MeerKAT S-band observations. \label{tab:polcal_results}}
\begin{tabular}{lcc}
\hline
CBID                           &  1688875155     &  1688961378         \\
\hline\hline
Reference frequency            &  2.412\,GHz     &   2.412\,GHz        \\
Stokes I                       &  6.36\,Jy       &   6.33\,Jy          \\
Polarised Intensity            &  0.60\,Jy       &   0.58\,Jy          \\
Fractional Polarisation $P/I$  &  9.4\%          &   9.2\%             \\
Polarisation Angle $\chi$      &  $-13.0\degree$ &   $-12.9\degree$    \\
\hline
\end{tabular}
\end{table}

While the polarisation properties measured from our data in Table~\ref{tab:polcal_results} are somewhat different than reference values of $P/I = 10.4\%$ and $\chi = -9\degree$ at 2.45\,GHz from \cite{PerleyButler2013} --- and see also Figure~\ref{fig:3c138_measurements} --- it is known that 3C\,138 has been undergoing a significant flare since 2021, and the frequency-dependent polarisation properties have been changing. The more recent MeerKAT L-band polarisation calibrator census performed by \cite{Taylor2024_MeerKAT_polcals} recovers fractional polarisation and intrinsic polarisation angle of $P/I = 7.71 \pm 0.002\%$ and $\chi = -15.1 \pm 0.2\degree$, respectively, from observations performed in 2019~August and 2020~August. 

As a further check, we imaged the calibrated data for 3C\,138 from two L-band CBIDs, 1639674087 and 1639848679 (project SCI-20210212-CR-01, P.I. Riseley), that were published by \cite{Riseley2024_A2142}. We refer the reader to that paper for the calibration procedure, but in summary those data were processed using the Containerized Automated Radio Astronomy Calibration (\texttt{CARACal}) pipeline\footnote{\url{https://github.com/caracal-pipeline/caracal}} \citep{Jozsa2020,Jozsa2021} with full-polarisation calibration enabled. The calibrated datasets were then imaged in full polarisation with the same settings as for our S-band data.

As seen in Figure~\ref{fig:3c138_measurements}, the imaged L-band data from CBIDs 1639674087 and 1639848679 show consistent polarisation properties to the measurements published by \cite{Taylor2024_MeerKAT_polcals}. By extrapolating from these L-band measurements to the frequencies of our S-band data, we suggest that our S-band measurements follow a consistent trend. The linearly-polarised flux density measurements are perhaps slightly below the extrapolation of the L-band trend (and thus the polarisation fraction may be slightly lower than expected) but the polarisation angle is consistent. As such we are satisfied with the consistency and accuracy of our polarisation calibration, although further work is motivated to understand the polarisation properties of 3C\,138 --- particularly at L-band, as the polarisation angle appears to change rapidly toward longer wavelengths --- and to incorporate S-band calibration capabilities into \texttt{CARACal}.

\end{appendix}

\end{document}